\documentclass[a4paper,11pt]{article}
\usepackage{graphicx} 
\usepackage{subcaption}

\pdfoutput=1 

\usepackage{jcappub} 

\usepackage{hanging}
\usepackage{orcidlink}
\usepackage[T1]{fontenc} 
\usepackage{multirow} 

\newcommand{\bq}{\boldsymbol q}
\newcommand{\bx}{\boldsymbol x}
\newcommand{\bk}{\boldsymbol k}
\newcommand{\bPsi}{\boldsymbol{\Psi}}
\newcommand{\bDelta}{\boldsymbol{\Delta}}
\newcommand{\bs}{\boldsymbol s}

\newcommand{\btheta}{\boldsymbol \theta}
\newcommand{\bChi}{\boldsymbol{\chi}}
\newcommand{\ihmpc}{\,h{\rm Mpc}^{-1}}
\newcommand{\gpcih}{\,h^{-1}{\rm Gpc}}

\newcommand{\kmax}{k_{\rm max}}

\newcommand{\velocileptors}{\texttt{velocileptors}}

\title{An analysis of parameter compression and Full-Modeling techniques with Velocileptors for DESI 2024 and beyond}

\affiliation{Affiliations are in Appendix \ref{sec:affiliations}}
\emailAdd{mark.maus@berkeley.edu}

\author[1,2]{{M.~Maus}\orcidlink{0000-0002-9020-911X},}
\author[3]{{S.~Chen}\orcidlink{0000-0002-5762-6405},}
\author[1,2]{{M.~White}\orcidlink{0000-0001-9912-5070},}
\author[2]{{J.~Aguilar},}
\author[5]{{S.~Ahlen}\orcidlink{0000-0001-6098-7247},}
\author[6,7]{{A.~Aviles}\orcidlink{0000-0001-5998-3986},}
\author[8]{{S.~Brieden}\orcidlink{0000-0003-3896-9215},}
\author[9]{{D.~Brooks},}
\author[2]{{T.~Claybaugh},}
\author[10]{{S.~Cole}\orcidlink{0000-0002-5954-7903},}
\author[11]{{A.~de la Macorra}\orcidlink{0000-0002-1769-1640},}
\author[12]{{Arjun~Dey}\orcidlink{0000-0002-4928-4003},}
\author[9]{{P.~Doel},}
\author[2,13]{{S.~Ferraro}\orcidlink{0000-0003-4992-7854},}
\author[14]{{N.~Findlay}\orcidlink{0009-0007-0716-3477},}
\author[15,16]{{J.~E.~Forero-Romero}\orcidlink{0000-0002-2890-3725},}
\author[17,14,18]{{E.~Gaztañaga},}
\author[19,17,20]{{H.~Gil-Mar\'in}\orcidlink{0000-0003-0265-6217},}
\author[2]{{S.~Gontcho A Gontcho}\orcidlink{0000-0003-3142-233X},}
\author[21]{{C.~Hahn}\orcidlink{0000-0003-1197-0902},}
\author[22,23,24]{{K.~Honscheid},}
\author[25]{{C.~Howlett}\orcidlink{0000-0002-1081-9410},}
\author[26]{{M.~Ishak}\orcidlink{0000-0002-6024-466X},}
\author[12]{{S.~Juneau},}
\author[2]{{A.~Kremin}\orcidlink{0000-0001-6356-7424},}
\author[25]{{Y.~Lai},}
\author[2]{{M.~Landriau}\orcidlink{0000-0003-1838-8528},}
\author[2]{{M.~E.~Levi}\orcidlink{0000-0003-1887-1018},}
\author[27,28]{{M.~Manera}\orcidlink{0000-0003-4962-8934},}
\author[29,28]{{R.~Miquel},}
\author[30]{{E.~Mueller},}
\author[31]{{A.~D.~Myers},}
\author[14]{{S.~Nadathur}\orcidlink{0000-0001-9070-3102},}
\author[32]{{J.~Nie}\orcidlink{0000-0001-6590-8122},}
\author[7,11]{{H.~E.~Noriega}\orcidlink{0000-0002-3397-3998},}
\author[33,2]{{N.~Palanque-Delabrouille}\orcidlink{0000-0003-3188-784X},}
\author[34,35,36]{{W.~J.~Percival}\orcidlink{0000-0002-0644-5727},}
\author[2,37,13]{{C.~Poppett},}
\author[11]{{S.~Ramirez-Solano},}
\author[38]{{M.~Rezaie}\orcidlink{0000-0001-5589-7116},}
\author[39,33]{{A.~Rocher}\orcidlink{0000-0003-4349-6424},}
\author[40]{{G.~Rossi},}
\author[41]{{E.~Sanchez}\orcidlink{0000-0002-9646-8198},}
\author[2]{{D.~Schlegel},}
\author[42,43]{{M.~Schubnell},}
\author[44]{{H.~Seo}\orcidlink{0000-0002-6588-3508},}
\author[12]{{D.~Sprayberry},}
\author[43]{{G.~Tarl\'{e}}\orcidlink{0000-0003-1704-0781},}
\author[11]{{M.~Vargas-Maga\~na}\orcidlink{0000-0003-3841-1836},}
\author[12]{{B.~A.~Weaver},}
\author[45]{{S.~Yuan}\orcidlink{0000-0002-5992-7586},}
\author[46]{{P.~Zarrouk}\orcidlink{0000-0002-7305-9578},}
\author[34,36]{{H.~Zhang}\orcidlink{0000-0001-6847-5254},}
\author[2]{{R.~Zhou}\orcidlink{0000-0001-5381-4372},}
\author[32]{{H.~Zou}\orcidlink{0000-0002-6684-3997},}

\abstract{In anticipation of forthcoming data releases of current and future spectroscopic surveys, we present the validation tests and analysis of systematic effects within \texttt{velocileptors} modeling pipeline when fitting mock data from the \texttt{AbacusSummit} N-body simulations. We compare the constraints obtained from parameter compression methods to the direct fitting (Full-Modeling) approaches of modeling the galaxy power spectra, and show that the ShapeFit extension to the traditional template method is consistent with the Full-Modeling method within the standard $\Lambda$CDM parameter space. We show the dependence on scale cuts when fitting the different redshift bins using the ShapeFit and Full-Modeling methods. We test the ability to jointly fit data from multiple redshift bins as well as joint analysis of the pre-reconstruction power spectrum with the post-reconstruction BAO correlation function signal. We further demonstrate the behavior of the model when opening up the parameter space beyond $\Lambda$CDM and also when combining likelihoods with external datasets, namely the Planck CMB priors. Finally, we describe different parametrization options for the galaxy bias, counterterm, and stochastic parameters, and employ the halo model in order to physically motivate suitable priors that are necessary to ensure the stability of the perturbation theory.}

\begin{document}
\maketitle
\flushbottom

\section{Introduction}
\label{sec:intro}

The large-scale structure (LSS) of the Universe is the observed, coherent spatial distribution of material on scales larger than the typical galaxy or halo scale, and provides a powerful observational tool for probing cosmic evolution. LSS observations allow us to study 3D volumes of the sky that span a long range of cosmic times, enabling us to study the initial conditions of the primordial universe as well as its evolution at later times. \cite{Pee80,Peacock99,Dodelson03,Baumann22}.

One of the primary methods of measuring the evolution of LSS is through galaxy redshift surveys that aim to probe the clustering of matter on a wide range of scales using galaxies as tracers. Spectroscopic galaxy surveys have had significant success over the years in scanning large regions of the sky. These include the 2dF \cite{2dF2001}, 6dF \cite{6dF2009}, GAMA \cite{GAMA}, WiggleZ \cite{Drinkwater10}, and most recently the completed Sloan Digital Sky Survey (SDSS), composed of data from SDSS, SDSS-II \cite{SDSS2000}, BOSS \cite{SDSSIII,BOSS_DR12,SDSSIII2015}, and eBOSS \cite{eBOSS2020,eBOSS2021,SDSSIV2020}. 
The next telescope surveys to further push the boundaries of LSS observations that have recently begun operations are the Euclid Satellite \cite{Euclid,Amendola18} and the ground-based Dark Energy Spectroscopic Instrument (DESI) \cite{DESI2016a.Science,DESI2016_2,DESI2022.KP1.Instr}. DESI  aims to cover over 14,000 deg$^2$ by the end of 5 years of observations, with target samples of stars from the Milky Way Survey (MWS), bright galaxies from the Bright Galaxy Survey (BGS, $0.0 < z < 0.4$), Luminous Red Galaxies (LRG, $0.4<z<1.1$), Emission Line Galaxies (ELG, $1.1<z<1.6$), and Quasars (QSO, $1.6 < z < 2.1$). Altogether the DESI survey will span an effective volume of about $20\,(h^{-1}\mathrm{Gpc})^3$ by the end of its 5 years of observation \cite{DESI2023}.

In anticipation of the upcoming Year-1 data release of DESI \cite{DESI2023b.KP1.EDR,DESI2024.I.DR1,DESI2024.II.KP3,DESI2024.III.KP4,DESI2024.V.KP5,DESI2024.IV.KP6,DESI2024.VI.KP7A,DESI2024.VII.KP7B,DESI2024.VIII.KP7C} (as well as later releases along with Euclid), it is important to characterize the performance of the current state-of-the-art models for analyzing the observed galaxy clustering 2-point statistics and the resultant cosmological constraints. The growth of large-scale structure is a competition between gravity, the dominant force on large scales, and the expansion of the universe.  Models must also include several other effects: First, galaxies are not perfect tracers of the underlying matter overdensity field, and thus a `biasing' scheme is needed in order to relate the matter power spectrum to the observed galaxy spectrum (see ref.~\cite{Des18} for a recent review). Second, since distances along the line-of-sight (LOS) are inferred from redshifts, components of galaxy peculiar velocities in the LOS direction influence the inferred distances and are a source of anisotropy in the observed clustering signal \cite{Kaiser87,Hamilton92}. This latter effect is known as redshift space distortions (RSD) and provides both a challenge to modeling while also giving direct access to information about the growth rate of LSS. Finally, nonlinear effects on small scales must be included.  We use perturbation theory to model the mildly non-linear regime, with additional parameters to account for the small-scale physics such that the models are not sensitive to the complicated processes e.g.\ involved with galaxy formation (sometimes known as Effective Field Theory or EFT terms \cite{CHS12,PorSenZal14,VlaWhiAvi15}). The model considered in this work, \texttt{velocileptors}\footnote{\href{https://github.com/sfschen/velocileptors/tree/2.0}{https://github.com/sfschen/velocileptors/tree/2.0}} \cite{Chen20,Chen21}, is one of the models that will will be used for analyzing the full-shape power spectra from the upcoming DESI survey data releases, the others being the Fourier space Eulerian PT codes \texttt{PyBird} \cite{,Pybird21,dAmico20,Colas20} and \texttt{FOLPS$\nu$} \cite{Noriega22} and the configuration space code \texttt{EFT-GSM} \cite{Ramirez23}. The purpose of this work is to characterize the performance of \texttt{velocileptors} and understand any systematic issues by comparing to a suite of simulated, or `mock', data. Similar tests are being performed with the other three models in addition to a comparison between models, and will be reported in companion publications\cite{KP5s3-Noriega,KP5s4-Lai,KP5s5-Ramirez,KP5s1-Maus}.  While \texttt{velocileptors} has been tested previously on simulations \cite{Chen21,Chen22,ChenVlahWhite19}, here we focus on DESI-like galaxies and redshift ranges, and also use the new AbacusSummit \cite{Maksimova21} suite of simulations produced for the DESI collaboration that is also used to test the other theory models.

Within the framework of the model, there are still various approaches to fitting data. One method, previously used by the BOSS and eBOSS collaborations, involved choosing a fiducial template for the linear power spectrum while compressing the observed power spectrum multipoles into three parameters: the amplitude of the redshift-space anisotropy $f\sigma_8$, and the two scaling parameters parallel and perpendicular to the line of sight, i.e.\ $\alpha_{\parallel}$ and $\alpha_{\perp}$.  This technique was meant to encode the intuition that, for currently popular cosmological models, primary CMB anisotropies fix the parameters determining the shape of the power spectrum but late-time effects such as non-trivial dark energy evolution or spatial curvature can affect the total growth and the distance-redshift relation.  These impacts are accounted for by the three parameters above and redshift surveys can constrain them well.  An extension to this standard ``template'' fit is to include another compressed ``ShapeFit'' parameter to allow a set of modifications to the shape of the linear power spectrum \cite{Briedan21}. The extra shape information of this method allows for tighter constraints on cosmological parameters when interpreting the compressed statistics in light of a given cosmological model without including CMB priors. This partially bridges the gap in constraining power between the traditional template fit and the direct fitting or ``Full-Modeling'' approach of directly varying the parameters of a specific cosmological model. In this paper we compare these three methods under a variety of conditions in order to better understand the advantages and disadvantages of the methods. A comparison of the template and Full-Modeling approaches was investigated in ref.~\cite{Maus23} on the BOSS DR12 dataset, specifically focusing on shifts in $f\sigma_8$ constraints between the two methods. Here we extend that analysis to include the ShapeFit method and compare the three methods for the range of different settings, parameterizations, and modeling choices.

This paper is organized as follows. We begin by describing the Abacus simulations in Sect.~\ref{sec: data} and give an overview of Lagrangian Perturbation Theory (LPT) and \texttt{velocileptors} in Sect.~\ref{sec: theory}. We describe the parameter compression and Full-Modeling fitting methods in more detail in Sect.~\ref{sec: fitmeth}. The results of our primary tests, namely the dependence on scale cuts, joint fitting of multiple redshift bins, post-reconstruction statistics, $w$CDM models, CMB priors, varying $n_s$, Lagrangian vs Eulerian (EPT) Perturbation Theory, and freeing $\sigma_8$ are presented in Sect.~\ref{sec: results}. We conclude the paper in Sect.~\ref{sec: conc}. We also provide a brief discussion of our method for analytic marginalization over the linear parameters in our model in Appendix~\ref{appendix: am} along with some further tests, namely the dependence of $\omega_b$ prior, inclusion of cubic bias, and inclusion of hexadecapole moment in Appendix~\ref{appendix: tests}. In Appendix~\ref{appendix: proj_priors} we discuss the issue of parameter projection effects and the dependence on priors within our model, a problem that also arises in many other areas of cosmology. We follow this up with a section dedicated to the halo model in Appendix~\ref{appendix: halo}, which allows us to estimate typical scales for stochastic parameters in our model and provide physical motivation for our prior choices. Appendix~\ref{appendix: emulator} explains our use of emulators based on Taylor series in order to speed up likelihood evaluations, and we show that they perform consistently with the direct theory predictions.

\section{Mock data}
\label{sec: data}

\begin{figure}
    \centering
    \resizebox{\textwidth}{!}{\includegraphics{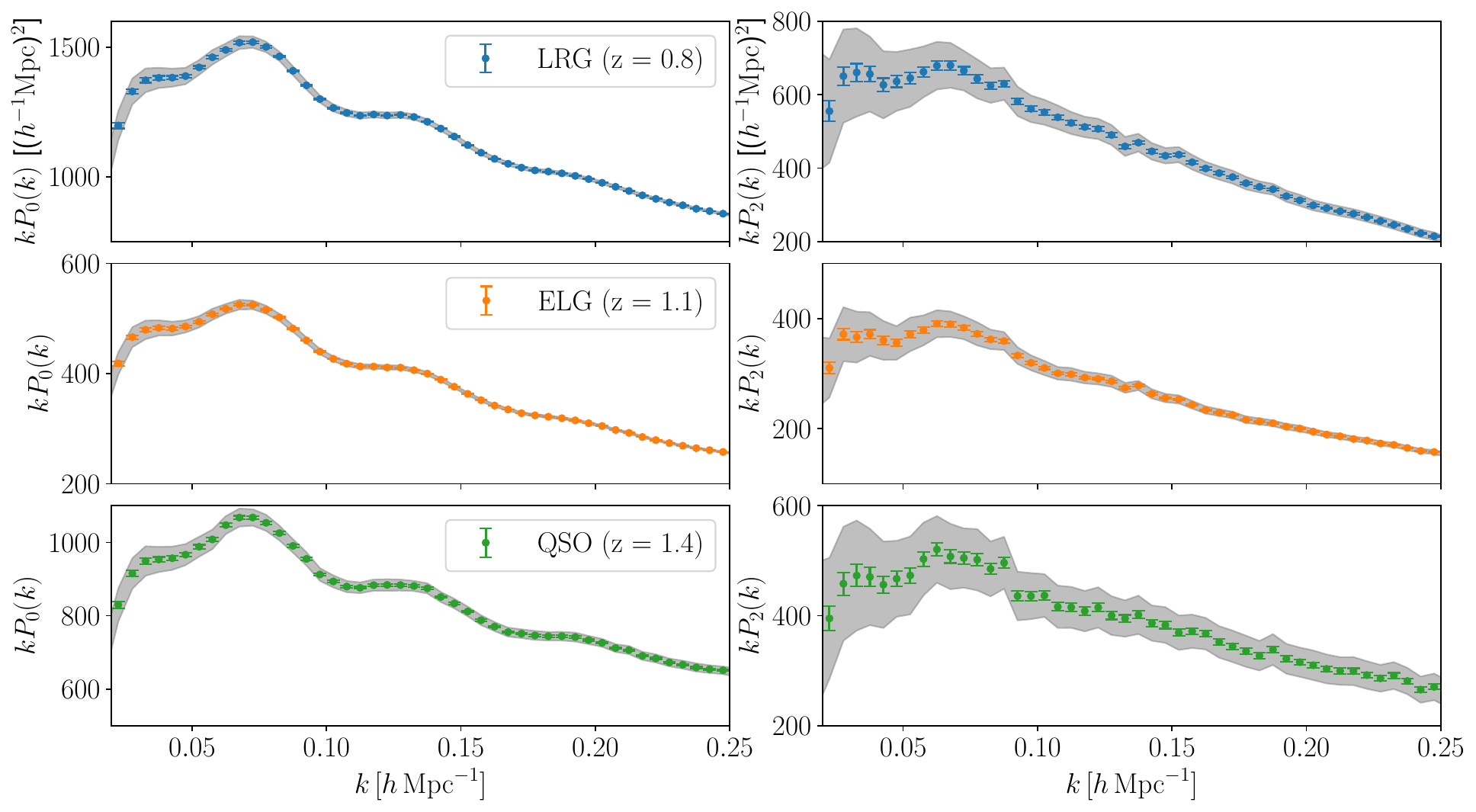}}
    \caption{Power spectrum monopole (left) and quadrupole (right) mock data for the LRG, ELG, and QSO tracers. For each tracer, the mean of the 25 N-body realizations is used. The error bars of the data correspond to the covariance re-scaled by the number of realizations, which represents a survey volume of 200 $h^{-3}$Gpc$^3$. The shaded regions show the error bars for a single cubic box, of volume $V = 8 \, h^{-3}$Gpc$^3$.}
\label{fig: Pkdata}
\end{figure}

To test our theory model we make use of the AbacusSummit \cite{Maksimova21} suite of N-body simulations in their native, cubic geometry.  These simulations were run with the \texttt{Abacus} \cite{Garrison21} N-body code on the Summit supercomputer at the Oak Ridge Leadership Computing Facility for use by the DESI collaboration. The simulations relevant to this work use a fixed cosmology\footnote{The Abacus fiducial cosmology has $h=0.6736$, $\omega_b = 0.02237$, $\omega_{\rm cdm} = 0.12$, $A_s = 2.0830\times 10^{-9}$, and $n_s= 0.9649$, with a corresponding BAO drag scale of $r_d = 99.08\, h^{-1}$Mpc}, with 25 boxes each with a different random number seed for the initial conditions run in a $(2\,h^{-1}\mathrm{Gpc})^3$ volume for a combined volume of 200$h^{-3}$Gpc$^3$. The mock galaxy catalogs have been produced for three types of tracers, each produced at a different redshift: Luminous Red Galaxies (LRGs) at $z=0.8$, Emission Line Galaxies (ELGs) at $z=1.1$, and Quasars (QSOs) at $z=1.4$.\footnote{The constraining power from a single redshift bin is similar to that expected for each tracer by year-5 of the DESI survey. While the real LRG data will actually be split into multiple redshift bins, the constraints from the joint analyses will be similar to those obtained from the single LRG bin in this work. We do not expect the conclusions in this paper to change significantly if the mocks had been produced in more redshift bins for each tracer. However, projection effects are expected to be more significant in extended models in Year-1 as the data is not as constraining yet as these mocks. This is discussed further in Appendix \ref{appendix: proj_desi}} For this study we ignore light-cone and evolution effects in order to better study the non-linear dynamics and biasing models. The RSD power spectrum data for each tracer is shown in Fig.~\ref{fig: Pkdata}.

The covariance we use for each tracer is calculated by Monte-Carlo from 1000 ``effective Zeldovich approximation'' (\texttt{EZmock} \cite{Chuang2015}) simulations of the same cosmology\footnote{Since these computationally efficient simulalions make use of the Zel'dovich approximation they may not be as accurate at small scales. As we will show later, our models are able to obtain unbiased constraints up to $k_{\rm max} = 0.2 \ihmpc$ but analytic covariances may be desirable in the future.}. We compute this covariance numerically via:
\begin{align}
    Cov[P(k)]_{ij} = \frac{1}{N-1} \sum_n^N [P_n(k_i) - \langle P(k_i) \rangle][P_n(k_j) - \langle P(k_j) \rangle]^\top
\end{align}
In principle, when using as data the mean of 25 cubic boxes the error bars of the data should also be re-scaled to reflect the increase in volume because $\sigma^2 \propto V^{-1}$. A proper treatment of the mean of 25 realizations would therefore involve re-scaling the covariance from the EZmocks by a factor of 1/25. However, we must be careful in interpreting results when the error bars of the data are so tight, as the ``survey volume'' of the simulations is orders of magnitude larger than any realistic survey will ever be able to achieve. For example, if we consider a future survey covering 18 000 deg$^2$ with tracers in a single redshift bin spanning $0.75<z<1.25$, then the comoving volume of that data would be about 24 ($h^{-1}$Gpc)$^3$, which is still much less than the 200 ($h^{-1}$Gpc)$^3$ volume of the simulations. The $8(h^{-1}\mathrm{Gpc})^3$ volume of a single box in our simulations is much closer to what we expect for any tracers/redshift bin by the end of five years of DESI observations.

The motivation for the large simulation volume is to detect systematic errors in the models relevant to the DESI Y5 data. If we define the detection of a systematic error as being larger than twice the statistical error $\sigma_{\rm sim}$ of the simulations and would like to keep systematic errors below some fraction $1/n$ of the Y5 data errors ($\sigma_{Y5}$), then this implies that we desire simulations with $\sigma_{\rm sim} \leq (2n)^{-1}\sigma_{\rm Y5}$. If $\sigma \propto 1/\sqrt{V}$, then for $n=3$ and a DESI Y5 volume of $~5 \, (h^{-1}\mathrm{Gpc})^3$, we would require a simulation volume of $~180 \, (h^{-1}\mathrm{Gpc})^3$. The Abacus simulations fulfill this requirement.  However the above argument fails to account for the systematic errors of the N-body simulations themselves. The fractional errors of the Abacus mock LRG monopole data with 25 box covariance (re-scaled by $1/25$) are roughly $0.15\%$ between $0.15<k<0.2 \, \ihmpc$. Ref.~\cite{Grove22} compared different cosmological N-body codes and found that RSD power spectra multipoles differed by $\approx 0.5\%$ in the same $k-$range, i.e.\ the simulations themselves do not agree to these levels of precision, even before uncertainties from initial condition generation, halo finding and additional physics are included \cite{Angulo22}. In addition to this, the large volume also reflects a level of precision that our models are not designed for, meaning that contributions from, e.g., two-loop terms that we don't include in our theory can result in poor fits. For all of these reasons, we will primarily focus on results using the un-rescaled covariance of the more reasonable single-box volume in the analysis of this paper, while only commenting briefly on the 25 box covariance results when relevant. Finally, when computing the covariance from a finite number of simulations, one should in principle include corrections such as the Hartlap factor\cite{Hartlap07}, which depends on the number of bins in the data vector versus the number of independent mock data sets used. Given the large number of EZmock simulations that we use, this factor is close to 1 and we therefore do not observe any noticeable change in constraints when including the correction. We also do not observe any significant bias in constraints arising from the finite number of mocks and therefore neglect the Hartlap correction in our analyses.

\section{Theory and Model}
\label{sec: theory}

The \texttt{velocileptors} code is based on the Lagrangian Perturbation Theory (LPT) approach to large-scale structure.
This approach treats dark matter as collisionless particles whose mapping from initial (Lagrangian) positions, $\bq$, to their final observed coordinates, $\bx$ is given by $\bx = \bq + \bPsi(\bq)$, where $\bPsi(\bq)$ is the displacement field. The dynamical equation, based on Newtonian gravity in an expanding spacetime, $\ddot{\Psi}+\mathcal{H}\dot{\Psi}=-\nabla_{\bx}\Phi$, is perturbatively expanded and solved as $\bPsi = \bPsi^{(1)}+ \bPsi^{(2)}+ \bPsi^{(3)} + ... $. The observed galaxy overdensity is derived from number conservation, with the inclusion of a bias functional in the initial conditions, $F[\delta_0(\bq)]$, that relates the tracer overdensity field to the linear matter field in the form of a Taylor series  \cite{Chen20,Chen21}. In Fourier space, this results in  
\begin{align}
   &1+\delta_g(\bk) = \int d^3\bq\ F[\delta_0(\bq)]e^{-i\bk \cdot(\bq + \bPsi(\bq))} \nonumber \\
   &F[\delta_0(\bq)] = 1 +  b_1\delta_0 + \frac{1}{2}b_2(\delta_0(\bq)^2 - \left\langle\delta_0^2\right\rangle)+b_s(s_0^2(\bq) - \left\langle s_0^2\right\rangle) + b_3 \mathcal{O}_3(\bq),
\end{align}
where $s_0= (\partial_i\partial_j/\partial^2 - \delta_{ij}/3)\delta_0$ is the initial shear tensor. The Lagrangian biases $b_O$ describe the response of galaxy formation to large-scale perturbations and are the free parameters of the theory---absent a complete model of galaxy formation at small scales their values must be measured directly from large-scale observables like the power spectrum, though rough estimates for their sizes can be made through toy models like halo occupation distributions. At 1-loop order there is only one non-degenerate cubic bias contribution which we include schematically as $\mathcal{O}_3$. Note that the Lagrangian bias parameters here are not equivalent to the Eulerian ones (for example the standard linear bias is $b = 1 + b_1$) but equivalent under a set of linear transformations (see e.g. ref.~\cite{Chen21}). Throughout most of this paper we will set $b_3 = 0$ under the assumption that the cubic nonlinearities in galaxy clustering are consistent with those from dynamical contributions alone \cite{Abidi18}. We test this assumption in Appendix~\ref{appendix: tests}.

The modeling of observed galaxy clustering statistics is complicated by the peculiar velocities of the galaxies, whose line-of-sight components introduce anisotropies in the clustering signal, an effect known as Redshift Space Distortions (RSD). In LPT, the transformation into redshift space amounts to a boost along the LOS direction, $\hat{n}$ so that the redshift space displacement field is 
\begin{align}
    \bPsi_s = \bPsi + \dot{\bPsi} = \bPsi + \frac{\hat{n}(\textbf{v}\cdot\hat{n})}{\mathcal{H}},
\end{align}
where $\textbf{v}$ is the galaxy peculiar velocity and $\mathcal{H}$ is the conformal Hubble parameter. We can simplify this relation with the Einstein-deSitter Approximation (EdS), such that
\begin{align}
    \bPsi_s^{(n)} = \bPsi^{(n)} + nf(\hat{n}\cdot\bPsi^{(n)}),
\end{align}
where $f$ is the linear growth rate. This can be expressed as a rotation of the real space field via the matrix $R^{(n)} = \delta_{ij}+nf\hat{n_i}\hat{n_j}$ such that $\bPsi^{s,(n)}=R^{(n)}\bPsi^{(n)}$. Defining the pairwise displacement field in redshift space as $\Delta_s = \Psi_s(\bq_1) - \Psi_s(\bq_2)$, the redshift-space galaxy power spectrum can be obtained from the cumulant expansion of
\begin{align}
    P_{s,g}(\bk) = \int d^3\bq \left\langle e^{i\bk \cdot (\bq + \Delta_s)}F(\bq_1)F(\bq_2) \right\rangle_{\bq = \bq_1-\bq_2}.
    \label{eq: VPint} 
\end{align}

In order to accurately capture the effects of long-wavelength (IR) linear displacements on the power spectrum, particularly with respect to their smearing of the BAO, it is necessary to include their effects beyond 1-loop order in perturbation theory \cite{CLPT,SenZal15,Blas16,Vlah16}. This class of techniques is known in the literature as ``IR resummation'': in our scheme the linear piece, i.e.\ the $A_{ij}^{s,(11)}$ component of $A_{ij}^s = \left<\Delta^s_i\Delta^s_j\right>$, is split into long- and short- wavelength components, $A_{ij}^{s,\rm lin} = A_{ij}^{s,<} + A_{ij}^{s,>}$, with a cutoff scale $k_{\rm IR}$, and we keep the $A_{ij}^{s,<}$ piece exponentiated while expanding all other contributions to 1-loop order. Due to the matrix transformation between the real and redshift space displacements, $\bPsi^{s,(n)}=R^{(n)}\bPsi^{(n)}$, both velocities and displacements contribute to the resummed $A_{ij}^{s}$. The expression for the power spectrum becomes \cite{Chen21}
\begin{align}
P_{s,g}^{PT}(\bk) & =\int d^3 \bq\ e^{i \bk \cdot \bq} e^{-\frac{1}{2} k_i k_j A_{i j}^{s,<}}\left\{1-\frac{1}{2} k_i k_j A_{i j}^{s,>}+\frac{1}{8} k_i k_j k_k k_l A_{i j}^{s,>} A_{k l}^{s,>}\right. \nonumber\\
& -\frac{1}{2} k_i k_j A_{i j}^{s, \text { loop }}+\frac{i}{6} k_i k_j k_k W_{i j k}^s 
 +2 i b_1 k_i\left(1-\frac{1}{2} k_i k_j A_{i j}^{s,>}\right) U_i^s-b_1 k_i k_j A_{i j}^{s, 10} \nonumber\\
& +b_1^2\left(1-\frac{1}{2} k_i k_j A_{i j}^{s,>}\right) \xi_{\text {lin }}+i b_1^2 k_i U_i^{s, 11}-b_1^2 k_i k_j U_i^{s, \operatorname{lin}} U_j^{s, \operatorname{lin}} \nonumber\\
& +\frac{1}{2} b_2^2 \xi_{\text {lin }}^2+2 i b_1 b_2 \xi_{\operatorname{lin}} k_i U_i^{s, \operatorname{lin}}-b_2 k_i k_j U_i^{s, \operatorname{lin}} U_j^{s, \operatorname{lin}}+i b_2 k_i U_i^{s, 20} \nonumber\\
& \left.+b_s\left(-k_i k_j \Upsilon_{i j}^s+2 i k_i V_i^{s, 10}\right)+2 i k_i b_1 b_s V_i^{s, 12}+b_2 b_s \chi+b_s^2 \zeta+2 i b_3 k_i U_{b_3, i}^s+2 b_1 b_3 \theta+\ldots\right\}.
\label{eq: vel_pk}
\end{align}
The other correlators appearing above ($\xi$, $W$, $V$, $U$, etc.) are defined in \cite{CLPT,VlaCasWhi16,Chen20,Chen21}.

We account for the sensitivity to small scales by introducing counterterms with coefficients, $\alpha_n$, that multiply the tree-level power spectrum. These coefficients describe couplings with short-wavelength modes whose sizes are not directly specified by perturbation theory. While their exact values (or even signs) are not known, we can put reasonable priors on them based on the size of gravitational nonlinearities seen in N-body simulations and expected nonlocalities induced by galaxy formation and baryonic physics, all of which contribute additively to the $\alpha_n$. Equivalently, the expected contribution of these effects dictates the scales on which our perturbative model is valid. We therefore put Gaussian priors on each counterterm centered at zero with widths set such that their corrections are perturbative at our chosen $k_{\rm max}$. We similarly include stochastic contributions which we parametrize with SN$_0=R_h^3$, SN$_2 = R_h^3\sigma_2$, and SN$_4 = R_h^3\sigma_4$, where $R_h^3$ is the typical galaxy or halo formation scale and the $\sigma_n$ arise from correlations of stochastic modes in densities and velocities, (e.g.\ $\langle \delta v\rangle, \langle v^2\rangle$, etc.). These stochastic terms again account for the small-scale modes missing in perturbation theory, whose signs and exact values are unknown, but whose rough size can be estimated based on our understanding of the small-scale distribution and velocities of galaxies in halos (see \S\ref{sec: FM} and Appendix~\ref{appendix: halo} and also Ref.~\cite{Schmittfull2021}). These contributions are added to the 1-loop power spectrum, $P_{s,g}^{PT}(\bk)$, above to give our final LPT prediction
\begin{align}
    P_{s,g}(\bk) = P_{s,g}^{PT}(\bk) &+ (b+f\mu^2)(b\alpha_0 + f\alpha_2\mu^2 + f\alpha_4\mu^4)k^2P_{{\rm s}, b_1^2}(\bk) \nonumber\\
    &+ (\text{SN}_0 + \text{SN}_2k^2\mu^2+\text{SN}_4k^4\mu^4),
\label{eq: cterms}
\end{align}
where $P_{{\rm s}, b_1^2}$ is the term containing $b_1^2\xi_{\rm lin}$ in Eq.~\ref{eq: vel_pk} evaluated to linear order outside of the exponential. This parameterization of the counterterms differs slightly from previous works using \velocileptors.  While giving consistent results, it makes it easier to interpret the counterterms as ``fractional corrections'' to the linear theory multipoles and motivates our choice of prior width on these parameters. For example, a value of $\alpha_{n} = 12.5\,h^{-2}\mathrm{Mpc}^2$ corresponds to a $50 \%$ correction to the $n^{\rm th}$ moment at $k_{\rm max} = 0.20 \ihmpc$. We also note that even though this parameterization may appear to introduce new degeneracies within the counterterms, we find no significant change in constraints or increased projection effects.

In computing the observed power spectrum, we assume a fiducial cosmology to convert $\mathbf{\theta}$ and $z$ to 3D distances using the fiducial distance-redshift relation. We need to account for distortions in $P(k)$ between assumed and true coordinates, the ``Alcock-Paczynski (AP) effect'' \cite{Alcock79}, in our modeling. We do this by rescaling the theoretical power spectrum in true cosmological coordinates to the observed coordinates by:
\begin{align}
    P^{\rm obs}_s(\bk_{\rm obs}) = q_\perp^{-2} q_\parallel^{-1} P_s(\bk)
    \quad , \quad
    k^{\rm obs}_{\parallel, \perp} =  q_{\parallel, \perp}\  k_{\parallel, \perp},
    \quad
\end{align}
with the scaling parameters above are defined by\footnote{Previously in BOSS analyses(e.g. \cite{Chen22,Maus23}) we have used the notation $\alpha_{\parallel,\perp}, \tilde{\alpha}_{\parallel,\perp}$ in place of $q,\alpha_{\parallel,\perp}$ but in this paper we use the latter in order to be consistent with the conventions of other DESI papers.}:
\begin{align}
    q_\parallel = \frac{H^{\rm ref}(z)}{H(z)}
    \quad , \quad
    q_\perp = \frac{D_A(z)}{D^{\rm ref}_A(z)} \quad.
    \label{eq: AP_geo}
\end{align}
$D_A(z)$ is the comoving angular diameter distance and the ``ref'' superscript labels the values from the fiducial cosmology. 

Finally, we use a Legendre transformation to compute the predicted power spectrum as multipoles,
\begin{align}
    P_{\ell}(k_{\rm obs}) = \frac{(2\ell + 1)}{2} \int_{-1}^1 d\mu\ P(k,\mu_{\rm obs})\mathcal{L}_{\ell}(\mu)
\end{align}
where ${\mathcal{L}}_{\ell}(\mu)$ is the Legendre polynomial of order $\ell$.

\section{Fitting methods}
\label{sec: fitmeth}

\subsection{Standard template and ShapeFit}

The traditional parameter compression method used originally by the BOSS/eBOSS collaborations involves choosing a reference cosmology, $\boldsymbol{\Theta}^{\rm ref}$, and keeping the resultant linear power spectrum, and by extension, the dependence on early-universe physics, fixed. The ``compressed'' parameters being varied are then the amplitude, $f\sigma_{s8}$ and the distance scalings transverse and along the line-of-sight, $\alpha_{\perp},\alpha_{\parallel}$; all of which are only dependent on late-time dynamics.  The quantity $f\sigma_{s8}$, which controls the ratio of monopole-to-quadrupole amplitudes, is a product of the growth rate, $f\simeq \Omega_m^{0.55}$ and the total amplitude, $\sigma_{s8}$, at $R = s\cdot8\,h^{-1}$Mpc scales. Here $s = r_{\rm d}/r_{\rm d}^{\rm fid}$ with $r_{\rm d}$ being the BAO scale at the drag epoch. We will comment on the $s$ scaling further below. The two distance scaling parameters are defined by,
\begin{equation}
    \alpha_\parallel = \frac{H^{\rm ref}(z)}{H(z)}\left(\frac{r_{\rm d}^{\rm ref}}{r_{\rm d}}\right) = q_\parallel\left(\frac{r_{\rm d}^{\rm ref}}{r_{\rm d}}\right) = \frac{q_\parallel}{s}
    \quad , \quad
    \alpha_\perp = \frac{D_A(z)}{D^{\rm ref}_A(z)}\left(\frac{r_{\rm d}^{\rm ref}}{r_{\rm d}}\right)  = q_\perp\left(\frac{r_{\rm d}^{\rm ref}}{r_{\rm d}}\right) = \frac{q_\perp}{s} \quad,
\label{eqn:AP}
\end{equation}
 We highlight that these parameters used in the template fitting are different from the scaling parameters defined in eq.~\ref{eq: AP_geo} by a factor of $\left(r_{\rm d}^{\rm ref}/r_{\rm d}\right)$\footnote{Technically, this ``ref'' is not necessarily the same as the ``ref'' in the definitions of $q_{\parallel,\perp}$. The one in $\left(r_{\rm d}^{\rm ref}/r_{\rm d}\right)$ refers to the reference template used in the standard template and ShapeFit fits, whereas in $q_{\parallel,\perp}$ it refers to the fiducial cosmology assumed when converting angles and redshift coordinates to physical distances when measuring the power spectrum. However, in practice it is simplest to choose the same cosmology for the template as was used for measuring the power spectrum from the data, so this distinction is not important.}. This is because in the template method we assume that most information comes from the BAO feature, and thus we account for the fact that both changes in $r_d$ and $q_{\parallel,\perp}$ induce stretching in the observed BAO signal.\footnote{See discussion in Appendix C of ref.~\cite{Chen_BOSSrecon2022}, where however the pure AP parameters are referred to as $\alpha$ and the BAO-rescaled ones are called $\tilde{\alpha}$.} In contrast, with a fitting method in which the underlying cosmology is directly being varied (see next subsection), the changes to $r_d$ affecting the BAO signal are automatically included in the linear power spectrum which is self-consistently varied. We must also emphasize that by including the factors of $s$ in our $\alpha$ scaling parameters we are implicitly assuming distances in units of the BAO scale, which motivates our use of the notation $f\sigma_{s8}$. This subtlety is discussed in detail in \S~3 of Ref.~\cite{Briedan21}.

Despite sacrificing constraining power through the lack of sensitivity to the early universe (the shape of the transfer function is held fixed by the reference cosmology), this ``template'' fitting method was sufficient at a time when the tightest constraints on early-time physics came from the CMB and LSS data was too noisy for direct fitting methods to be feasible without significant priors from Planck. The advantages of the template fitting method include the model-independence that allows for mapping the compressed parameter constraints to a cosmological model of one's choosing. Furthermore, computing the linear power spectrum using a Boltzmann code such as \texttt{CLASS} or \texttt{CAMB} at every step of a Markov Chain Monte Carlo (MCMC) sampler, in addition to calculating nonlinear perturbation theory (PT) corrections, is computationally very expensive. Fixing the linear power spectrum avoids this step, allowing for a faster fitting procedure without needing to train an emulator. 

The ``ShapeFit'' method is an extension to the standard template-fit compression, and was conceived as a way to partially bridge the gap in constraining power between the standard template and direct/full modeling methods, while preserving some of the model-independence of the former technique~\cite{Briedan21}. This is achieved by allowing modifications to the shape of the linear power spectrum via a multiplicative factor, 
\begin{align}
    P^{\prime}_{\rm lin}(\bk) = P_{\rm lin}^{\rm ref}(\bk)\ \exp\left\{ \frac{m}{a}\tanh \left[a\ln\left(\frac{k}{k_p}\right) \right] + n\ln\left(\frac{k}{k_p}\right) \right\}, 
    \label{eq: plin_sf}
\end{align}
where $P_{\rm lin}^{\rm ref}(\bk)$ is the template power spectrum produced by \texttt{CLASS} and is fixed throughout the fit. The form of this scaling was an ansatz chosen to best replicate the effect of varying $\omega_b, \omega_m$, and $n_s$ on the shape of the power spectrum (logarithmic slope and small/large scale limits), which would otherwise be captured in the transfer function when running \texttt{CLASS}. The modified power spectrum $P^{\prime}_{\rm lin}(\bk)$ is what we provide to \texttt{velocileptors} to produce the full 1-loop prediction for a given $(f\sigma_8,\alpha_{\parallel},\alpha_{\perp},m)$. For simplicity we keep fixed the second shape parameter, $n=0$. Allowing this parameter to vary accounts variations of the template emulating a spectral index effect, which in this paper we do not consider.  Following the original ShapeFit paper \cite{Briedan21} we choose for $a$ and $k_p$ their proposed values, $a=0.6$ and $k_p=0.03\, h\,{\rm Mpc}^{-1}$. With this modification to the classic template analysis, ShapeFit is now able to capture more information from the early universe without sacrificing its model independence. As a drawback, the freedom given by the ShapeFit parametrization in the linear power spectrum may not be sufficient to reproduce the exact shape of the transfer function as modeled by the Direct/Full-Modeling Fit technique (see next subsection) when 1) the fiducial cosmology is very different from the true cosmology, and 2) when the statistical errors of the data are very small. In Ref.~\cite{KP5s3-Noriega} (Fig. 2) this effect is quantified for the power spectrum, as well as in an upcoming paper (Ref.~\cite{KP5s8-Findlay}, in prep) focused on DESI Y1 geometry. On another hand, this effect could also be important if the ShapeFit compression technique is applied to higher-order statistics, such as the bispectrum, but this has not been yet quantified, as it goes beyond the scope of this paper.

\subsection{Full modeling: $\Lambda$CDM and extensions}
\label{sec: FM}

The alternative modeling technique to parameter compression is a more conventional forward-modeling approach that involves directly varying the underlying parameters of a cosmological model and making a theoretical prediction for the observed quantities. While the $\Lambda$CDM model depends on six parameters: ($\omega_b$, $\omega_{cdm}$, $H_0$, $\log(10^{10}A_s)$, $M_{\nu}$ and $n_{\rm s}$), some of these parameters are not constrained by galaxy clustering analyses independently.  For these quantities we use priors derived from e.g.\ Big-Bang Nucleosynthesis (BBN) and/or CMB anisotropies. We initially fix the spectral tilt and neutrino mass to the Abacus fiducial values of ($n_s$,$M_\nu$) = (0.9649,0.06) -- though see Section~\ref{ssec: free_ns}. For the baryon abundance we adopt a narrow gaussian BBN prior of $\mathcal{N}[\mu=0.02237,\sigma=0.00037]$ \cite{Cooke2018} (though see discussion in Appendix \ref{appendix: tests}). 
Within these constraints, in this ``Full-Modeling'' approach the shape of the linear power spectrum is able to change at each step of the MCMC as the shape of the transfer function is dependent on the $\Lambda$CDM parameters being varied. If done directly, this method is more computationally expensive because the linear power spectrum must be calculated using a Boltzmann code such as \texttt{CLASS} or \texttt{CAMB} in addition to the \texttt{Velocileptors} PT corrections. However, through the use of an emulator we can efficiently and accurately approximate the predictions for a given set of $\Lambda$CDM parameters. Under the assumption that the predicted power spectrum multipoles are a smooth function of the underlying parameters when close to some reasonably chosen values, we can use an emulator based on a Taylor series expansion in the relevant parameter space \cite{Colas20,Chen22}.\footnote{In the event that the data require a significantly different parameter space the analysis can be iterated with the Taylor series recomputed closer to the best fit, assuming the data are sufficiently constraining.} We find that the emulator agrees well with the direct LPT prediction when going to fourth order in the Taylor expansion. After employing such an emulator both for the Full-Modeling and template/ShapeFit methods, the MCMC chains converge (Gelman-Rubin $|R-1|<0.01$) within roughly 1-2 hours\footnote{This is when using 8 parallel chains on a single node}. By analytically marginalizing of stochastic and counterterm contributions (see Appendix~\ref{appendix: am}), the MCMC converges in 5-10 minutes for all methods. Therefore, the improved computational efficiency of a compression is no longer relevant in our setup.

The advantage of the Full-Modeling approach is that it is sensitive to both the early-universe physics that determines the shape of the transfer function, as well as late-time dynamics/geometry. Parameters such as $\omega_b$, $\omega_{cdm}$, and $H_0$ affect both the early- and late- universe dynamics, and are thus expected to be more tightly constrained in the Full-Modeling approach, when compared to the methods employing a template that fixes the early-universe dependence. On the other hand, the Full-Modeling approach requires choosing a specific cosmological model from the start, and a new MCMC fit is needed for any other model being employed. The parameter compression methods, however, only require one fit, and afterwards the results can be reused and mapped to any model of choice, though the model of choice must be sufficiently close to the template cosmology unlike in the Full-Modeling approach which does not suffer from this requirement. 

We show in Table~\ref{tab: param_priors} the parameters and priors used for the Full-Modeling and ShapeFit methods. We show the priors on bias parameters for three parametrizations. The standard setting in this paper is the ``intermediate'' freedom case for which the cubic bias is fixed to zero while $(1+b_1)\sigma_8$, $b_2\sigma_8^2$, and $b_s\sigma_8^2$ are varied with Gaussian priors applied to the latter two. The other parameter choices are discussed in Appendix~\ref{appendix: tests}. We analytically marginalize over the parameters controlling the stochastic and counterterm contributions, and refer readers to Appendix~\ref{appendix: am} for further details and validation of this method.

Finally we remark that in order to make contact with earlier work, and in particular with our companion papers, we use $\log(10^{10}A_\mathrm{s})$ as the ``normalization'' of the power spectrum throughout.  This choice, being the normalization of the curvature power spectrum at $k=0.05\,\mathrm{Mpc}^{-1}$, is actually better motivated for CMB surveys than galaxy redshift surveys. Most of the constraining power of our data comes from quasi-linear scales and we better constrain the matter power spectrum than the curvature (or potential) power spectrum.  In this respect a better choice for normalization may be $\sigma_8$.  We will discuss constraints on $\sigma_8$ later. We also reiterate that the Full-modeling method does not require any re-scaling of distances by $s = r_{\rm d} / r^{\rm ref}_{\rm d}$, and therefore the amplitude being constrained here is $\sigma_8$ not $\sigma_{s8}$. 

\begin{table}[t!]
\centering
\begin{tabular}{c|c|ccc|c}
Full-Modeling & ShapeFit & \multicolumn{3}{c|}{Bias}  & Stoch/Counter \\ 
              &          & Min. F. & Int. F.* & Max. F.         &               \\ \hline \hline
H$_0$        & $f\sigma_8$ & \multicolumn{3}{c|}{$(1+b_1)\sigma_8$} & $\tilde\alpha_0$ \\
$\mathcal{U}[55,79]$ & $\mathcal{U}[0,2]$ & \multicolumn{3}{c|}{$\mathcal{U}[0.5,3.0]$} & $\mathcal{N}[0,12.5]$ \\
\hline
$\omega_{\mathrm{b}}$ &  $\alpha_{\parallel}$ & \multicolumn{3}{c|}{$b_2\sigma_8^2$} & $\tilde\alpha_2$ \\
$\mathcal{N}[0.02237,0.00037]$ & $\mathcal{U}[0.5,1.5]$ & $\mathcal{N}[0,5]$ & $\mathcal{N}[0,5]$ & $\mathcal{N}[0,5]$ & $\mathcal{N}[0,12.5]$ \\
\hline
$\omega_{\mathrm{cdm}}$ &  $\alpha_{\perp}$ & \multicolumn{3}{c|}{$b_s\sigma_8^2$} & SN${}_0$ \\
$\mathcal{U}[0.08,0.16]$ & $\mathcal{U}[0.5,1.5]$ & 0 & $\mathcal{N}[0,5]$ & $\mathcal{N}[0,5]$ & $\mathcal{N}[0,\mathcal{O}(1/\bar{n}_g)]$ \\
\hline
$\log(10^{10} A_\mathrm{s})$& $m$ & \multicolumn{3}{c|}{$b_3\sigma_8^3$} & SN${}_2$ \\
$\mathcal{U}[2.03,4.03]$ & $\mathcal{U}[-3.0,3.0]$& 0 & 0 & $\mathcal{N}[0,5]$ & $\mathcal{N}[0,\mathcal{O}(f_{\rm sat} \sigma_v^2/\bar{n}_g)]$ \\
\\
\end{tabular}
\caption{Velocileptors LPT priors on parameters used in the Full-Modeling ($\Lambda$CDM) and ShapeFit fitting methods. The $\Lambda$CDM model involves H$_0$, $\Omega_{\mathrm{b}}$, $\omega_{\mathrm{cdm}}$, $\log(10^{10} A_\mathrm{s})$ and all of the bias, stochastic, and counterterms. The ShapeFit method fits $f\sigma_8$, $\alpha_{\parallel}$, $\alpha_{\perp}$, $m$ as well as the same bias, stochastic and counterterms.  The entries  $\mathcal{U}[{\rm min,max}]$ and $\mathcal{N}[\mu,\sigma]$ refer to uniform and Gaussian normal distributions, respectively. For the bias terms we show both minimal, intermediate (standard), and maximal freedom cases, defined in Appendix~\ref{appendix: tests}. For the two counterterms we report the priors within the parameterization for which the counterterms scale relative to the linear theory multipoles.  The priors on the stochastic terms are given in Table \ref{tab: stoch_priors} and discussed in the text.}    
\label{tab: param_priors} 
\end{table}

\begin{table}
\centering
\begin{tabular}{c|c|c|c|c|c|c|c|c}
Tracer &$z_{\rm eff}$& $1/\bar{n}_g$ & $f_{\rm sat}$ & $\log_{10}\bar{M}_h$  & $\sigma_v^{\rm est.}$ &SN$_0$ & SN$_2$ & SN$_4$ \\
\hline
LRG & 0.8 & 1000 & 0.1  & 13.3 & 7.8 & 2000 & $5.0\times10^4$ & $1.0\times10^6$ \\
ELG & 1.1 & 300 & 0.1  & 11.9 & 2.9 & 1000 & $2500$ & $2.5\times10^4$ \\
QSO & 1.4 & 8000 & 0.03  & 12.7 & 5.7 & $1.5\times10^4$ & $5.0\times10^4$ & $1.0\times10^6$
\end{tabular}
\caption{Relevant quantities used for the prior widths of stochastic parameters (see text). The typical halo mass, $\log_{10}\bar{M}_h$, per galaxy is expressed in units of $h^{-1}M_\odot$ and $1/\bar{n}_g$ is expressed in $h^{-3}\,\mathrm{Mpc}^3$. The characteristic velocities, $\sigma_v^{\rm est.}$ are in $h^{-1}$Mpc. Motivated by these numbers, the last three cloumns show the widths of the Gaussian priors (centered on 0 and in $h^{-1}$Mpc units) that are used in this paper for each stochastic parameter within each redshift bin.  The results do not depend upon the precise values chosen.}
\label{tab: stoch_priors}
\end{table}

\subsection{Cosmological inference from compressed statistics}

In order to interpret the ShapeFit and standard template results, we must do so in the context of a chosen cosmological model such as $\Lambda$CDM.
While it is simple to take a set of $\Lambda$CDM parameters and compute the distances, $H(z)$, $D_{\rm A}(z)$, and $r_{\rm d}$ using \texttt{CLASS} or \texttt{CAMB}, in order to compute compressed parameters assuming a certain fiducial cosmology, it is more tricky in reverse \cite{Briedan21}. Instead we must fit $\Lambda$CDM parameters to the results of a fixed template fit with another MCMC. We take the chains in the compressed parameters that were obtained from the initial template fits, and compute the parameter mean vector and covariance matrix, i.e. $\bar{\bf \Theta} = (\bar{f\sigma_8}, \bar{\alpha_{\parallel}},\bar{\alpha_{\perp}},\bar{m})$ and \textbf{C}$_{4 \times 4}$. Treating $\bar{\bf \Theta}$ and \textbf{C}$_{4 \times 4}$ as a ``data'' vector and associated covariance, we can now sample in $\Lambda$CDM parameters so that for each proposed set of ($\omega_b$, $\omega_{cdm}$, $h$, $\log A_s$) we compute the corresponding vector $\boldsymbol{\Theta_{\rm thy}} = (f\sigma_8, \alpha_{\parallel},\alpha_{\perp},m)_{\rm thy}$. Assuming all compressed parameters are Gaussian, we then use an MCMC to sample from the likelihood, 
\begin{equation}
    \mathcal{L} \propto \exp \left\{-\frac{1}{2}(\boldsymbol{{\Theta}_{\rm thy}} - \boldsymbol{\bar{\Theta}})^T \mathbf{C}_{4 \times 4}^{-1} (\boldsymbol{{\Theta}_{\rm thy}} - \boldsymbol{\bar{\Theta}})\right\}.
\end{equation}

When inferring cosmological constraints from the ShapeFit parameters, care must be taken in interpreting the amplitude $f\sigma_{s8}$ appropriately, as the slope rescaling via the $m$ parameter also changes $\sigma_{s8}$. As noted in refs.~\cite{Briedan21,Brieden23}, the parameter $f$ that is varied in ShapeFit analyses is actually $fA \equiv f(A_{sp}/A_{sp}^{\rm ref})^{1/2}$, where  $A_{sp} = s^{-3} P_{\rm no-wiggle}^{\rm lin}(k_p/s, \boldsymbol{\Theta})$ is the amplitude of the no-wiggle power spectrum at the pivot scale, $k_p \simeq 0.03 h$Mpc$^{-1}$. The parameter $s$ describes the scaling of lengths relative to the BAO and is defined to be the ratio $r_{\rm d} / r^{\rm ref}_{\rm d}$. In order to generate the model 1-loop power spectrum multipoles, we must provide \texttt{velocileptors} with the linear power spectrum $P^{\prime}_{\rm lin}(\bk)$ from Eq.~\ref{eq: plin_sf} and the growth factor $f$. Defining \texttt{LPT\_RSD} as the function that produces 
the power spectrum multipoles, the nearly exact degeneracy between $f$ and the power spectrum amplitude (see \S~\ref{sec: fsig8_degen}) implies that
\begin{align}
    \texttt{LPT\_RSD}\left[f\times\left(\frac{A_{sp}}{A_{sp}^{\rm ref}}\right)^{1/2};
    P^{\prime}_{\rm lin}(\bk)\right] \leftrightarrow \texttt{LPT\_RSD}\left[f; \left(\frac{A_{sp}}{A_{sp}^{\rm ref}}\right)\times P^{\prime}_{\rm lin}(\bk)\right],
\end{align}
and thus the true $f\sigma_{s8}$ is given by
\begin{align}
    f\sigma_{s8} &= f\times  \left[\left(\frac{A_{sp}}{A_{sp}^{\rm ref}}\right) \int \frac{dk}{2\pi^2}k^2 \tilde W_R^2(kR)P^{\prime}_{\rm lin}(\bk)\right]^{1/2} \nonumber \\
    &= f\times\left(\frac{A_{sp}}{A_{sp}^{\rm ref}}\right)^{1/2} \left[\int \frac{dk}{2\pi^2}k^2 \tilde W_R^2(kR)P_{\rm lin}(\bk)\exp\left\{ \frac{m}{a}\tanh \left[a\ln\left(\frac{k}{k_p}\right) \right] \right\} \right]^{1/2} \\
    &\simeq \frac{(f\sigma_{s8})^{\rm ref}}{(fA_{sp}^{1/2})^{\rm ref}} fA_{sp}^{1/2} \times \exp\left\{\frac{m}{2 a}\tanh\left[ a \ln\left(\frac{r_d^{\rm fid}}{\rm R}\right)\right]  \right\}.
\end{align}
Here $R$ is the smoothing scale of the amplitude parameter $\sigma_{R}$ and is chosen to be $R=8\,h^{-1}$Mpc by convention. There are now two ways in which one could use ShapeFit chains in order to infer about cosmological parameters: one can use the above equations (either the exact or approximate forms) to transform the sampled $fA$ chain into $f\sigma_{s8}$, and then use \texttt{CLASS} to compute $f\sigma_{s8}$ for every set of $\Lambda$CDM parameters at the interpretation step; or one can directly perform the interpretation on $fA$ by always computing $f$ and $A_{sp}$ while sampling in $\Lambda$CDM parameters. We find that the two approaches give consistent constraints in the $\Lambda$CDM parameter space.

Finally, the $m$ parameter in ShapeFit that controls the shape of the linear power spectrum can be computed from $\Lambda$CDM parameters through the ratio \cite{Briedan21}
\begin{equation}
    m = \left.\frac{d}{dk} \left( \ln \left[  \frac{T(k_p/s,\boldsymbol{\Theta})}{T(k_p,\boldsymbol{\Theta}^{\rm ref})} \right] \right) \right\vert_{k=k_p} \quad , \quad
    T(\boldsymbol{\Theta},k) = \frac{P_{\rm no-wiggle}^{\rm lin}(k,\boldsymbol{\Theta})}{\mathcal{P}_{\mathcal{R}}(k,\boldsymbol{\Theta})},
\end{equation}
with primordial power spectrum $\mathcal{P}_{\mathcal{R}}$.

\section{Results}
\label{sec: results}

\begin{figure}
    \centering
    \resizebox{\textwidth}{!}{\includegraphics{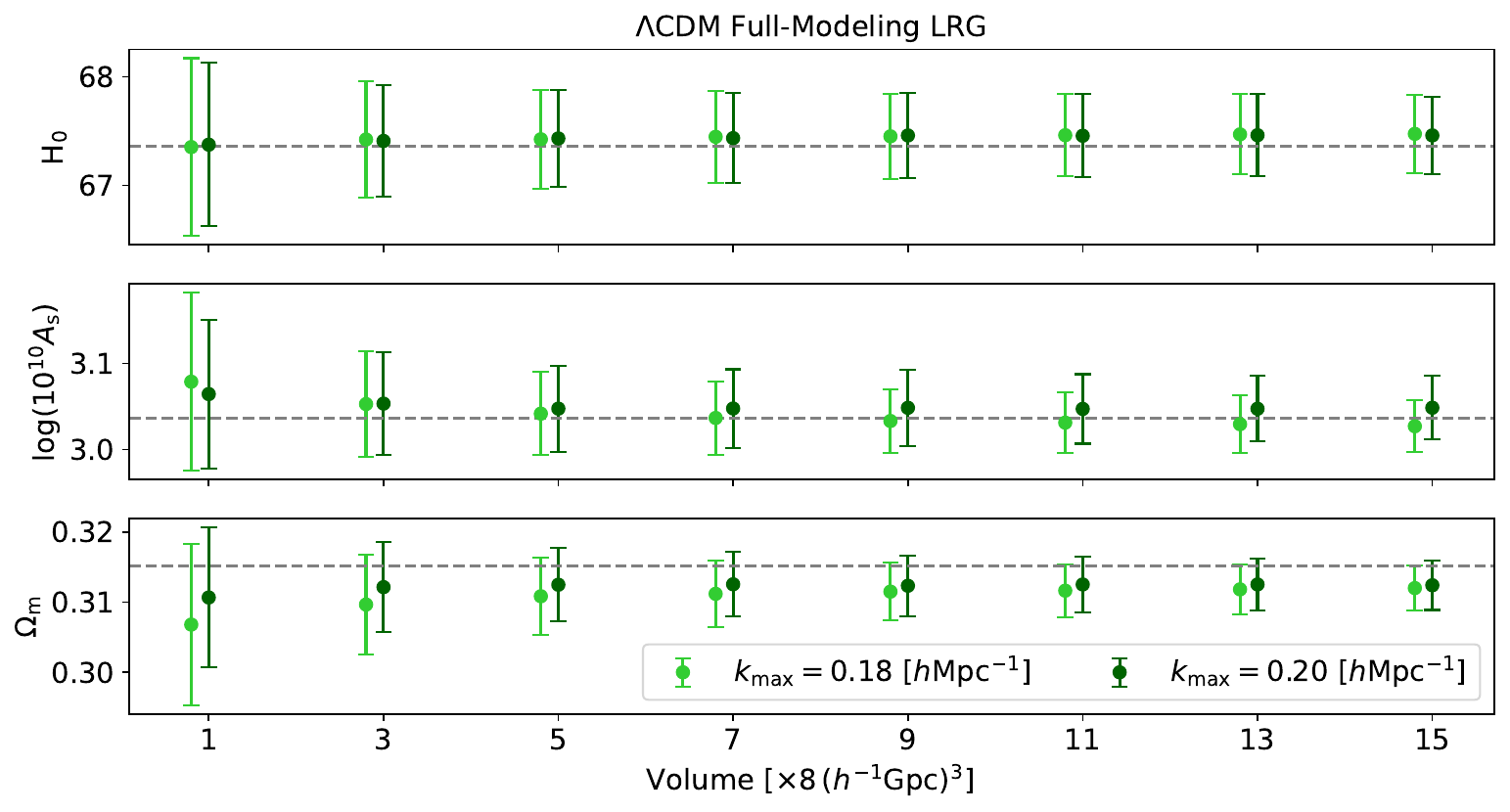}}
    \caption{1D posteriors from the Full-Modeling fit as the covariance volume is varied from that of a single box (8 $[\gpcih]^3$) to 15 boxes ((120 $[\gpcih)^3$])}
\label{fig: FM_vol}
\end{figure}

Before we present the results from the various systematic tests of \texttt{velocileptors} and the different modeling methods, we first revisit the issue of covariance volume. In Fig.~\ref{fig: FM_vol} we present 1D posterior constraints from the Full-Modeling fit to LRG mock data as a function of covariance volume, i.e.\ multiples of the single-box volume such that the covariance is rescaled by $1/n, \, n=1, 3, 5, \cdots, 15$. We show results for fits using two different $k$-ranges, $0.02\leq k\,[\ihmpc]\leq0.18$ and $0.02\leq k\,[\ihmpc]\leq0.20$ (which will be our `standard' range). We find that as the volume is increased, the constraints in  $\Omega_{\rm m}$ shift towards the truth as the error bars tighten, which is indicative of a prior volume effect. For $H_0$ and $\log(10^{10}A_s)$ the constraints remain mostly stable as the volume is increased, with small shifts increasing with volume that likely relate to the increasing sensitivity to two-loop effects that are not included in the model. For similar reasons, we observe a divergence in constraints between $k_{\rm max} =  0.18$ and $k_{\rm max} =  0.20 \ihmpc$ that grows as the volume is increased. This shows that when using an ultra-tight covariance such as that of the $200 \gpcih$ simulation volume, one can expect $1-2\sigma$ offsets in constraints arising purely from theoretical errors due to the limited number of terms included in the 1-loop power spectrum model. In addition, as mentioned earlier, the N-body simulations themselves have systematic errors that become important at these volumes and can contribute to the shifts we observe.

\subsection{Baseline Comparison}
\label{sec: Standard conditions}

\begin{figure}
\captionsetup[subfigure]{labelformat=empty}
\begin{subfigure}{.5\textwidth}
\centering
\includegraphics[height=8cm]{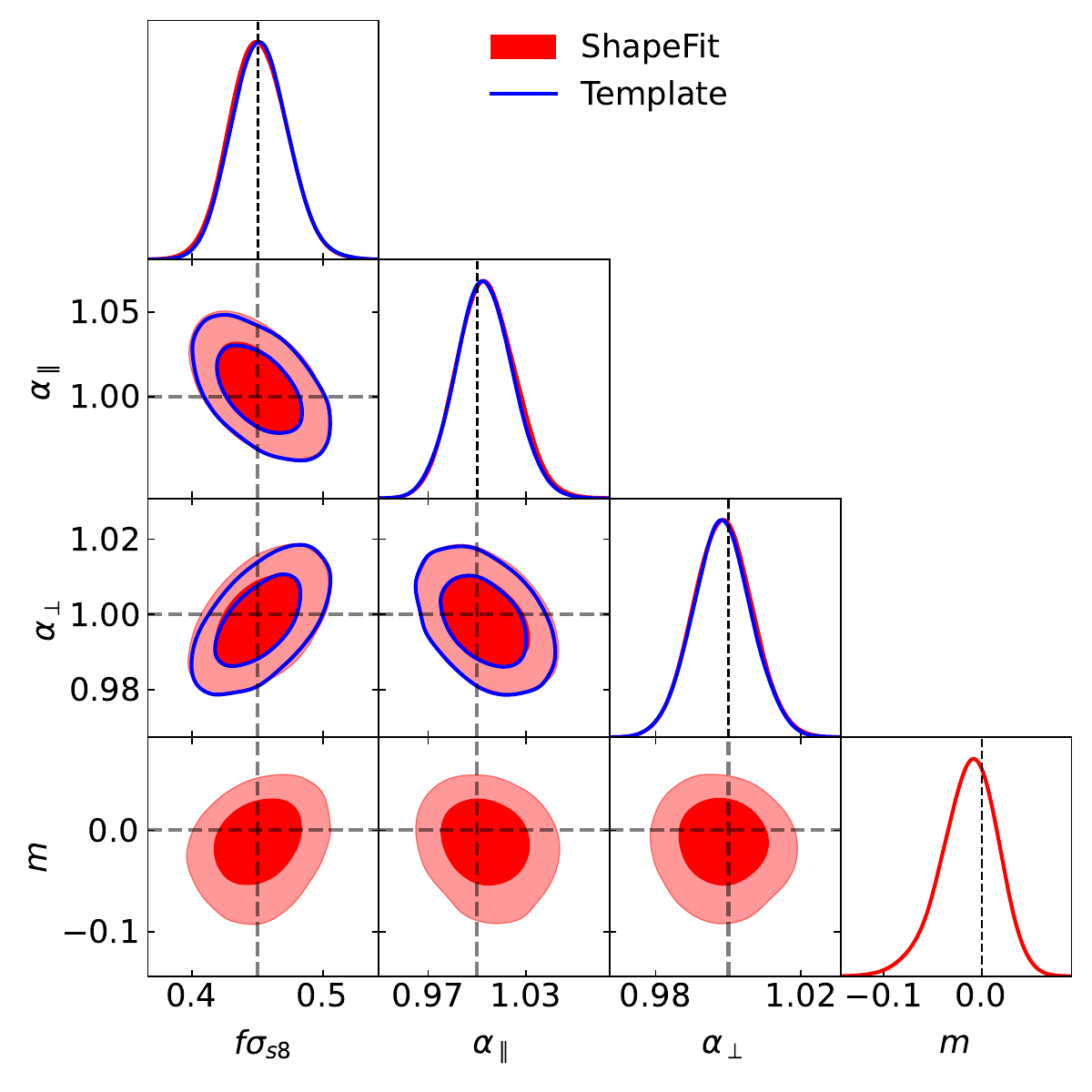}
\end{subfigure}%
\begin{subfigure}{.5\textwidth}
\centering
\includegraphics[height=8cm]{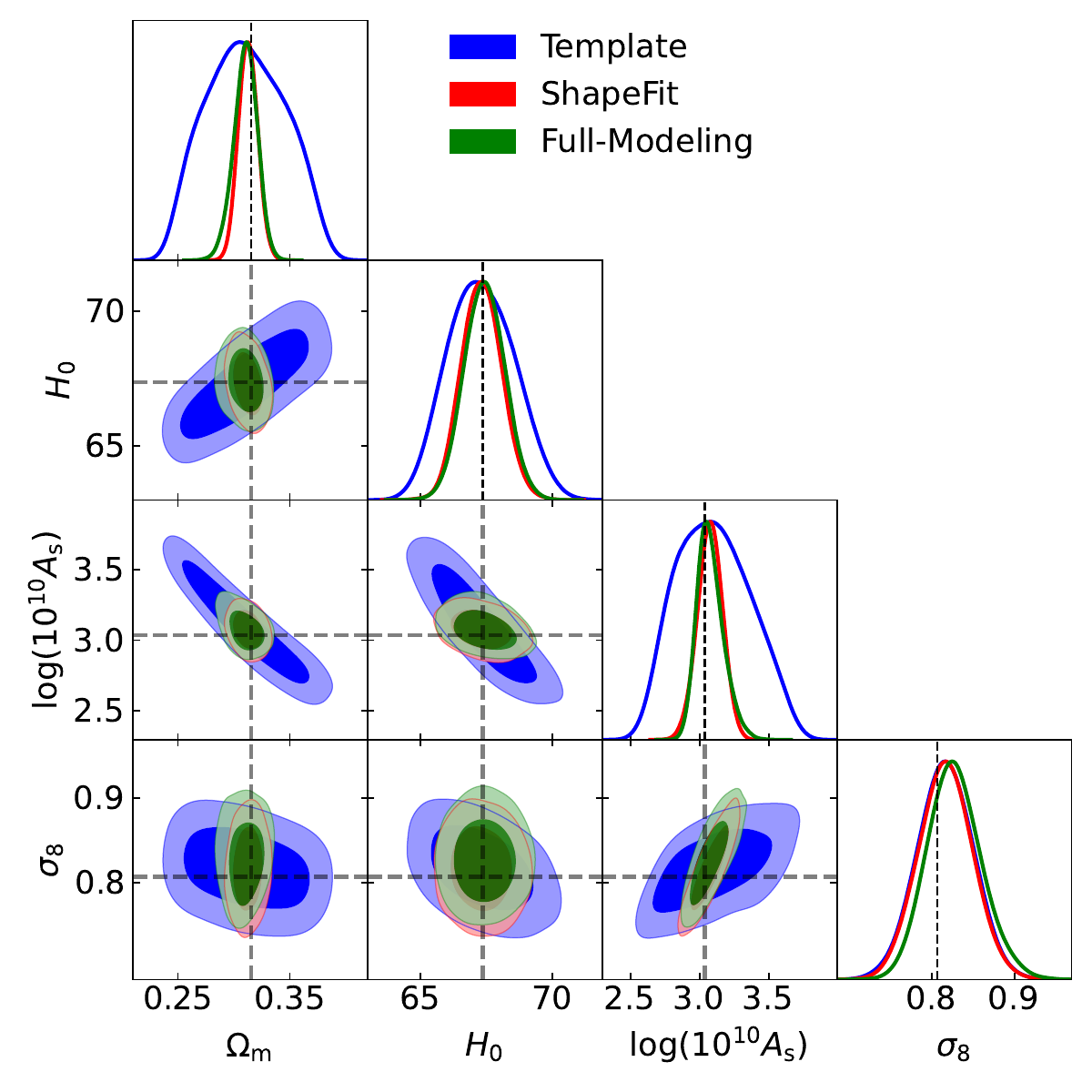}
\end{subfigure}%
\caption{\textit{(left):} Comparison of constraints of compressed parameters for the standard template method vs.\ ShapeFit. \textit{(right):} Comparison of constraints on $\Lambda$CDM parameters for the standard template, ShapeFit, and full modeling methods. 
The single-box covariance is used for these results, with our `standard' $k_{\rm max}=0.2\,h\,\mathrm{Mpc}^{-1}$ (see \S\ref{sec: kmax} for the discussion of $k_{\rm max}$ dependence).}
\label{fig: standard_both}
\end{figure}

We begin with comparing constraints in the compressed parameter space between the standard template and ShapeFit approaches, using the single-box covariance, as shown in the left panel of Fig.~\ref{fig: standard_both}. We see that the posterior means of the two methods agree very closely, with slightly smaller contours for the standard template due to varying fewer parameters. Since the reference template used in these fits is the true Abacus cosmology, we expect $\alpha_{\parallel,\perp}=1$ and $m=0$. In both cases, the means of all parameters are within 1$\sigma$ of the expected values. When interpreting these results in terms of a $\Lambda$CDM cosmology, however, we see a significant difference in the constraints from the two compression methods (right panel of Fig.~\ref{fig: standard_both}). While both methods give unbiased constraints on $\Lambda$CDM parameters (within 1$\sigma$ of truth) the error bars for all parameters are significantly larger for the template case due to the lack of information from the power spectrum shape in the template approach. This is expected, as the template method was traditionally combined with external data sets under the assumption that the parameters determining the shape are not as well constrained from LSS data than e.g. CMB anisotropies, but in our setup we rely purely on the LSS data alone(but see \S \ref{sec: planck priors}). Meanwhile, when comparing the constraints between the ShapeFit and Full-Modeling methods, we find a very close agreement in the shape and orientations of the contours, showing that the ShapeFit method is able to match the constraining power of direct model fitting, at least for the $\Lambda$CDM case for which it was designed. We do observe mild differences in the tightness of constraints between the ShapeFit and Full-Modeling methods. These could be due to a combination of various approximations in the ShapeFit method, such as controlling the shape of $P_{\rm lin}$ with only one parameter and assuming the compressed parameters to be perfectly Gaussian in the interpretation step.

\subsection{Dependence on $k_{\rm max}$}
\label{sec: kmax}

\begin{figure}
\centering
\resizebox{0.99\columnwidth}{!}{\includegraphics{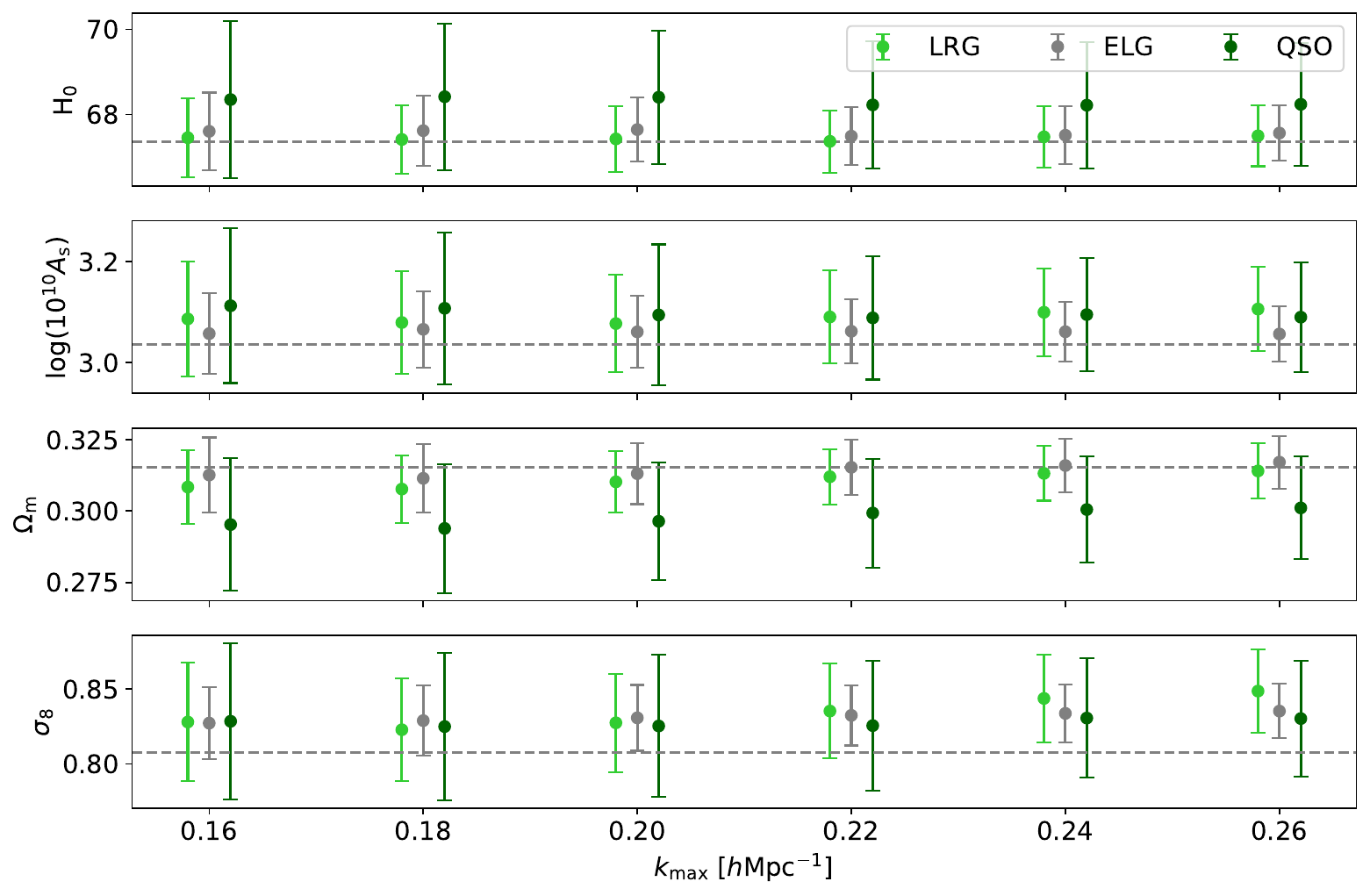}}
\resizebox{0.99\columnwidth}{!}{\includegraphics{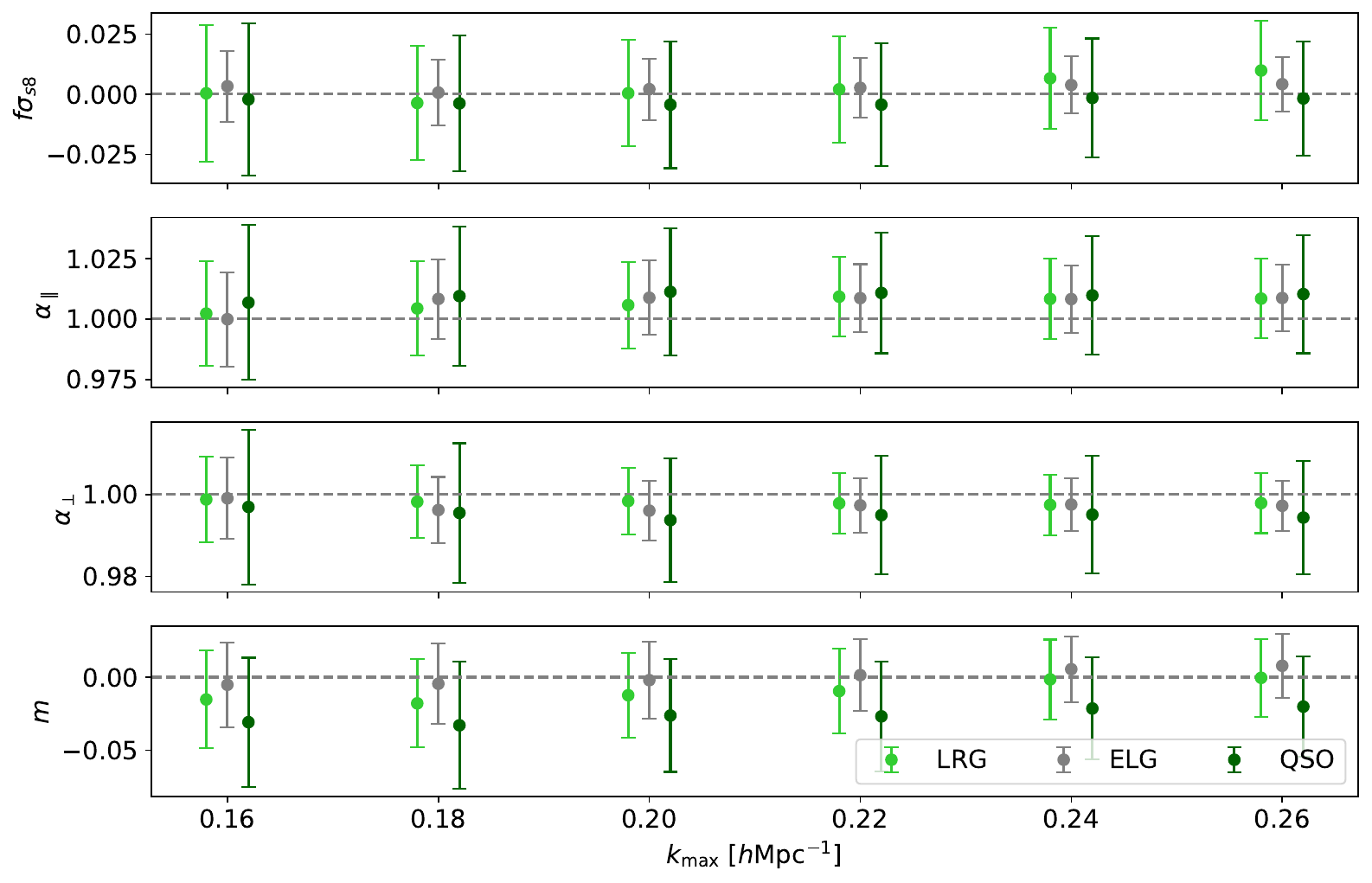}}
\caption{Dependence on $\kmax$ of the Full-Modeling (upper panel) and ShapeFit (lower panel) methods for the three tracer types: LRG ($z=0.8$). ELG ($z=1.1$), and QSO ($z=1.4$) (points slightly offset for clarity). The single-box covariance is used for all of these fits.
\label{fig: kmax_1d_alltracers}}
\end{figure}

We next test the dependence on scale cuts of our model, for the different methods. In all cases we fix the lower bound of the $k$-range to $0.02 h$Mpc$^{-1}$. This is fully in the linear regime so the the stability of the theory is not affected by the specific value chosen, but this choice simply removes points too close to the fundamental mode of the cubic box ($k=0.003 \ \ihmpc$).  We then run our fits with upper bounds of $k_{\rm max}=0.16 - 0.26 \ihmpc$. The results are shown for Full-Modeling and ShapeFit in Fig.~\ref{fig: kmax_1d_alltracers} for the LRG, ELG, and QSO tracers. The higher $k$-modes, above $\sim 0.2\,h\,\mathrm{Mpc}^{-1}$, correspond to smaller scales which are more sensitive to nonlinear effects and galaxy/halo formation physics, which are not well-understood and therefore difficult to model.  Our model includes non-linearities only at the 1-loop level and bias only up to cubic order. We therefore expect biases to worsen as higher $k$-modes are included in the fit. For the single-box volume we find the two methods to remain relatively stable as $k_{\rm max}$ is increased as the observational errors match or exceed the theoretical or modeling errors, however we do observe $\gtrsim 1\sigma$ offsets in the $\sigma_8$ constraints for LRG and ELG tracers in the Full-modeling method when $k_{\rm max} \geq 0.22 \ \ihmpc$. We additionally find that for the ELG sample we get more of an tightening of constraints in many parameters as $k_{\rm max}$ is increased than for the other samples. This could be due to the redshift coverage and higher number density of the mock ELG sample. 

\begin{figure}
\captionsetup[subfigure]{labelformat=empty}
\begin{subfigure}{1\textwidth}
\centering
\includegraphics[width=0.99\textwidth]{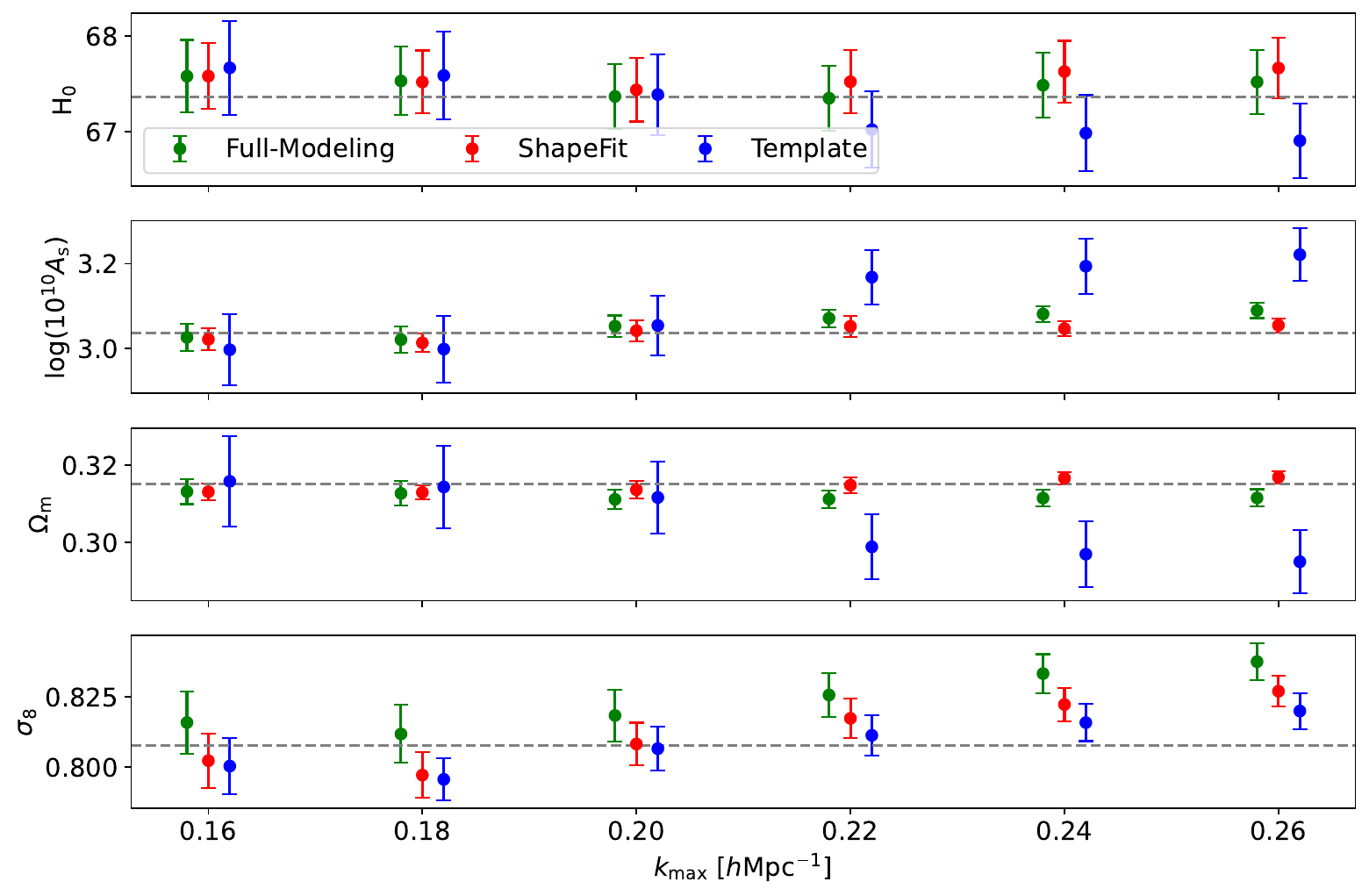}
\end{subfigure}%
\\
\begin{subfigure}{1\textwidth}
\centering
\includegraphics[width=0.99\textwidth]{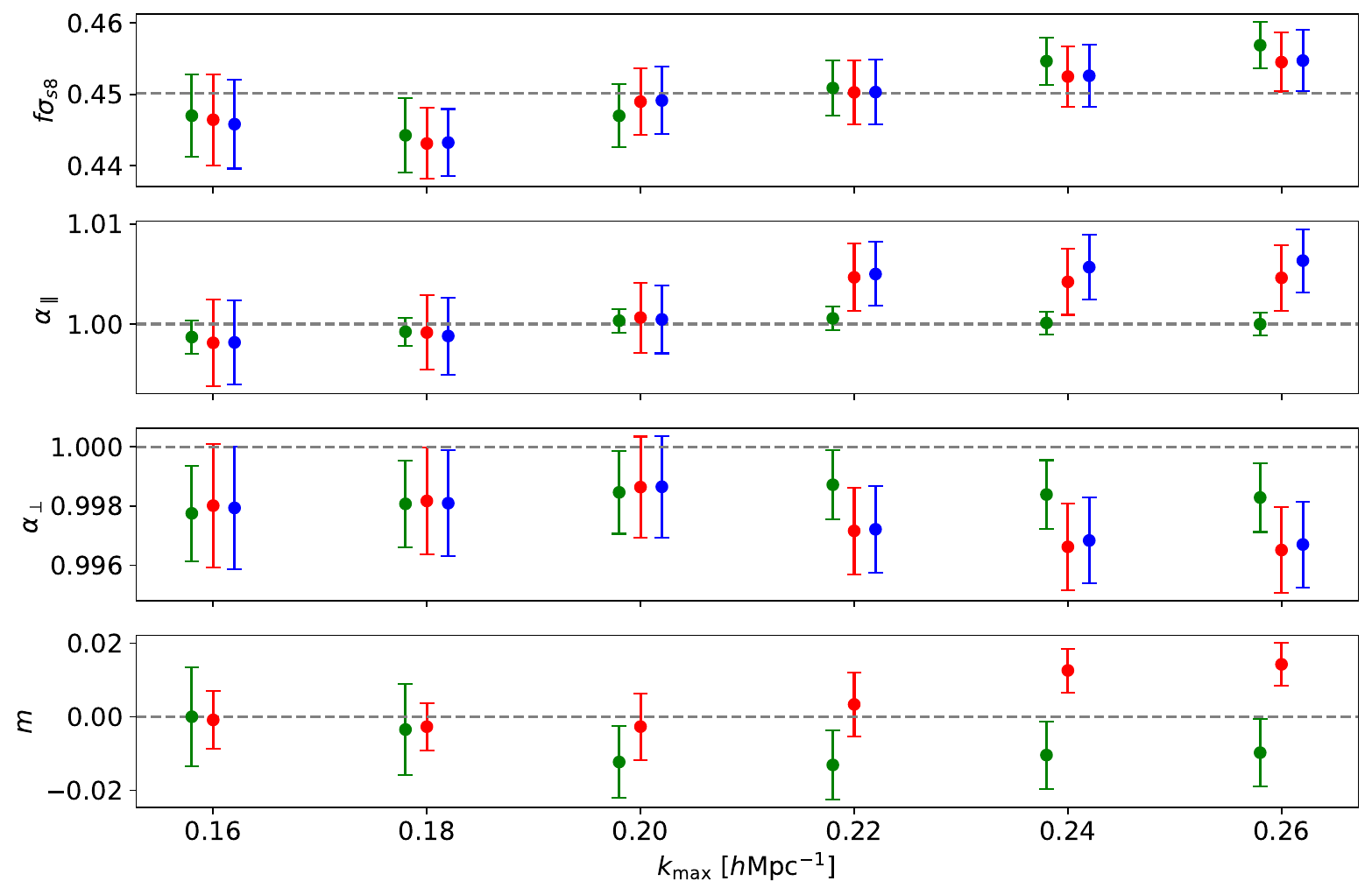}
\end{subfigure}%
\caption{Constraints on $\Lambda$CDM and compressed parameters from the three modeling methods with varying $k_{\rm max}$ between $0.16 \leq k_{\rm max} \leq 0.26 \ \ihmpc$(points slightly offset for clarity). These results are obtained with a covariance appropriate to the 25 box volume, fitting to the LRG cubic mock data.}
\label{fig: kmax_1d}
\end{figure}

In Fig.~\ref{fig: kmax_1d}  we repeat this test for the LRG tracers but using the 25 box covariance. We show constraints in the $\Lambda$CDM as well as ShapeFit parameter spaces. In this case we obtain significantly biased constraints when $k_{\rm max} > 0.2 \ihmpc$. In the $\Lambda$CDM parameters, we find a mild improvement in constraining power of Full-Modeling at k$_{\rm max}=0.18\ihmpc$ versus our usual setting of $k_{\rm max}=0.20\ihmpc$. This worsening of constraints when $k_{\rm max}$ is increased is likely due to a sensitivity to higher-order effects that our theory does not adequately describe, and which become increasingly important with increasing $k$. When using an extremely tight covariance, the additional high-$k$ points push the fit towards incorrect models and away from the constraints coming from low-$k$ data points. In the compressed parameter space we observe slightly more significant offsets ($\gtrsim 1.\sigma$) in the $\alpha_{\perp}$ and $f\sigma_{s8}$ constraints for ShapeFit at k$_{\rm max}=0.18 \ihmpc$. When deriving summary statistics from the Full-Modeling constraints, the $\alpha_{\parallel}$, and $\alpha_{\perp}$ parameters are significantly more tightly constrained than in the ShapeFit and Template methods because the $\Lambda$CDM priors in Full-Modeling restrict the allowable values that the scaling parameters can take \cite{Maus23}. We use the results from Figs.~\ref{fig: kmax_1d_alltracers}-\ref{fig: kmax_1d} to motivate a choice of $k_{\rm max}=0.20 \ \ihmpc$ as our baseline analysis setting, as this is the largest $k_{\rm max}$ for which all three modeling methods are acceptably close to truth ($\lesssim 1\sigma$ offsets) in the $\Lambda$CDM parameter space in both the single-box and full covariance volume cases.

\begin{figure}
    \centering
    \resizebox{\textwidth}{!}{\includegraphics{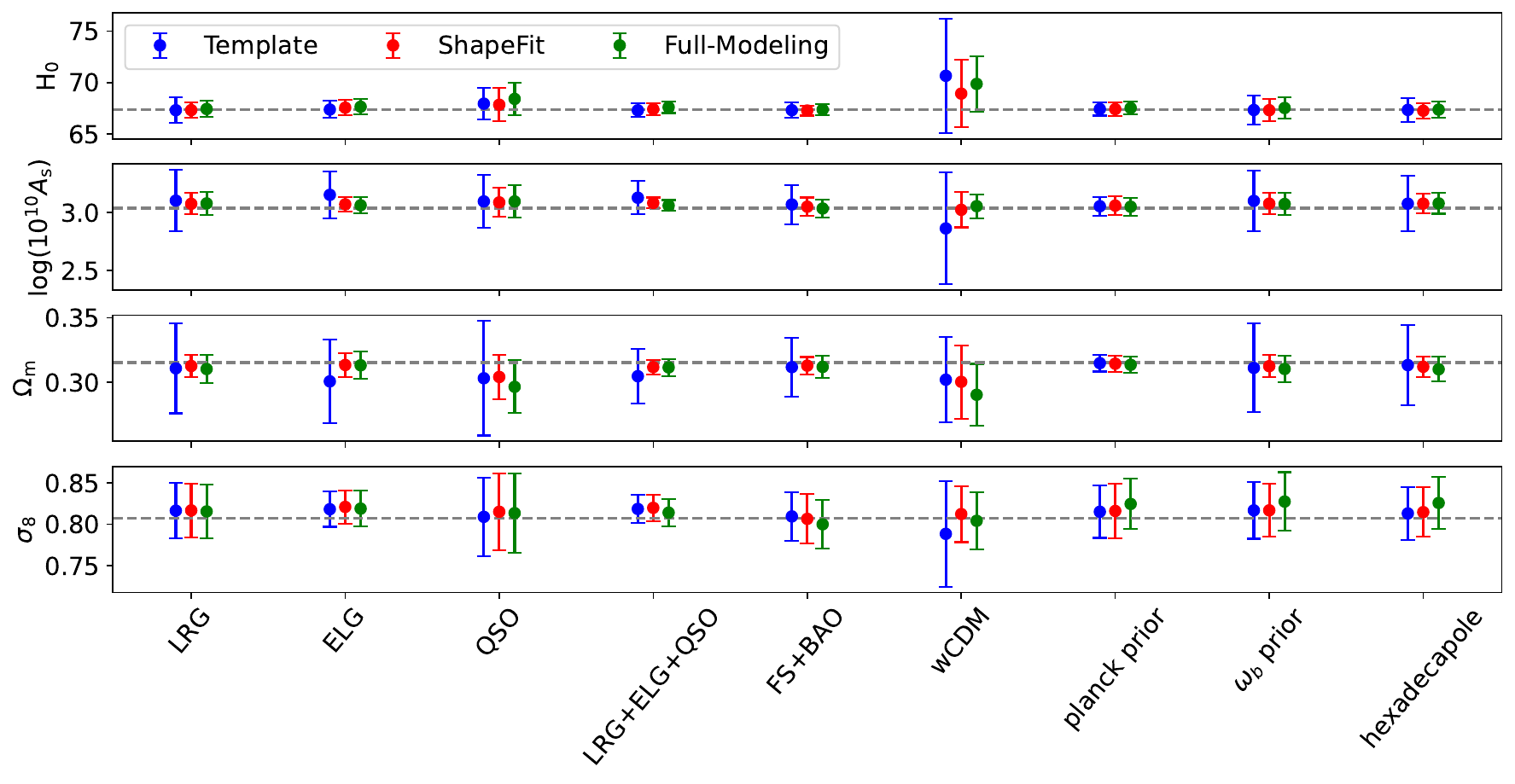}}
    \caption{Constraints on $\Lambda$CDM parameters for the three modeling methods for a variety of different fit settings and data sets to be discussed in the text. The results are obtained with the covariance for a single box, 8 $(\gpcih)^3$, volume and $k_{\rm max}=0.2\,h\,\mathrm{Mpc}^{-1}$. In many of the cases $\Omega_{\rm m}$ appears slightly below the truth, which is in part due to projection effects. Here ``standard'' refers to our baseline result on the LRG mocks. 
    }
\label{fig: sum_tests}
\end{figure}

As we proceed to the remainder of tests presented in this paper, we refer readers to Fig.~\ref{fig: sum_tests} for a summary figure of 1D constraints on $\Omega_{\rm m}, H_0,$ and $\log 10^{10}A_s$ obtained from each of the tests.

\subsection{Joint fitting of LRG, ELG, and QSO mocks}

We now turn to the joint fitting of data samples from different tracers and redshift bins. The three tracers are Luminous Red Galaxies (LRG, $z=0.8$), Emission Line Galaxies (ELG, $z=1.1$), and Quasars (QSO, $z=1.4$). For the Full-Modeling case, we still sample in $\Lambda$CDM parameters as usual but compute separate $P_{\ell}(k)$ models for each redshift bin and the likelihood is computed from all data sets, i.e. the data vector becomes $\boldsymbol{d} = (P_0^{LRG},P_2^{LRG},P_0^{ELG},P_2^{ELG}, P_0^{QSO},P_2^{QSO})$. This results in a total effective volume of 600 $(\gpcih)^3$. We do not assume any correlation between tracers at different redshifts\footnote{The mean data vectors for the LRG, ELG, and QSO tracers actually came from the same 25 realizations and therefore share initial conditions. In principle this means that the redshift bins are not truly uncorrelated, but we assume so in this work for simplicity.}, so the total joint covariance matrix has zeros in the indices corresponding to cross correlations between different tracers. This ensures that contributions to the log-likelihood such as $\Delta P_i^{LRG} C_{ij}^{-1}\Delta P_j^{ELG} = 0$. We use a separate set of nuisance parameters for each type of tracer. 
For the standard template and ShapeFit fits, the free parameters ($f\sigma_{s8}$,$\alpha_{\parallel,\perp}, m$) are in general redshift dependent. While in principle one could use a single $f\sigma_{s8}(z=0)$ as a free parameter and then rescale by the fiducial growth factor $D(z,\Omega_m = \Omega_m^{\rm fid})$ in order to get the corresponding parameter for the different samples, the redshift dependence of the $\alpha$'s and $m$ parameters is not as obvious. Instead we perform the parameter compression separately for the LRG, ELG, and QSO samples and obtain three sets of $(f\sigma_{s8}$,$\alpha_{\parallel,\perp}, m)_{z}$ to be used as ``summary statistics'' of each tracer sample. It is in the cosmological interpretation step that we can either infer $\Lambda$CDM parameters from a single sample or from the combination of $(f\sigma_{s8}$,$\alpha_{\parallel,\perp}, m)_{z}$ sets of multiple tracer samples. 

\begin{figure}
\captionsetup[subfigure]{labelformat=empty}
\begin{subfigure}{0.5\textwidth}
\centering
\includegraphics[height=8cm]{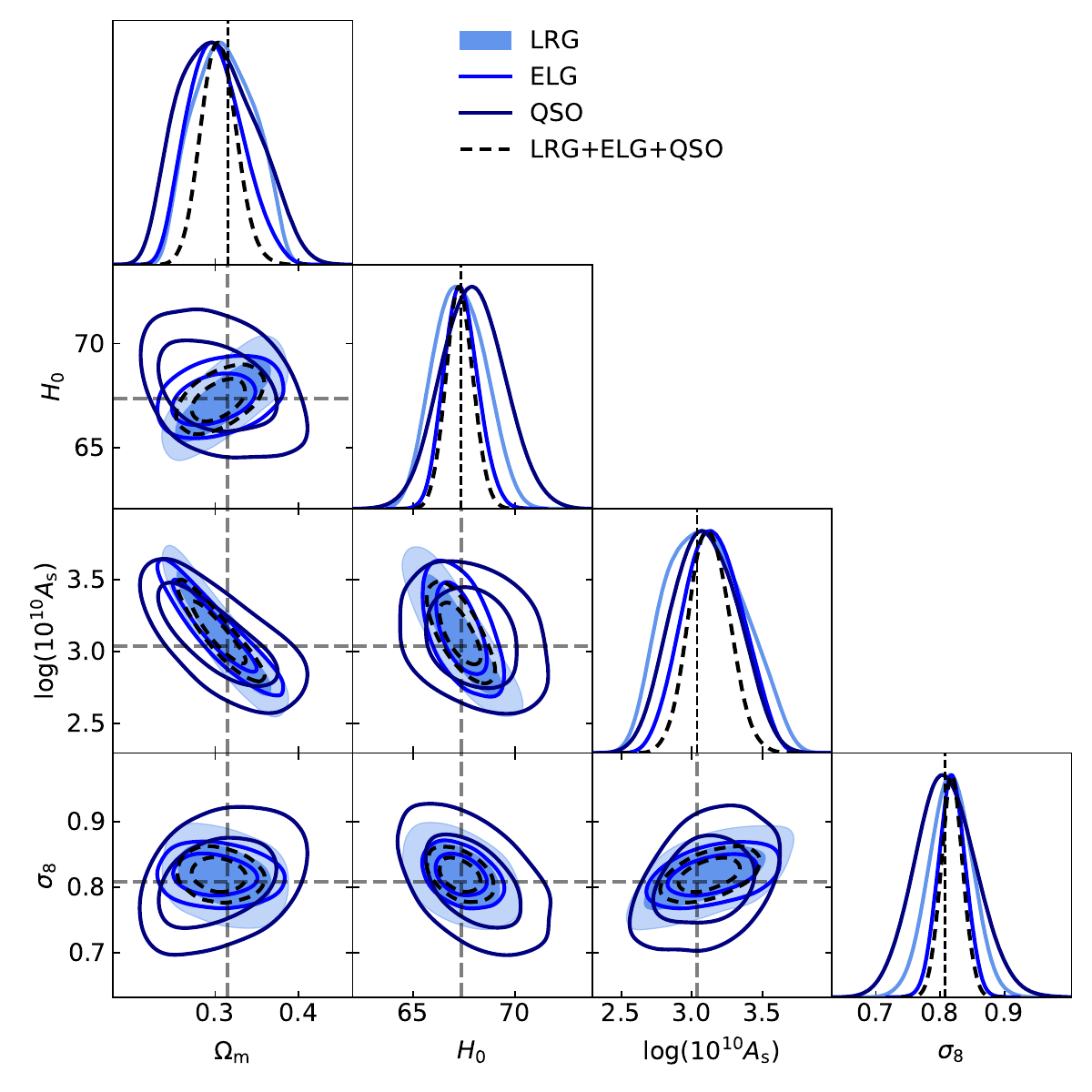}
\caption{Standard Template}
\end{subfigure}%
\begin{subfigure}{0.5\textwidth}
\centering
\includegraphics[height=8cm]{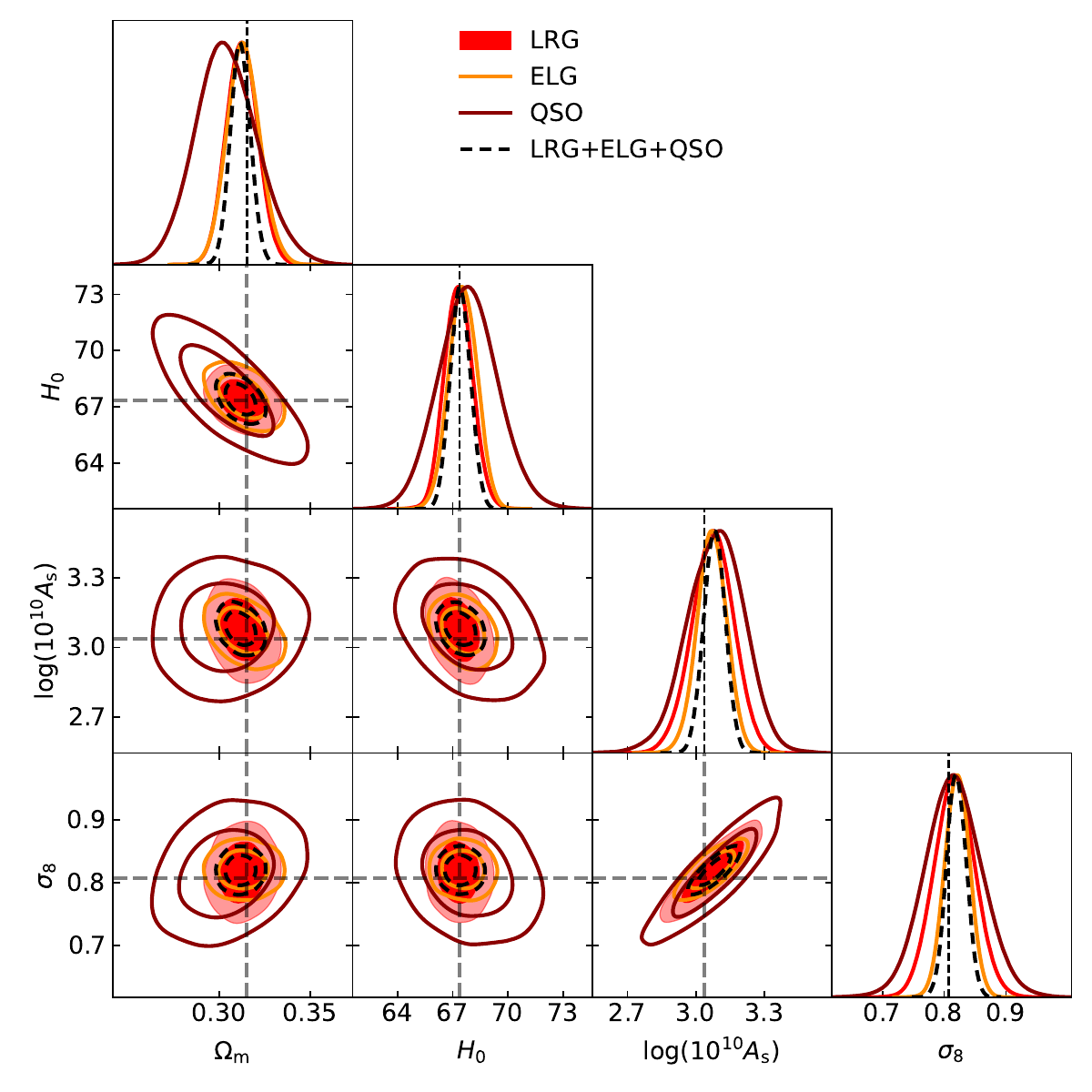}
\caption{ShapeFit}
\end{subfigure}%
\\
\begin{subfigure}{0.5\textwidth}
\centering
\includegraphics[height=8cm]{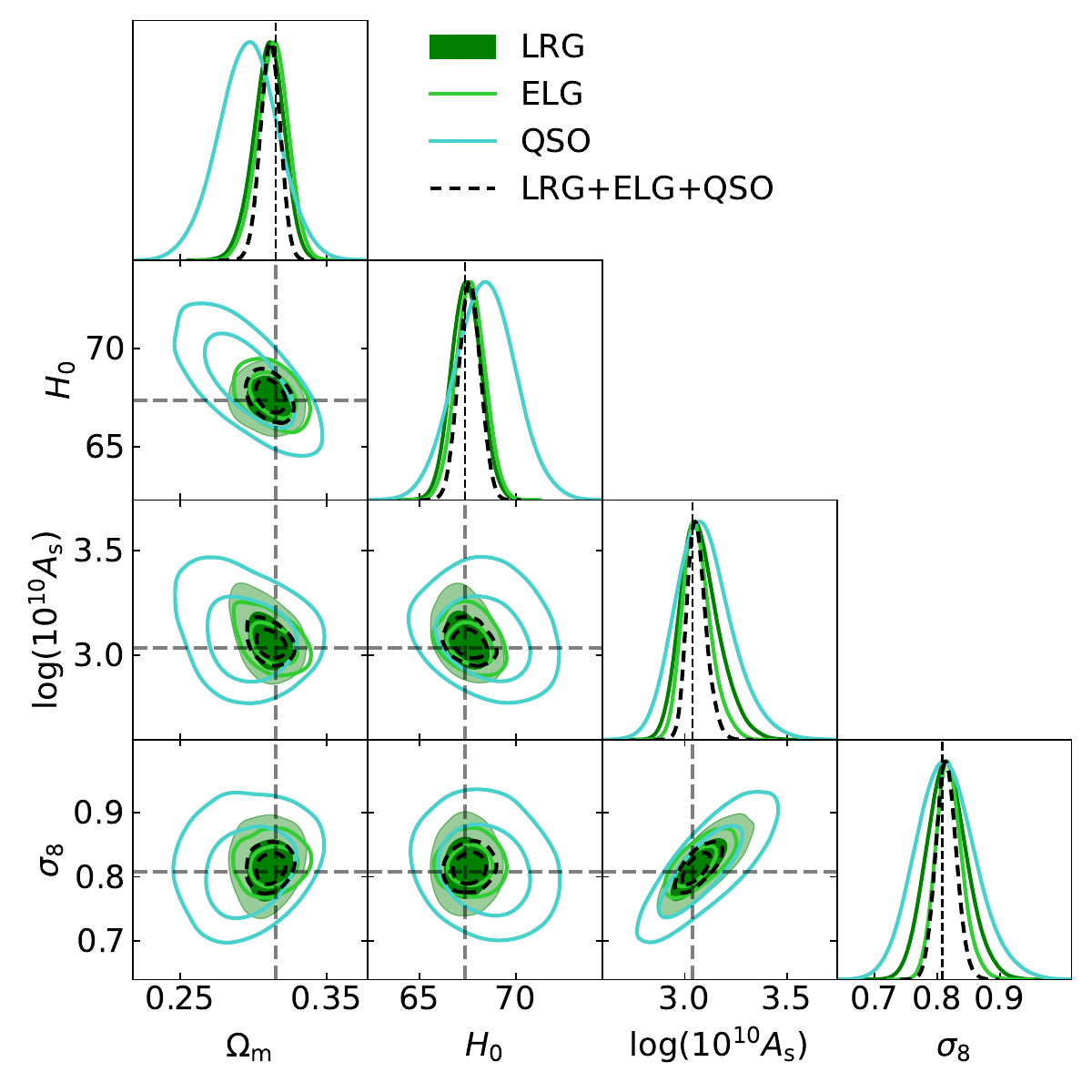}
\caption{Full-Modeling}
\end{subfigure}%
\begin{subfigure}{0.5\textwidth}
\centering
\includegraphics[height=8cm]{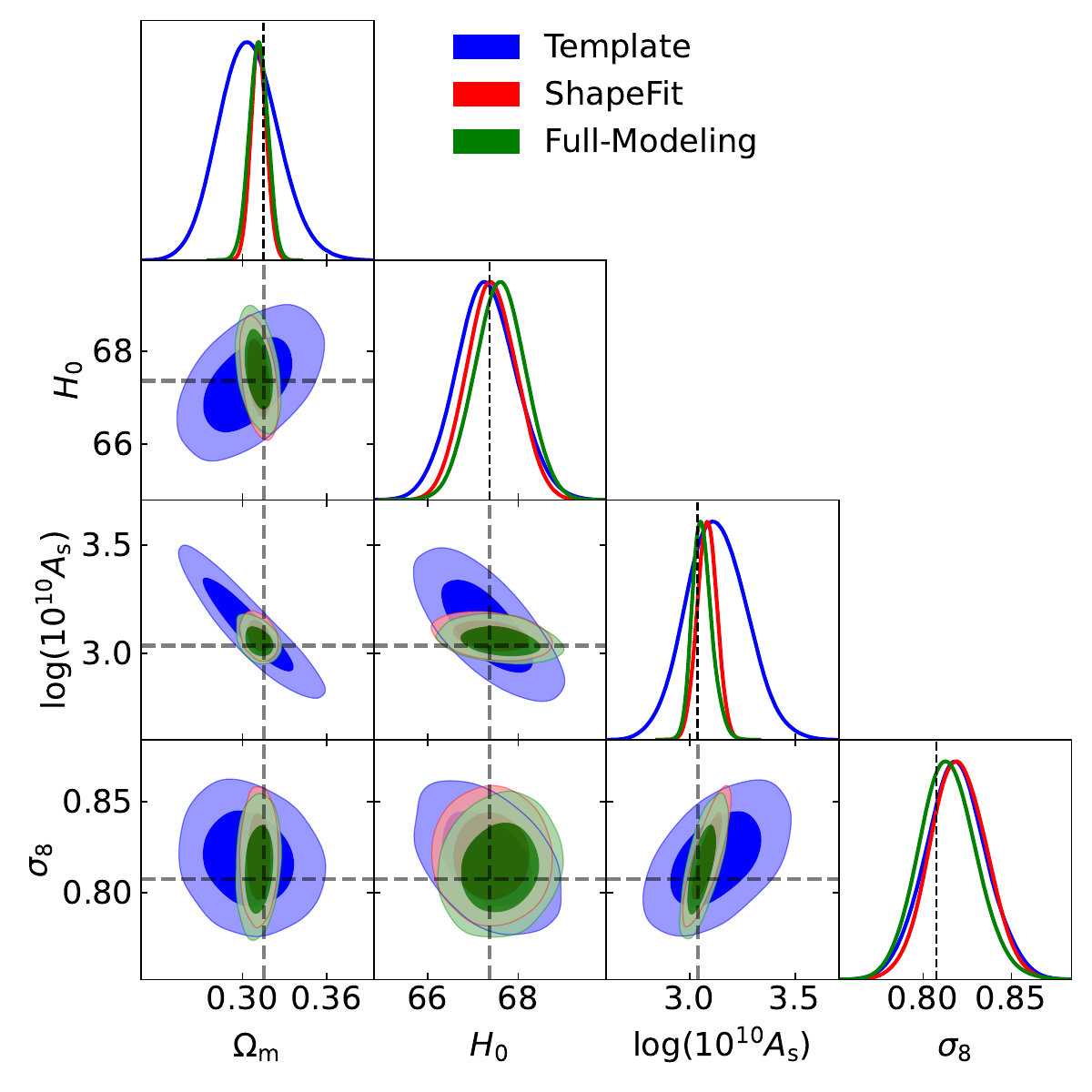}
\caption{All Methods}
\end{subfigure}%
\caption{Comparison of cosmological constraints of different tracers (LRG, $z=0.8$; ELG, $z=1.1$; QSO, $z=1.4$) for the three fit methods. Here we only show the results with the covariance of the single-box volume of $(2 \gpcih)^3$ and $k_{\rm max}=0.2\,h\,\mathrm{Mpc}^{-1}$.}
\label{fig:LRG_ELG_QSO}
\end{figure}

In the three panels of Fig.~\ref{fig:LRG_ELG_QSO} we show a comparison between results of fitting a single sample versus joint fits of multiple tracers, for the standard template, ShapeFit, and Full-Modeling methods respectively. We observe that in each method, the ELG data is significantly more constraining than the LRG sample, and thus the joint fitting constraints appear to be dominated by the ELG sample. The QSO mocks are the least constraining data set, due to the lower number density of Quasars from which the power spectrum is measured. Therefore the error bars at each Fourier mode are larger than those of the ELG and LRG data, resulting in significantly poorer constraints in the model parameters governing the power spectrum shape, i.e. $\Omega_{\rm m}$ and $H_0$. Meanwhile, the amplitude parameter $\log A_s$ is not as sensitive to the type of tracer and we observe smaller differences in constraint between the tracer types. Overall, the tightest constraints on all parameters are obtained in the joint analysis of LRG+ELG+QSO, but with an almost negligible improvement coming from the inclusion of QSO data.

\subsection{Full-shape + BAO Reconstruction}
\label{sec: BAO}

In addition to fitting the full-shape power spectra using our model, we can gain extra constraining power through a joint analysis with the reconstructed BAO correlation function. The BAO reconstruction procedure aims to undo some of the damping of the BAO signal due to nonlinear structure growth in order to sharpen its peak, allowing for a better measurement of the cosmological distance-redshift relation via the well-defined drag horizon scale (see e.g.\  refs.~\cite{ESSS07,Noh_recon2009,Padmanabhan_recon2009,White_recon2015,Chen_recon2019,KP4s2-Chen}). This procedure begins by smoothing the observed clustering signal by a Gaussian filter $S(k) = \exp(-(kR_s)^2/2)$, which serves to filter out small-scale modes. Next, we use this smoothed density to estimate the smoothed Zel'dovich displacement, $\bChi \approx S(k)\bPsi_{\rm Zel}$, which we subtract from the observed galaxy field as well as from a random matter density field in order to preserve large-scale power. The reconstructed galaxy density field is then $\delta_{\rm rec} = \delta_{\rm d}-\delta_{\rm s}$, with $\delta_{\rm d}$ and $\delta_{\rm s}$ being the displaced galaxy and shifted random fields, respectively. Moving to redshift space once again amounts to a rotation of the real-space field, $\bChi_s = R\bChi$ with matrix $R$ defined in Sect.~\ref{sec: theory}. In the literature one commonly encounters two methods for reconstructions in redshift space: \textbf{RecSym} \cite{White_recon2015} and \textbf{RecIso} \cite{ESSS07,Padmanabhan_recon2009}. The first applies the transformation into redshift space equally to both $\delta_{\rm d}$ and $\delta_{\rm s}$, whereas the latter method keeps the shifted field in real-space (see ref.~\cite{KP4s2-Chen} for further discussion). For the DESI simulations considered in this work, the \textbf{RecSym} procedure is applied to produce the post-reconstruction mock data.   

We model the damping of the BAO feature in the reconstructed power spectrum, $P_{\rm rec} = P_{\rm dd}+P_{\rm ss}-2P_{\rm ds}$ within the Zel'dovich approximation by splitting the linear theory predictions into the wiggle and no-wiggle components\footnote{There are numerous methods for performing this split. Here we use the method described in Appendix D of ref.~\cite{Wallisch2018} that uses a sine transform to identify the BAO feature in real space and subtracts it before transforming back to Fourier space to produce a wigge-free power spectrum.} and apply an exponential damping factor\footnote{Previous works studying BAO reconstruction have sometimes derived different damping factors for $P_{dd}$, $P_{ds}$ and $P_{ss}$. This results from a $1^{\rm st}$ order approximation in LPT, and a more consistent approach has the randoms damped by the same factor. This subtlety is described in detail in ref.~\cite{KP4s2-Chen}, as well as in ref.~\cite{Sugiyama24} for a slightly different reconstruction scheme. However, we find that the difference between the old and new methods results in negligible effects to the fit posteriors.} to the wiggle part \cite{KP4s2-Chen}
\begin{align}
    P(k,\mu) = (b+f\mu^2)^2\left(P^{NW}(k) + e^{-\frac{1}{2}k^2\Sigma^2(\mu)}P_{ab}^{W}(k)\right), 
\label{eq: P_BAO}
\end{align}
where the $\Sigma^2$ in the damping factor is the isotropic component of the linear pairwise displacement $A_{ij}^{dd} = \left<\Delta_i^{dd}\Delta_j^{dd}\right>$, of the displaced density field at $|\bq| = r_d$, i.e.
\begin{align}
    \Sigma^2(\mu) &=\left. \frac{1}{3}\delta_{ij}A_{ij}(\bq)\right\vert_{q=r_d} \\
    &= \left[ 1+f(2+f)\mu^2\right] \left[ 2\tilde\Sigma^2(0) - 2\tilde\Sigma^2(r_d)\right]  \\
    \tilde\Sigma^2(q) &= \frac{1}{3}\int\frac{dk}{2\pi^2}(1-\mathcal{S})^2j_0(kq)P_{\rm lin}(k)
    \quad .
\end{align}

Finally, after generating the reconstructed power spectrum, we use a Fourier transform to obtain the reconstructed correlation function. We limit our model to linear bias as it has been found in previous works that the IR damping of the BAO feature dominates over other nonlinear effects such as mode-coupling which are largely cancelled by reconstruction. Following Ref.~\cite{KP4s2-Chen} we employ a new method for modeling the broadband that is not degenerate with the BAO signal, which in Fourier space involves using a basis of cubic splines. When fitting the correlation function in configuration space this is equivalent to setting a minimum scale, $r_{\rm min}$, with the exception of two Hankel transformed basis functions that are included in the quadrupole:
\begin{align}
    \mathcal{Q}_{2,n}(r\Delta) = \frac{i^2}{2\pi^2} \int dk k^2 W_3\left(\frac{k}{\Delta}-n\right)j_2(kr), \quad n=0,1
\end{align}
where $W_3$ is the piecewise cubic spline kernel \cite{Hockney88,Jeong10}, $j_2$ is a $\nu=2$ spherical Bessel function, and we choose $\Delta = 0.06 \ihmpc$ for the separation scale of the splines. We additionally include a template of polynomials in even powers of $r$ for the monopole and quadrupole moments, truncated at quadratic order, to marginalize over contamination by large-scale systematics below some $k_{\rm min}$. The broadband model in configuration space is thus \cite{KP4s2-Chen}:
\begin{align}
    &\mathcal{B}_{0}(r) = a_{0,0} + a_{0,1}\left(\frac{r k_{\rm min}}{2\pi}\right)^2 \nonumber \\
    &\mathcal{B}_{2}(r) = a_{2,0} + a_{2,1}\left(\frac{r k_{\rm min}}{2\pi}\right)^2 + \Delta^3(a_{2,2}\mathcal{Q}_{2,0}(r\Delta) + a_{2,3}\mathcal{Q}_{2,1}(r\Delta))
\end{align}
where $k_{\rm min} = 0.02 \ihmpc$ and the parameters \{$a_{0,0},a_{0,1},a_{2,0},a_{2,1},a_{2,2},a_{2,3}$\} can be analytically marginalized over. We use broad Gaussian priors centered at 0 with widths of $5\times10^5$ for all of these broadband parameters. Finally, we note that one should also include some more flexibility in the damping factor by introducing parameters $\Sigma_{\parallel,\perp}$ in the exponent in Eq.~\ref{eq: P_BAO} to marginalize over the effects of nonlinearities. However, we did not find this necessary in the tests presented here, and so the damping factors vary only as $f, P_{\rm lin},$ and $r_{\rm d}$ change in Full-modeling and likewise with ShapeFit through the $f\sigma_{s8}$ and $m$ parameters.

The joint covariance matrix is computed numerically using the reconstructed correlation function realizations of the \texttt{EZmock} simulations. So the joint data vector is now $\textbf{d} = \{P_0^{\rm pre}(k),P_2^{\rm pre}(k), \xi_0^{\rm post}(r), \xi_2^{\rm post}(r)\}$ with cross-correlations between $P_\ell^{\rm pre}(k)$ and $\xi_\ell^{\rm post}(r)$ accounted for as nonzero off-diagonal elements in the joint covariance matrix. (e.g. see Fig.~3 of \cite{Chen22})

\begin{figure}
\captionsetup[subfigure]{labelformat=empty}
\begin{subfigure}{0.5\textwidth}
\centering
\includegraphics[height=8cm]{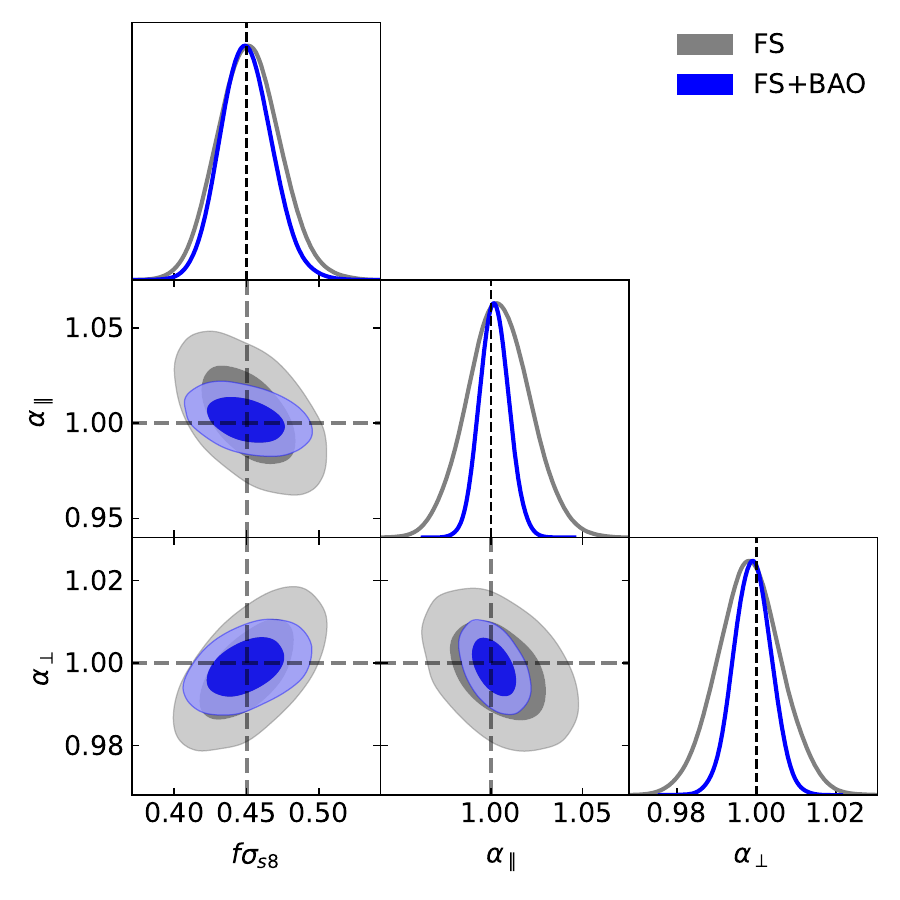}
\caption{Standard Template}
\end{subfigure}%
\begin{subfigure}{0.5\textwidth}
\centering
\includegraphics[height=8cm]{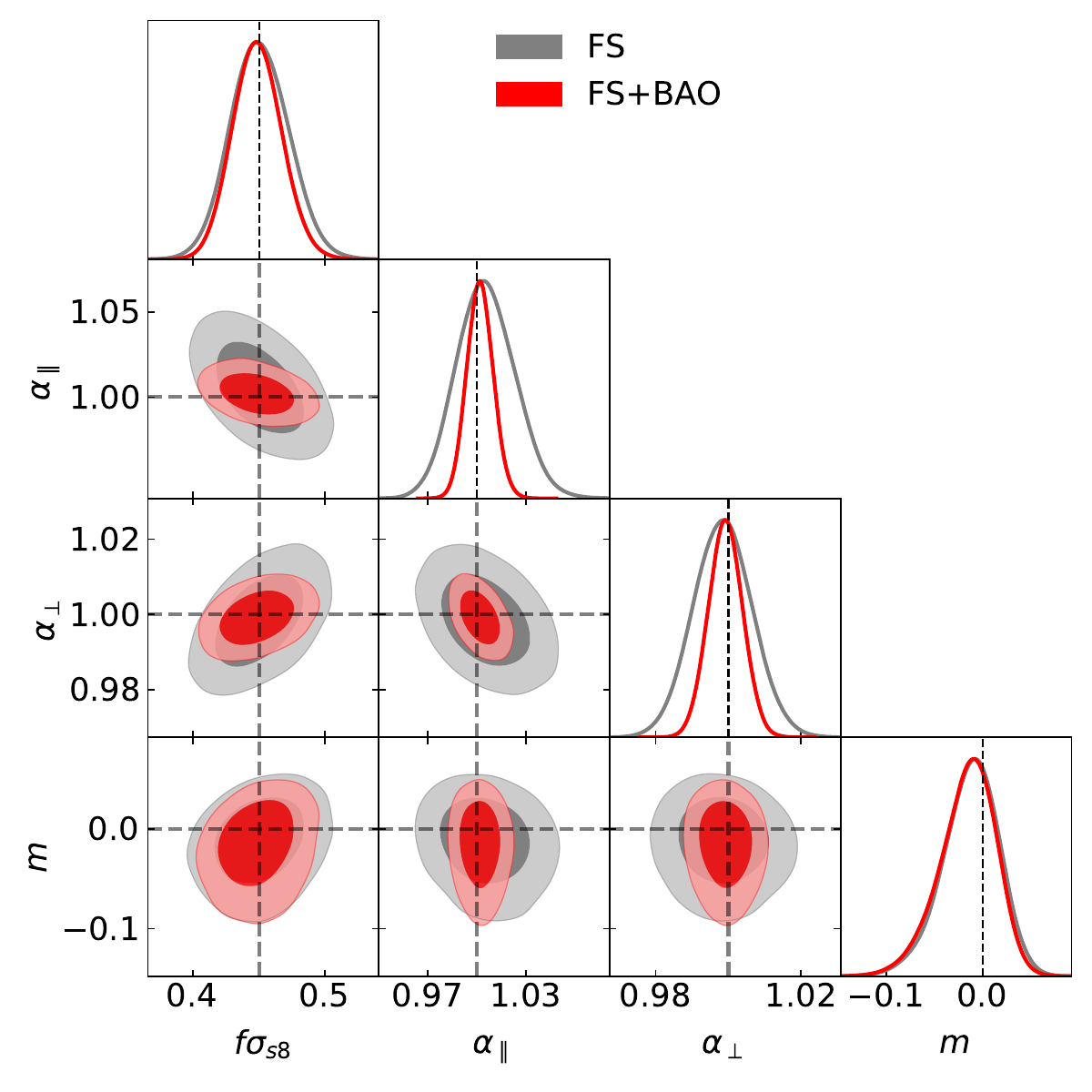}
\caption{ShapeFit}
\end{subfigure}%
\\
\begin{subfigure}{0.5\textwidth}
\centering
\includegraphics[height=8cm]{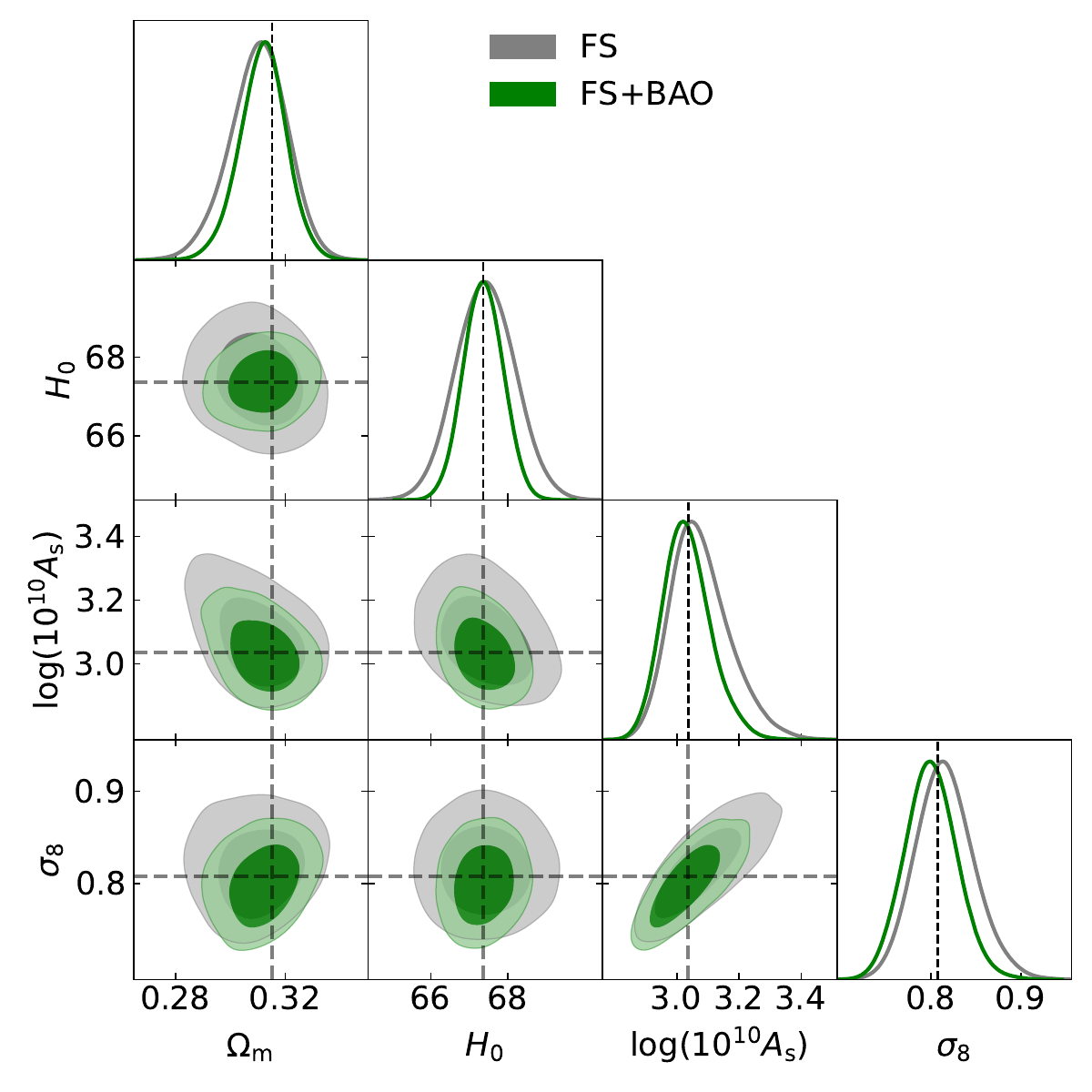}
\caption{Full-Modeling}
\end{subfigure}%
\begin{subfigure}{0.5\textwidth}
\centering
\includegraphics[height=8cm]{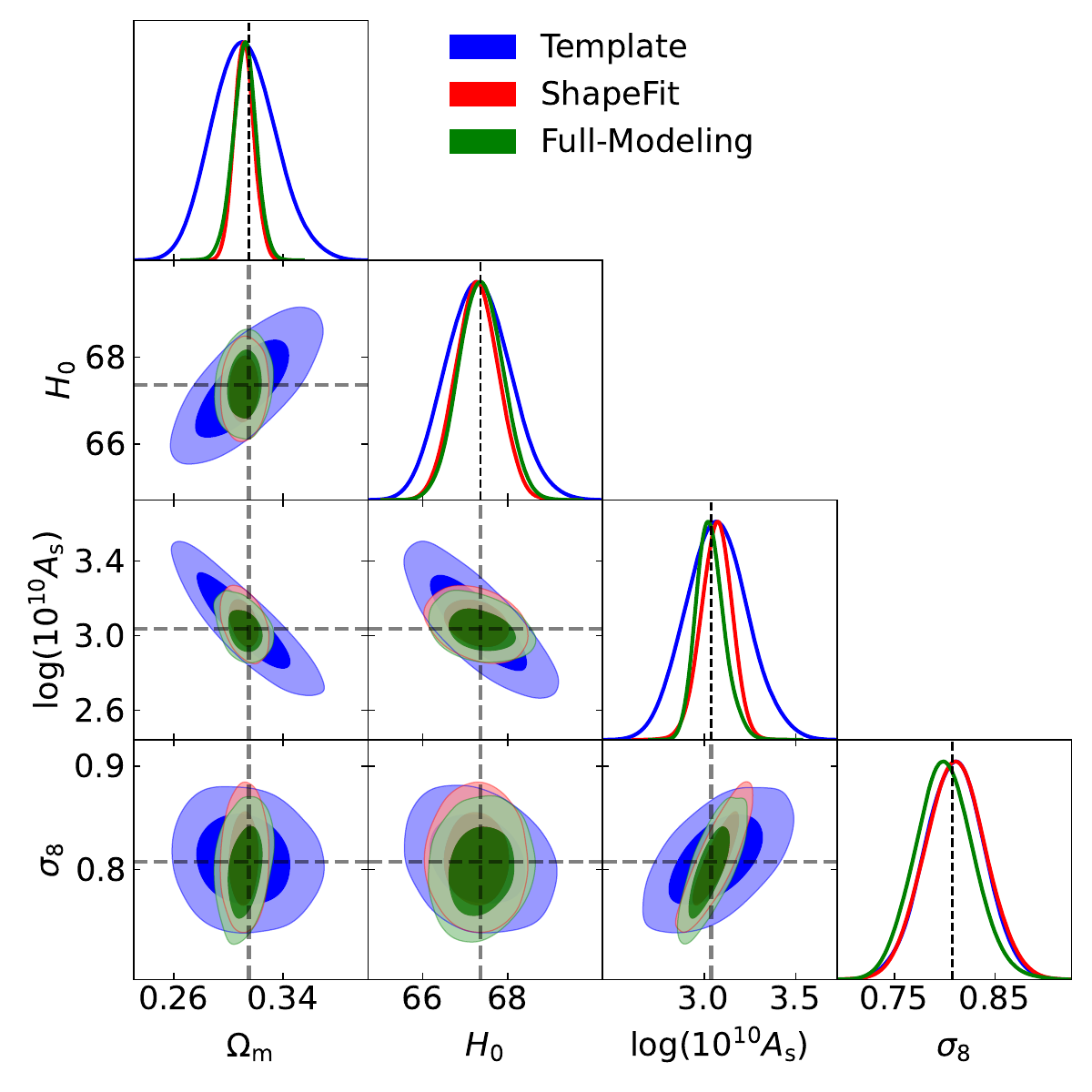}
\caption{Post Recon (all methods)}
\end{subfigure}%
\caption{Comparison of cosmological constraints with and without BAO reconstruction for each modeling method. The bottom right panel compares the post-reconstruction constraints of the three methods. For all plots above, we present results using the covariance appropriate to the single-box volume. In the legends, ``FS'' refers to the pre-reconstruction full-shape power spectrum data, and ``BAO'' refers to the BAO signal in the post-reconstruction correlation function.}
\label{fig: FS_BAO}
\end{figure}

We show in Fig.~\ref{fig: FS_BAO} comparisons of the cosmological constraints pre/post BAO reconstruction. We find that for all three modeling methods there is significant improvement in constraints when joint-fitting with the post-recon correlation function, most significantly in $H_0$ as the cleaner measurement of BAO scale from the sharpened peak allows for better calibration of the distance-redshift relation that constrains Hubble's constant. When comparing all methods we find consistent constraints between ShapeFit and Full-Modeling that are both tighter than those of the standard template. 

\subsection{Beyond $\Lambda$CDM: $w$CDM model}

\begin{figure}
\captionsetup[subfigure]{labelformat=empty}
\begin{subfigure}{.5\textwidth}
\centering
\includegraphics[height=8cm]{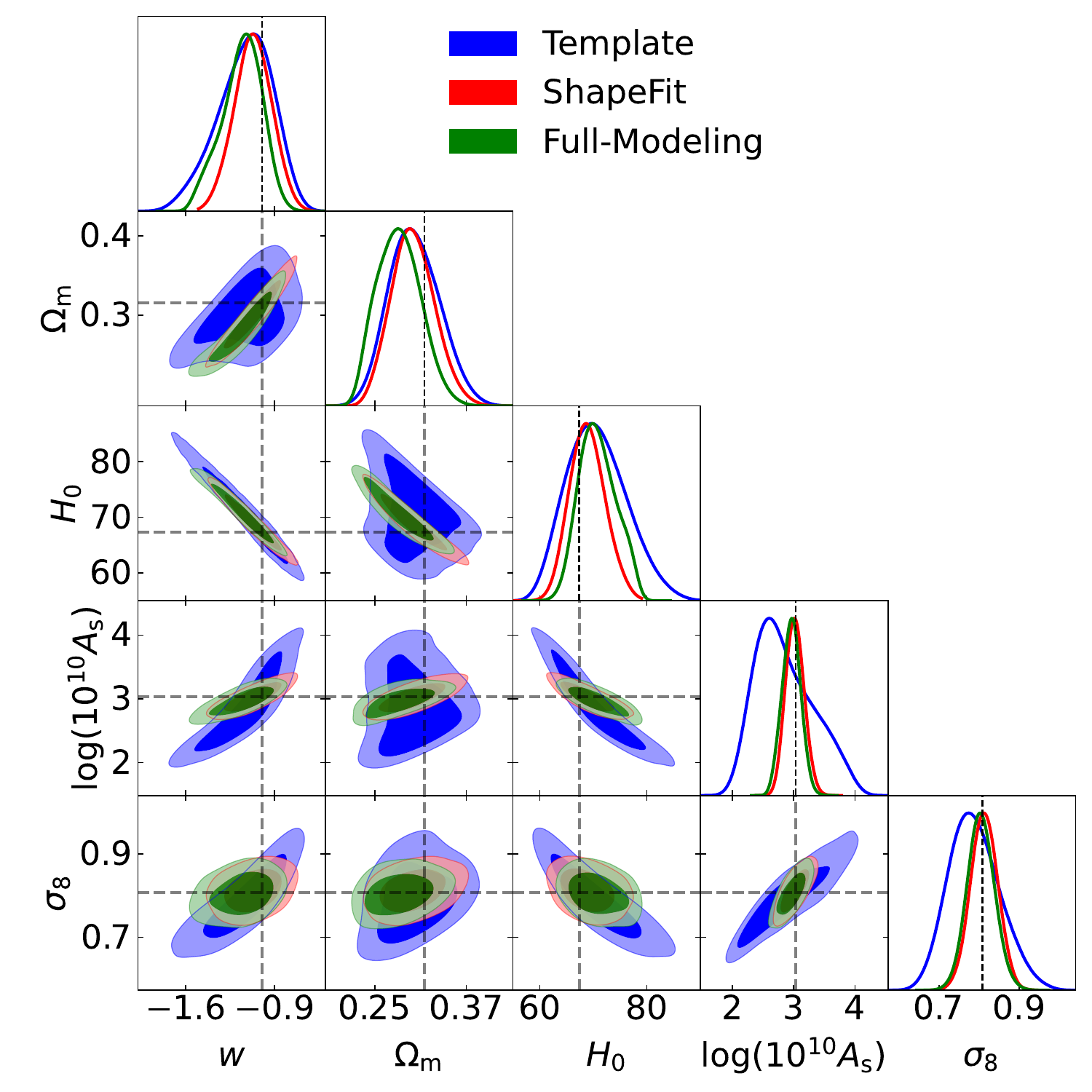}
\caption{All methods}
\end{subfigure}%
\begin{subfigure}{.5\textwidth}
\centering
\includegraphics[height=8cm]{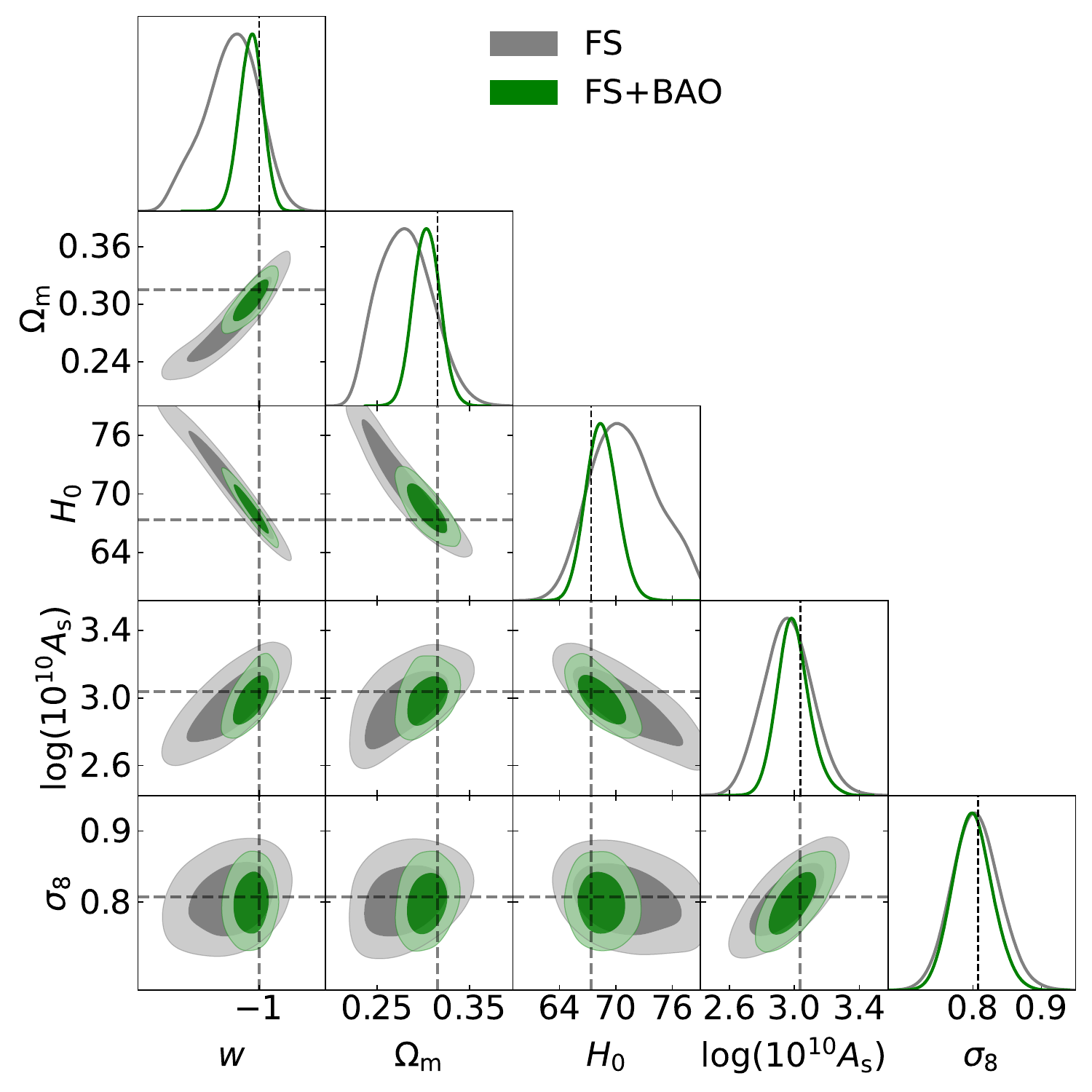}
\caption{Full-Modeling}
\end{subfigure}%
\caption{On the left we show a comparison of cosmological constraints on the $w$CDM parameters for the three modeling methods. On the right we show $w$CDM Full-Modeling fits with and without the inclusion of post-reconstruction BAO data. In both plots we show the results with the single-box $(2\gpcih)^3$  volume. In the legends, ``FS'' refers to the pre-reconstruction full-shape power spectrum data, and ``BAO'' refers to the BAO signal in the post-reconstruction correlation function.
\label{fig: wcdm}
}
\end{figure}

With the expected improvement in cosmological parameter estimation from future galaxy redshift surveys, we hope to place better constraints on parameters not just underlying the standard $\Lambda$CDM model, but also departures from it. From the Friedmann equations, the energy density of a specific component of the Universe is related to the scale factor, $a$, by
\begin{align}
    \rho \propto a^{-3(1+w)}
\end{align}
where $w = p/\rho$ is the equation of state parameter. One of the simplest extensions to $\Lambda$CDM involves allowing the dark energy equation of state to differ from the value of $-1$ that corresponds to a cosmological constant ($\Lambda$) as the energy density is constant in that case. On the other hand, ``quintessence'' models have $w \neq -1$ such that dark energy is a dynamic quantity in the Universe.\footnote{If dark energy is described by a scalar field, $\phi$, with a canonical kinetic term then the equation of state can be interpreted in terms of kinetic and potential energies via,
\begin{align}
    w = \frac{\frac{1}{2}\dot{\phi}^2 - V(\phi)}{\frac{1}{2}\dot{\phi}^2 + V(\phi)} \quad .
\end{align}
Under this assumption the equation of state is usually expected to lie between $-1<w<1$, with values $w<-1/3$ leading to cosmic acceleration. However, more exotic models exist that do allow for negative kinetic energies.}

Fig.~\ref{fig: wcdm} shows in the left panel the constraints on $w$CDM parameters obtained from each of the three modeling methods, for the covariance of the single-box volume. Since the Abacus cosmology assumes a cosmological constant for dark energy, the expected value is $-1$. We find that the ShapeFit and Full-Modeling methods both give constraints on $w$ that are within $1\sigma$ of the expected equation of state. Meanwhile the parameters in the template method are very poorly constrained when $w$ is varied.  When changing the properties of dark energy away from the cosmological constant the universe's expansion history and geometry are significantly altered, thus affecting the $\alpha_{\parallel,\perp}$ parameters and $f\sigma_{s8}$. This results in the observed degeneracies between $w$ and the other parameters (which also determine $f\sigma_{s8}$, and $\alpha_{\parallel,\perp}$). If those three parameters are the only information we have from the data, as is the case in the template fit, then this results in very poor constraints. However, moving far along those degeneracy directions also significantly affects the shape of the power spectrum, which the ShapeFit and Full-Modeling methods \emph{are} sensitive to. Therefore these two methods do not suffer from the degeneracies as much as the template fit. Comparing ShapeFit to Full-Modeling, we find that the constraints on parameters from the ShapeFit method are a bit wider than in Full-Modeling. This is likely because all of the shape information is contained in a single parameter, which then needs to be interpreted as constraints on three different cosmological parameters ($w, \omega_{\rm m},$ and $H_0$), as these all control the shape of the power spectrum. Thus, a poorer measurement of $m$ results in more sensitivity to the degeneracies in shape that the template fit also suffered from. Finally, we also note that projection effects (see Appendix~\ref{appendix: proj_priors}) in Full-Modeling cause close-to 1$\sigma$ offsets in the $w, \Omega_{\rm m},$ and $H_0$ parameters. While these shifts are not huge for this dataset, we also are interested into what extent including more data can mitigate projection effects. We show in the right panel of Fig.~\ref{fig: wcdm} a comparison of Full-Modeling fits with and without the inclusion of reconstructed BAO data. We find that including BAO results in noticeable improvements in the constraints by shifting the posteriors closer to the truth. These projection effects are not as significant in the ShapeFit method, which suggests that the extra information that Full-modeling obtains w.r.t. ShapeFit may come from regions of the power spectrum that are degenerate with counterterm and/or stochastic parameters. A similar effect was observed and reported in Ref.~\cite{Maus23} when comparing $f\sigma_{s8}$ constraints between Full-Modeling and standard template methods in BOSS data.

\subsection{Priors from CMB}
\label{sec: planck priors}

\begin{figure}
\captionsetup[subfigure]{labelformat=empty}
\begin{subfigure}{.5\textwidth}
\centering
\includegraphics[height=8cm]{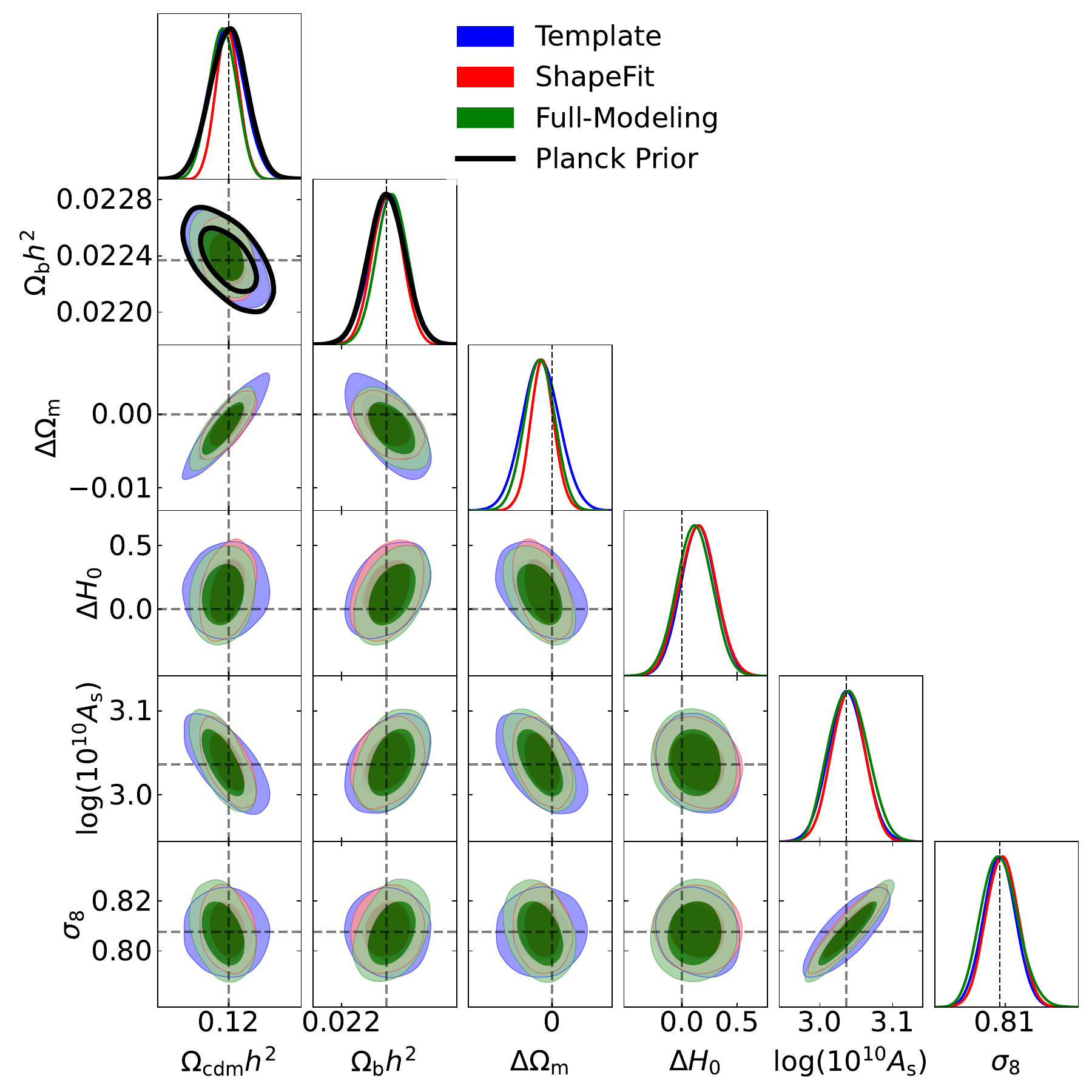}
\caption{Full Covariance}
\end{subfigure}%
\begin{subfigure}{.5\textwidth}
\centering
\includegraphics[height=8cm]{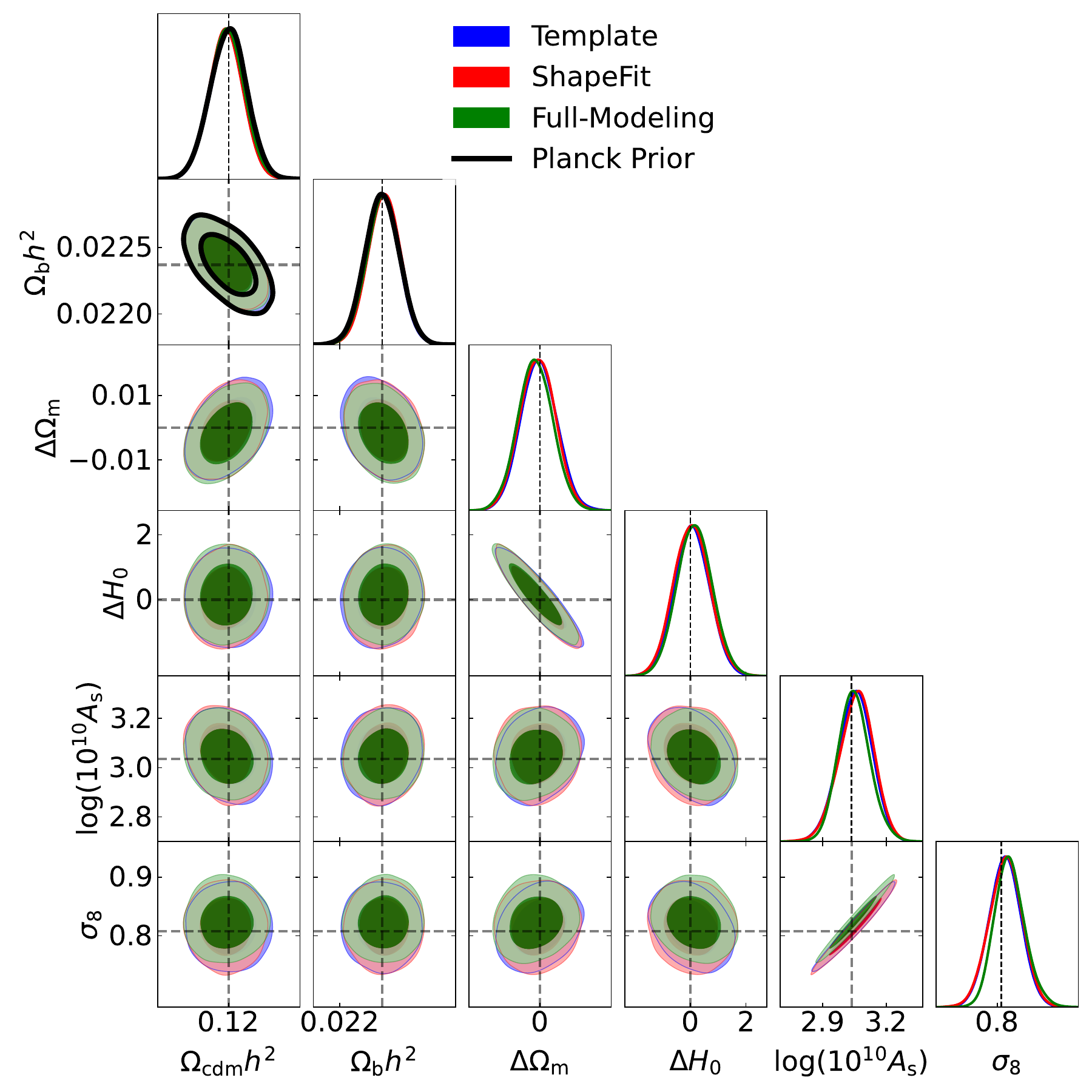}
\caption{Single Box Covariance}
\end{subfigure}%
\caption{Comparison of $\Lambda$CDM constraints from the template fit(blue),  Full-Modeling (green), and ShapeFit (red) with Planck priors, on the Abacus LRG ($z=0.8$) mock data. On the left we show the fits with the covariance for the full ($25 \times (2\gpcih)^3$) volume while the right figure shows the results with the single-box $(2\gpcih)^3$  volume. In the $\Omega_{\rm cdm}h^2 - \Omega_{\rm b}h^2$ panels we include black lines showing the Planck prior.
}
\label{fig: planck_prior}
\end{figure}

The `standard' template method was conceived at a time when the data from galaxy redshift surveys was not constraining enough on early-universe physics to be competitive with constraints from probes such as Planck that modeled CMB anisotropies.  In particular, data from CMB anisotropies tightly constrain the $\Lambda$CDM parameters that determine the shape of the power spectrum \cite{Planck18-I}, and this shape is left unaltered by late-time physics such as dark energy or spatial curvature.  These constraints are tighter than those from the galaxy survey themselves.  In such a scenario, the primary degrees of freedom to be constrained by galaxy surveys are late time growth and the late-time distance-redshift relation.  The template method was intended to be used in conjunction with the other probes, such that most of the information on $P(k)$ shape came from strong priors using results from e.g.\ Planck. To demonstrate this, we repeat the cosmological inference of the template results, but including an additional likelihood derived from the Planck 2018 results \cite{Aghanim2020}. We do this by taking the chains from the baseline model of the Planck Legacy Archive, ``base plikHM TT lowl lowE'', and compute the covariance matrix, $C_{\rm pl}$, from the ($\omega_{\rm b}$, $\omega_{\rm cdm}$) samples. We do not apply a prior on $A_s$ or $H_0$ as we are interested in how information from galaxy clustering constrains the late-time growth compared to Planck. When we sample in these $\Lambda$CDM parameters we now include the additional likelihood 
\begin{align}
    \mathcal{L}_{\rm pl} \propto \exp\{-\frac{1}{2}\boldsymbol{\Delta\Theta}^T C_{\rm pl}^{-1}\boldsymbol{\Delta\Theta}\},
\end{align}
where $\boldsymbol{\Delta\Theta}$ is the difference between the sampled ($\omega_{\rm b}$, $\omega_{\rm cdm}$) and the values in the Abacus cosmology. Because we are including the CMB prior on $\omega_b$, we remove the BBN prior that we usually use in our standard analyses. We show these results, comparing the template, Shapefit, and Full-Modeling methods with Planck priors, in Fig.~\ref{fig: planck_prior}, using the LRG ($z=0.8$) mock data within the standard $\Lambda$CDM model. We see that the inclusion of Planck priors significantly tightens the constraints on $\Omega_{\rm m}$. Despite us not applying any prior on $H_0$ and $\log A_s$, we still observe a shift to the truth and tightening in those parameter constraints for all three methods, with the $\log A_s$ posterior slightly narrower for the Full-Modeling approach. Overall, all three methods agree very closely in all of the parameters when including these priors, suggesting that the difference in constraining power of these methods is almost entirely due to shape information (which is better determined by the CMB than the galaxy survey). 

\subsection{Varying $n_s$}
\label{ssec: free_ns}

\begin{figure}
\captionsetup[subfigure]{labelformat=empty}
\begin{subfigure}{.5\textwidth}
\centering
\includegraphics[height=8cm]{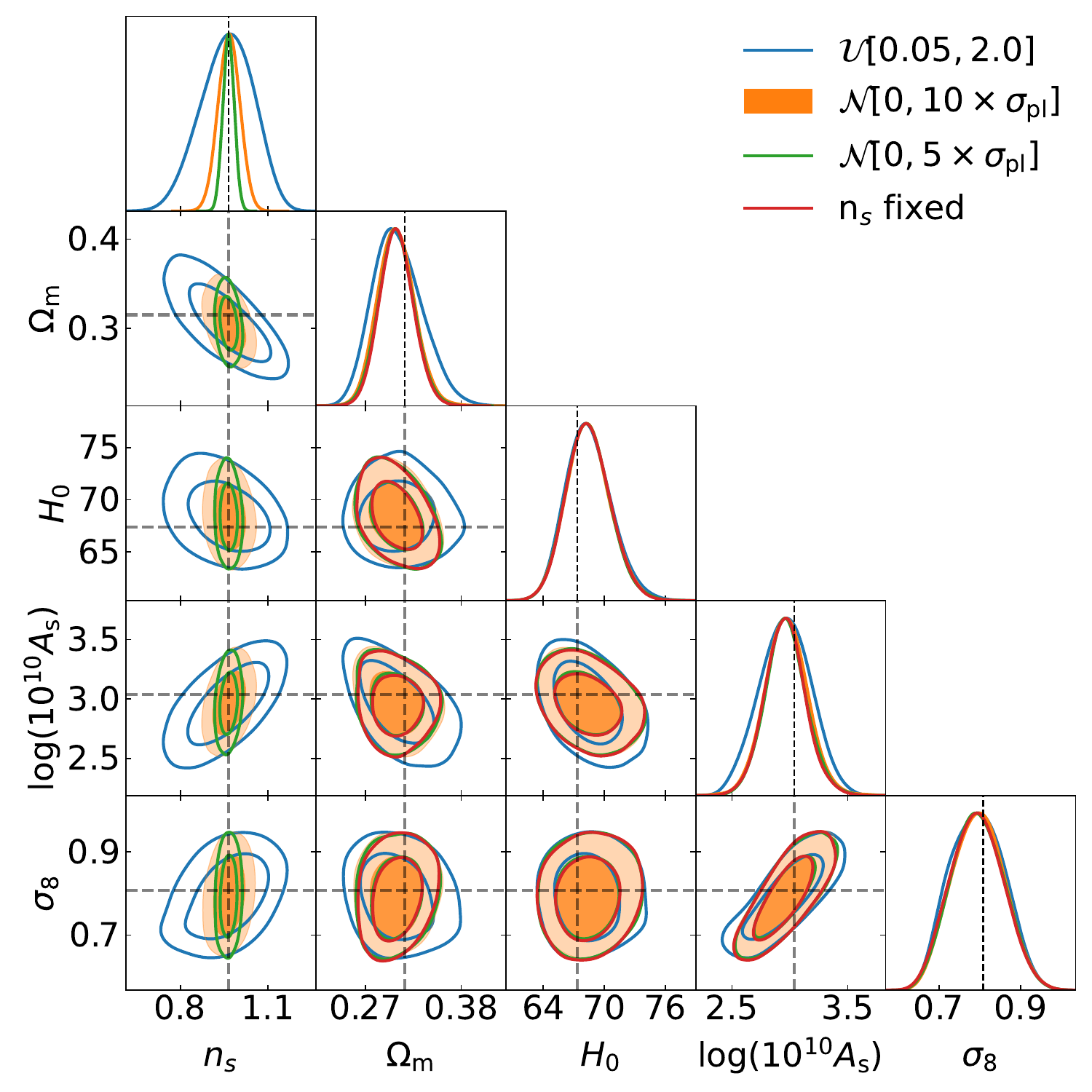}
\caption{LRG ($0.4<z<0.6$)}
\end{subfigure}%
\begin{subfigure}{.5\textwidth}
\centering
\includegraphics[height=8cm]{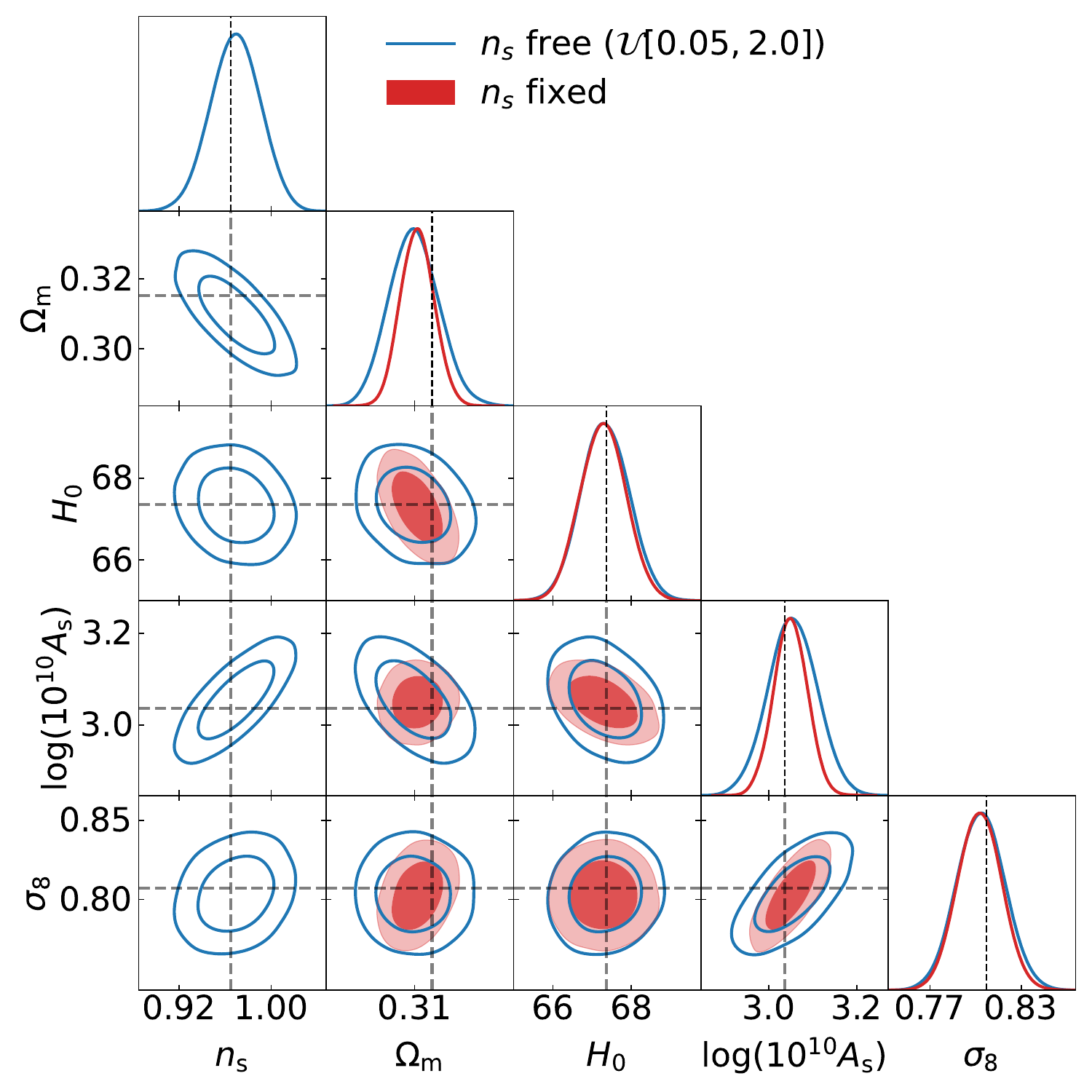}
\caption{All Y1 bins}
\end{subfigure}%
\caption{The left panel shows Full-Modeling fits to the mean of Cutsky mocks in the DESI Y1, LRG $0.4<z<0.6$ redshift bin, with different priors on $n_s$. We use the notation $\mathcal{U}[\mathrm{min},\mathrm{max}]$ and $\mathcal{N}[\mu,\sigma]$ to denote uniform and Gaussian priors, respectively. In the right panel we compare Full-Modeling $n_s$ fixed versus free using the synthetic mocks created with velocileptors simulating the the full Y1 footprint: BGS ($0.1<z<0.4$), LRG ($0.4<z<0.6$), LRG ($0.6<z<0.8$), LRG ($0.8<z<1.1$), ELG ($0.8<z<1.1$), ELG ($1.1<z<1.6$), and QSO ($0.8<z<2.1$). We show constraints with uniform priors on $n_s$ in the free case.
}
\label{fig: ns_free}
\end{figure}

For previous fullshape analyses from spectroscopic surveys, it was common/necessary to fix (or impose tight priors) on several of the $\Lambda$CDM parameters such as $\omega_b$, $M_\nu$, and $n_s$, using information from the CMB and BBN. With the increasing constraining power of DESI and future surveys it is of interest to see how much we can untangle fullshape analyses from other probes. While a tight prior on $\omega_b$ (see Appendix~\ref{appendix: tests}) is still necessary, the improved constraining power of DESI may allow us to free $n_s$ and/or $M_{\nu}$\footnote{In this paper we only perform tests with $n_{\rm s}$ free and refer readers to Ref.~\cite{KP5s3-Noriega} for a discussion on varying $M_\nu$.}. To investigate the impact of uncertainty in $n_s$ on our analysis given the statistical uncertainties in Y1, we chose mock data from one of the DESI Y1 redshift bins (LRG; $0.4<z<0.6$) with an appropriate analytic covariance. We compare constraints on $\Lambda$CDM parameters with various prior choices on $n_s$, including a uniform prior, Gaussian with widths of 10$\times$ and 5$\times$ Planck 2018 constraints ($\sigma_{ns} = 0.004$)\cite{Aghanim2020}, and with $n_s$ fixed. These results are shown in the left panel of Fig.~\ref{fig: ns_free} for the Full-Modeling method. We find that for both the 10$\times$ and 5$\times$ priors on $n_s$ the constraints on $\omega_{\rm cdm}$, $H_0$, and $\log A_s$ are identical to those when $n_s$ is fixed, suggesting that the Full-Modeling constraints on $\Lambda$CDM parameters are robust even if the $n_s$ constraints from the CMB are systematically off by 10$\sigma$. 
In order to see how well $n_s$ can be constrained completely independently from Planck we additionally fit to noiseless synthetic mock data vectors simulating all seven DESI Y1 redshift bins: BGS ($0.1<z<0.4$), LRG ($0.4<z<0.6$), LRG ($0.6<z<0.8$), LRG ($0.8<z<1.1$), ELG ($0.8<z<1.1$), ELG ($1.1<z<1.6$), and QSO ($0.8<z<2.1$) using the appropriate Y1 analytic covariance for each redshift bin. We compare the case with uniform priors on $n_s$ to the case with $n_s$ fixed. These results are shown in the right panel of Fig.~\ref{fig: ns_free}. We find that despite the slight degradation in $\Omega_{\rm m}$ constraint with the flat prior on $n_s$, we are able to measure $n_s$ to a 3\% precision. 


\subsection{Comparison of LPT and EPT}
\label{appendix: EPT}

In addition to the LPT model that we primarily focus on in this paper, \velocileptors\ also has an Eulerian perturbation theory module. The EPT kernels are constructed from the Lagrangian kernels while setting the IR resummation scale, $k_{IR}$, to zero. The Eulerian and Lagrangian theories differ in their treatment of cold dark matter, the first describing dark matter as a perfect pressureless fluid, and the latter describing it as collisionless particles. The overdensities derived from both theories agree order-by-order except when particle trajectories cross. The EPT model in \velocileptors\ employs the galaxy bias scheme described in Ref.~\cite{McDRoy09}. The mapping between the Lagrangian and Eulerian bias bases can be achieved within \velocileptors\ via the transformations \cite{ChenVlahWhite20}:
\begin{align} 
    b_1^E &= 1 + b_1^L \nonumber \\
    b_2^E &= b_2^L + \frac{8}{21}b_1^L 
    \quad , \quad
    b_s^E = b_s^L -\frac{2}{7}b_1^L  \nonumber \\
    b_3^E &= 3b_3^L + b_1^L. 
\end{align}

\begin{figure}
    \centering
    \resizebox{\columnwidth}{!}{\includegraphics{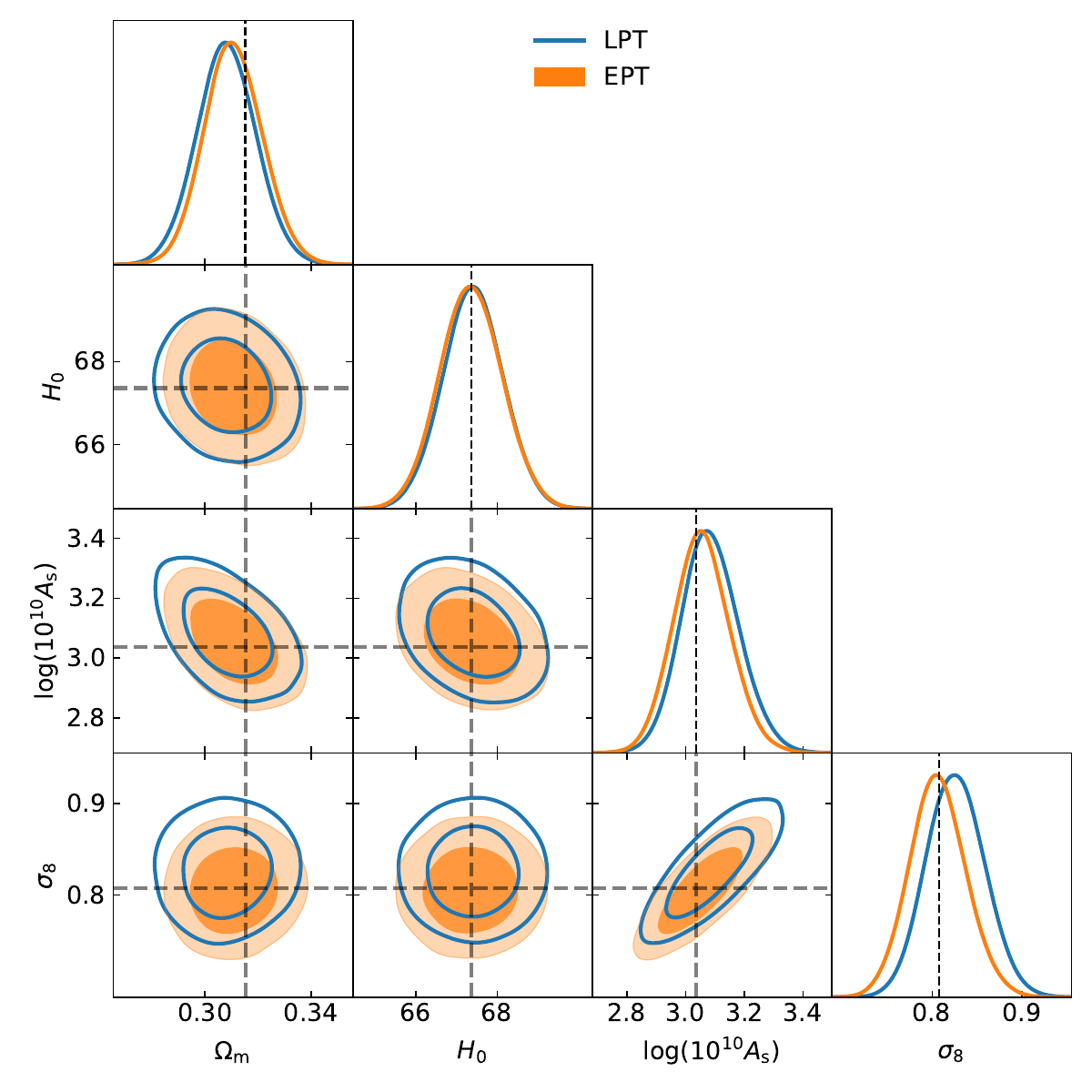}}
    \caption{Full-Modeling fits to the mean of LRG Cubic mocks using the LPT and EPT models within \velocileptors\ }
\label{fig: LPT_EPT}
\end{figure}

Lastly, the IR resummation in EPT is performed by splitting the wiggle and no-wiggle parts of the power spectrum, using the same method as is employed in modeling the poste-reconstruction BAO correlation function (\S~\ref{sec: BAO}) and applying a damping factor to the wiggle component. We refer readers to Ref.~\cite{ChenVlahWhite20} for full details of the Eulerian model and how it compares to LPT. We show in Fig.~\ref{fig: LPT_EPT} a comparison of Full-Modeling constraints when fitting the LRG cubic mocks using LPT and EPT. We see that the constraints agree to within fractions of a $\sigma$. A more detailed comparison between the two models, including fits to the ELG and QSO mocks for ShapeFit and Full-Modeling, is presented in Ref.~\cite{KP5s1-Maus} along with comparisons to other EFT models on the market.

\subsection{Varying $f$ and $\sigma_8$ separately}
\label{sec: fsig8_degen}

\begin{figure}
    \centering
    \resizebox{\textwidth}{!}{\includegraphics{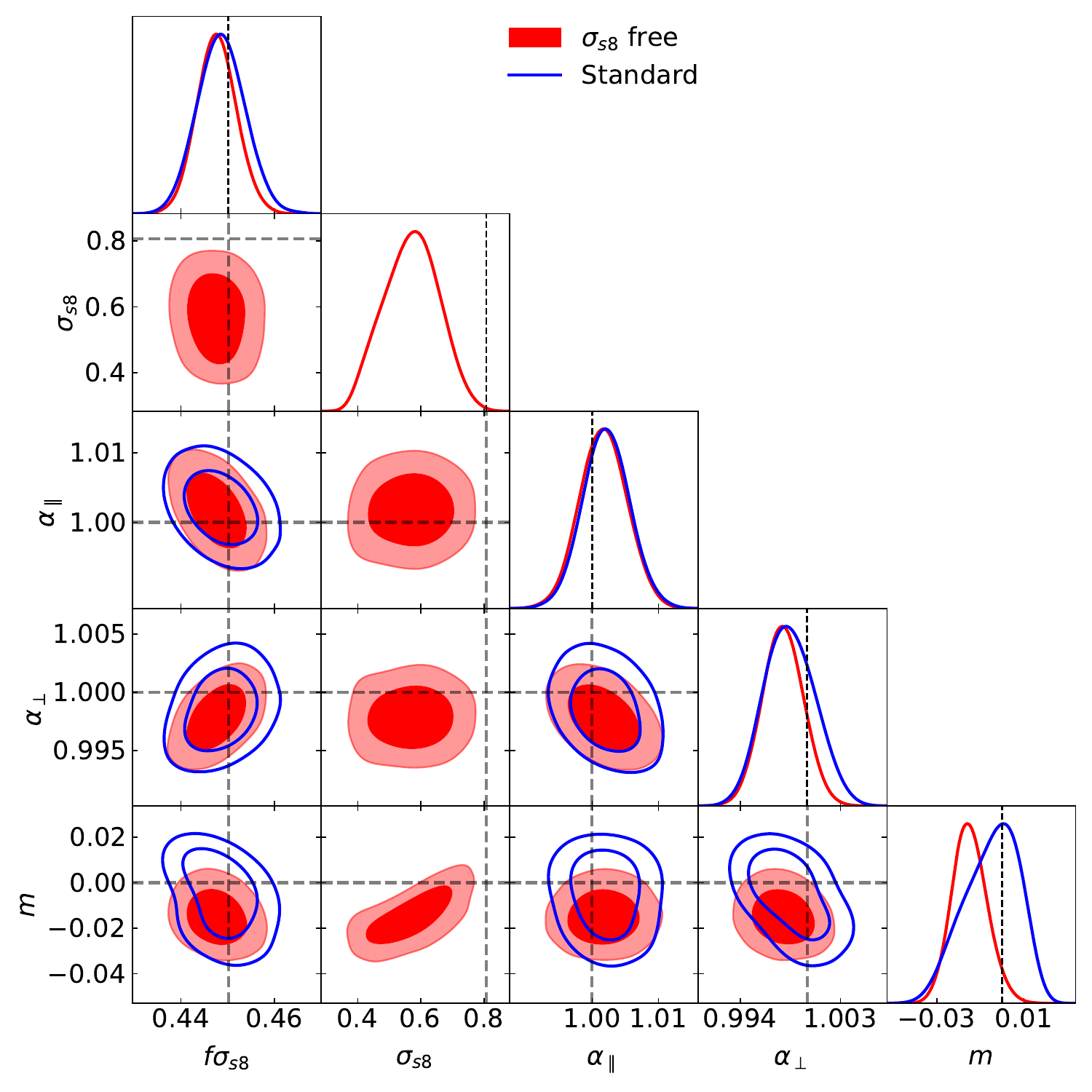}}
    \caption{ShapeFit constraints, where in the red data we allow $\sigma_{s8}(z=0)$ to vary independently of $f(z)$. This test is performed using the covariance for the 25 box volume of 200 $(\gpcih)^3$.}
\label{fig: sig8_free}
\end{figure}

The ``standard'' method of compression involves varying $f$ while keeping $\sigma_{s8}$ fixed to the fiducial value $\sigma_8^{\rm ref}$, and then reporting the product as $f^{\rm true}\sigma_{s8}^{\rm true}$. In principle, one should be able to vary $f$ and $\sigma_{s8}$ independently and present the result as $f^{\rm true}\sigma_8^{\rm true}$. This is because the degeneracy between $f$ and $\sigma_{s8}$ is broken in the 1-loop terms of the power spectrum. In order to test the ability to constrain $\sigma_{s8}$, we run a fit in which $\sigma_{s8}(z=0)$ is a free parameter in addition to $f(z)$ and the other compressed parameters. We vary $\sigma_{s8}(z=0)$ by re-scaling the linear power spectrum by:
\begin{align}
    P_{\rm lin}^{\prime}(k) = \left( \frac{\sigma_{s8}}{\sigma_8^{\rm fid}} \right)^2\times P_{\rm lin}(k),
\end{align}
Where $\sigma_8^{\rm fid} = 0.8076$ for the Abacus fiducial cosmology. The reported $f\sigma_{s8}$ is then $f\sigma_{s8} = f(z)\sigma_{s8} D(z)$
where the growth factor $D(z)$ is computed from the fiducial value of $\Omega_{\rm m} = 0.315$. We show these results in Fig.~\ref{fig: sig8_free}. We observe that even though $f\sigma_{s8}$ agrees with that obtained from the standard method, the $\sigma_{s8}$ constraint of $0.570 \pm 0.087$ is significantly below the true value of $0.8076$. This implies a growth rate $f(z)\sim [\Omega_{\rm m}(z)]^{0.55} > 1$ which is unphysical.  While it is unfortunate that the 1-loop corrections to the power spectrum can not sufficiently constrain $f$ and $\sigma_{s8}$ independently, we reiterate that our constraint on $f\sigma_{s8}$ remains robust. We also note a slight degeneracy between $\sigma_{s8}$ and $m$. While $m$ is designed to change the shape of the power spectrum, $\sigma_{s8}$ is an integrated quantity that is also mildly affected by changes in the shape.

\section{Conclusion}
\label{sec: conc}

Observations are probing the Universe and its evolution with unprecedented precision, allowing for significant improvements in measurements of fundamental parameters. The increased constraining power of these data also increases the sensitivity of our results to systematic effects present in models and analysis methods. The largest galaxy redshift survey to date, the Dark Energy Spectroscopic Instrument (DESI), is currently under way with its first year of Fullshape data being unblinded in the spring of 2024. To prepare for unblinding we must have a detailed understanding of the sources of systematic and theoretical error when fitting observations, the flexibility and limitations of our models, and the performance of different analysis methods. In this paper we presented tests of these effects using the public effective-perturbation-theory code {\tt velocileptors}, fitting data from the the AbacusSummit suite of simulations. Our focus will be on cosmological constraints using the Lagrangian Perturbation Theory (LPT) module in {\tt velocileptors}, though we also explore fits using its Eulerian Perturbation Theory (EPT) counterpart. In particular, we fit LRG, ELG, and QSO mock data at effective redshifts of $z=0.8,1.1,1.4$ respectively, consisting of clustering measurements from 25 cubic boxes of 8 ($h^{-1}$Gpc)$^3$ each for a total volume of 200 ($h^{-1}$Gpc)$^3$ for each tracer type. Companion papers to this one, using other effective perturbation theory codes {\tt Folps$\nu$ and PyBird}, are scheduled to appear concurrently (Refs.~\cite{KP5s3-Noriega,KP5s4-Lai}, including in addition a comparison paper (Ref.~\cite{KP5s1-Maus}) showing that all three effective-theory pipelines and models behave very similarly when the underlying assumptions and settings are consistent.

In this paper we discussed three modeling methods: (1) the standard Template fit, the default method used in previous BOSS and eBOSS analyses, that compresses observed multipoles into three summary statistics, ($f\sigma_{s8}$,$\alpha_{\parallel}$,$\alpha_{\perp}$) while keeping the linear power spectrum fixed; (2) the ShapeFit method which introduces an additional compressed parameter $m$ to the standard Template that modulates the shape of the linear template power spectrum which depends on early universe physics; and (3) the Full-Modeling method which directly samples in the parameter space of a cosmological model in order to fit the data. The first two methods are model-agnostic and so the compression only needs to be performed once, after which the obtained summary statistics can be mapped to any cosmological model ($\Lambda$CDM or extensions) of ones choosing. Despite the Full-Modeling method technically requiring a Boltzmann code to compute the linear power spectrum at every step of an MCMC, the use of Taylor series expansion emulators make the difference in computational cost/time negligible when compared to the compressed analyses.

We showed throughout the paper that the increased information from the shape of the linear power spectrum results in significant improvements in cosmological constraints in ShapeFit when compared to the standard Template analysis, when CMB data are not included. Compared to the Full-Modeling approach, ShapeFit provides consistent results on $\Lambda$CDM (and $w$CDM) parameters with minimal loss in constraining power. In varying the upper bound of the fitting range, we found that the models give unbiased constraints for scale cuts up to $k_{\rm max} \leq 0.2\ihmpc$. When including priors from Planck in order to constrain early universe information, all three methods give consistent results. Since the upcoming data will include tracers from different redshifts, we tested the ability of our pipelines in fitting simultaneously the tracers from three redshift bins, finding the joint analysis to improve the constraints without any noticeable systematic effects. 

Because one of the most powerful sources of cosmological information in LSS that DESI can detect is the Baryon Acoustic Oscillation (BAO) signal, whose well-defined scale can be used as a standard ruler to constrain the distance-redshift relation, we combined our fullshape analyses with post-reconstruction BAO correlation function, finding significant improvements in constraints for each modeling method. Finally, we also show how each method performs when extending the parameter space beyond the standard $\Lambda$CDM model by varying the dark energy equation of state parameter $w$. The ShapeFit and Full-Modeling methods are both able to obtain consistent and unbiased constraints within the wCDM model, whereas the standard template suffers greatly from degeneracies that can not be broken without shape information. 

In addition to the \velocileptors LPT model, the pipeline also has a module based on Eulerian perturbation theory (EPT). We show that these two theoretical frameworks provide consistent constraints, in agreement with the more extensive comparisons along with other PT pipelines, \texttt{FOLPS$\nu$} and \texttt{PyBird}, presented in Ref.~\cite{KP5s1-Maus}. 


We conclude by summarizing the optimal setup for \velocileptors\ for DESI Y1 fullshape analyses. The scaling of the biases with $\sigma_8$ appears to be a more natural choice of parameterization that is closer to the constraints from the data and can ameliorate shifts to lower $\sigma_8$ in the posteriors when the data is not sufficiently constraining. We recommend against the use of the partial Jeffrey's prior in attempts to reduce projection effects, due to it being a highly informative prior in the cosmological parameters. Our counterterm parameterization that scales relative to linear theory allows for a more intuitive choice of priors on the $\alpha_n$ parameters as ``fractional corrections to linear theory''. When fitting the hexadecapole we strongly suggest restricting the $k-$range in $P_4$ to a $k_{4,\rm max} \sim 0.1 \ \ihmpc$ as this minimizes the model's sensitivity to higher orders in perturbation theory and non-linear effects such as Fingers of God. For the monopole and quadrupole a scale cut of $k_{\rm max} = 0.20 \ \ihmpc$ has been found to perform well. Finally, we also suggest the use of physically motivated Gaussian priors on the stochastic parameters that can be justified based on the characteristic physical scales in the system (as captured, for example, in the halo model).

\section{Data availability}

Data from the plots in this paper are available on Zenodo as part of DESI’s Data Management Plan (DOI: \href{https://doi.org/10.5281/zenodo.10951714}{10.5281/zenodo.10951714}).
The data used in this analysis will be made public along the Data Release 1 (details in \href{https://data.desi.lbl.gov/doc/releases/}{https://data.desi.lbl.gov/doc/releases/})

\section*{Acknowledgements}
We thank Arnaud de Mattia, Pat McDonald, and other members of the Galaxy and Quasar Clustering working group within DESI for helpful discussions pertaining to this work. SC thanks Misha Ivanov and Matias Zaldarriaga for useful discussions on velocity stochasticities. MM and MW are supported by the DOE. SC acknowledges the support of the National Science Foundation at the Institute for Advanced Study.
This material is based upon work supported by the U.S. Department of Energy (DOE), Office of Science, Office of High-Energy Physics, under Contract No. DE–AC02–05CH11231, and by the National Energy Research Scientific Computing Center, a DOE Office of Science User Facility under the same contract. Additional support for DESI was provided by the U.S. National Science Foundation (NSF), Division of Astronomical Sciences under Contract No. AST-0950945 to the NSF’s National Optical-Infrared Astronomy Research Laboratory; the Science and Technology Facilities Council of the United Kingdom; the Gordon and Betty Moore Foundation; the Heising-Simons Foundation; the French Alternative Energies and Atomic Energy Commission (CEA); the National Council of Humanities, Science and Technology of Mexico (CONAHCYT); the Ministry of Science and Innovation of Spain (MICINN), and by the DESI Member Institutions: \url{https://www.desi.lbl.gov/collaborating-institutions}. Any opinions, findings, and conclusions or recommendations expressed in this material are those of the author(s) and do not necessarily reflect the views of the U. S. National Science Foundation, the U. S. Department of Energy, or any of the listed funding agencies.

The authors are honored to be permitted to conduct scientific research on Iolkam Du’ag (Kitt Peak), a mountain with particular significance to the Tohono O’odham Nation.

\appendix

\section{Analytic Marginalization}
\label{appendix: am}

We can substantially speed up our MCMC fits by analytically marginalizing over the linear nuisance parameters in our model, i.e.\ the parameters of the stochastic and counterterm contributions ($\alpha_0$, $\alpha_2$, $\alpha_4$, $SN_0$, $SN_2$, $SN_4$). By reducing the number of sampled parameters our chains are able to converge in under 10 minutes instead of an hour or two. The procedure for marginalizing over the linear parameters $b_i$ involves splitting the theoretical prediction, into the piece dependent on the nonlinear parameters $\boldsymbol{a}$ that we sample in  and the ``template'' piece that is multiplied by the linear parameters: $\bPsi = \bPsi_{0}(\boldsymbol{a}) + \sum_i \theta_i\bPsi_{t,i}$. The likelihood distribution marginalized over the linear nuisance parameters is given by\cite{Bridle02,Taylor10}

\begin{figure}
    \centering
    \resizebox{\textwidth}{!}{\includegraphics{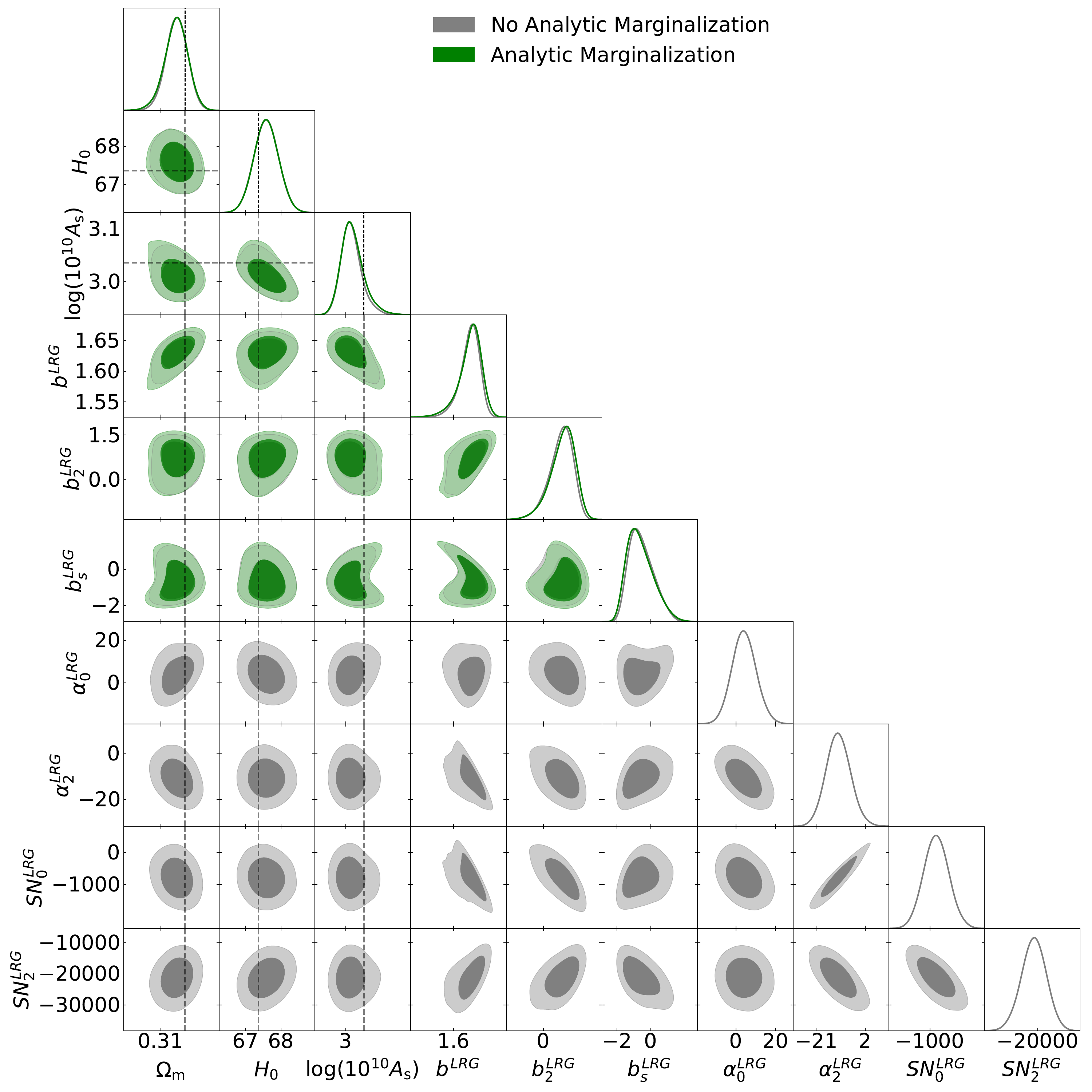}}
    \caption{Comparison of Full-Modeling constraints with and without analytic marginalization of linear parameters: $\alpha_0$, $\alpha_2$, SN$_0$, SN$_2$. Here we show results for the full $25\times (2 \gpcih)^3$ volume to show that our analytical marginalization method is robust even at very large volumes.
    }
\label{fig: FM_am}
\end{figure}

\begin{align}
    P(\bPsi_{d}|\bPsi_{0},\bPsi_{t},\sigma_\theta) = \int d\btheta\ \mathcal{L}(\bPsi_{d}|\bPsi_{0},\bPsi_{t},\btheta)P(\btheta),
\end{align}
where $\bPsi_{d}$ is the data and $P(\btheta)$ denotes the priors on parameters $\theta_i$, which we choose to be Gaussian (centered at zero) with widths $\sigma_{\theta,i}$:
\begin{align}
    P(\theta_i|\sigma_{\theta,i}) = \frac{1}{\sqrt{2\pi\sigma_{\theta,i}^2}}\exp\left(-\frac{\theta_i^2}{2\sigma_{\theta,i}^2}\right) 
\end{align}
The model likelihood in the integrand is 
\begin{align}
    \mathcal{L}(\bPsi_{d}|\bPsi_{0},\bPsi_{t},\btheta) &= (2\pi)^{-n/2} \left| \mathcal{C}^{-1}\right| \nonumber \\
    &\times e^{-\frac{1}{2}\left[\bPsi_d-\left(\bPsi_0 +\sum_i \theta_i\bPsi_{t,i}\right)\right]^{\rm T}\mathcal{C}^{-1}\left[\bPsi_d-\left(\bPsi_0 +\sum_i \theta_i\bPsi_{t,i}\right)\right]}.
\end{align}
Defining $\bDelta = \bPsi_d-\bPsi_0$ and $\log \mathcal{L}_0 = -\frac{1}{2}\bDelta^{\rm T}\mathcal{C}^{-1}\bDelta$ we get
\begin{align}
    P(\bPsi_{d}|\bPsi_{0},\bPsi_{t},\sigma_\theta) &\propto \mathcal{L}_0 \int d\btheta\ e^{-\frac{1}{2}\sum_{i,j}\theta_i\theta_j\left( \bPsi_{t,i}^{\rm T}\mathcal{C}^{-1}\bPsi_{t,j} + \frac{1}{\sigma_{i}\sigma_{j}}\delta_{ij}\right) + \sum_i\bDelta^{\rm T}\mathcal{C}^{-1}\theta_i\bPsi_{t,i} } \nonumber \\
    &= \mathcal{L}_0 \int d\btheta\ \exp \left[-\frac{1}{2}\left(\btheta^{\rm T}L\btheta  - V^{\rm T}L^{-1}V\right)\right] \nonumber \\
    &= \frac{(2\pi)^{n/2}}{\sqrt{|L|}}\mathcal{L}_0 e^{\frac{1}{2}V^{\rm T}L^{-1}V},
\end{align}
where we completed the square in the second line and defined the matrices $L_{ij} = \bPsi_{t,i}^{\rm T}\mathcal{C}^{-1}\bPsi_{t,j} + \delta_{ij}/(\sigma_i\sigma_j)$ and $V_i = \bPsi_{t,i}^{\rm T}\mathcal{C}^{-1}\bDelta$ before taking the multivariate Gaussian integral. So then the log-likelihood consists of the four terms
\begin{align}
    \log P = \log\mathcal{L}_0 + \frac{1}{2}V^{\rm T}L^{-1}V - \frac{1}{2}\log|L| +\frac{n}{2}\log(2\pi).
\label{eq: logP}
\end{align}
Despite analytically marginalizing over the linear parameters, we can always recover their distribution using the chain containing non-linear parameters. At each step of the chain, the nonlinear parameters are fixed and the likelihood is a Gaussian function of the linear parameters with known mean and variance, i.e.\ for step $n$ in the MCMC, the likelihood depends on linear parameter $\theta_i$ like:
\begin{equation}
    \log\mathcal{L}_{n,i} = (\theta_i - \bar{\theta}_i)^{\rm T}\mathcal{N}^{-1}(\theta_i - \bar{\theta}_i) + \mathrm{const}
\end{equation}
with variance $\mathcal{N}$ and the mean $\bar{\theta}_i$ determined by the (fixed) non-linear parameters. Reconstructing the distribution of parameter $\theta_i$ simply amounts to averaging over all of these Gaussians. This allows us to still be able to e.g.\ check the effects of our priors or to identify any degeneracies between linear parameters and others in the model that could be driving projection effects.  
We show in Fig.~\ref{fig: FM_am} a comparison of constraints from the Full Modeling method with and without analytic marginalization of the linear parameters. For the parameters that are being sampled in both cases, we find consistent behavior in the contours as expected. In order to make sure that the analytic marginalization is also correctly handling the parameters that we marginalize over, we maximize the first two terms in \ref{eq: logP} (the latter terms describe the volume/width of the likelihood surface). This gives us the best-fitting values for the nonlinear parameters. From the maximized posterior, the corresponding best-fit points of the analytically marginalized parameters can then be directly calculated: 
\begin{align}
    \theta_j^{\rm bf} = \sum_i V_i L_{ij}^{-1}.
\end{align}
Once we have found the best-fitting nonlinear parameters and by extension $\bPsi^{\rm bf} = \bPsi_0^{\rm bf} + \sum_i \theta_i^{bf}\bPsi_{t,i}^{\rm bf}$, the maximum log-likelihood is just:
\begin{align}
    \log P^{\rm max} = -\frac{1}{2}[\bPsi^{\rm bf}]^{\rm T}\mathcal{C}^{-1}\bPsi^{\rm bf} +\log |\mathcal{C}^{-1}| - \frac{n}{2}\log(2\pi).
\end{align}

In Table~\ref{tab:am} we show the best-fitting parameter values from Full-Modeling fits with and without analytic marginalization. We see that the parameters that we marginalize over are well behaved and on the same order as they take when being sampled. 

We also note the third term in Eq.~\ref{eq: logP}, $-(1/2)\log |L|$, which is the log of the determinant of the (linear parameter) part of the Fisher matrix. One prior choice that one can very easily implement is a ``partial Jeffrey's prior'' which removes this term from the likelihood. This prior can cause significant shifts in constraints in cases where parameter projection effects are noticeable, as the Jeffrey's prior removes some of the phase space volume from the likelihood. We discuss the implications of such a prior in Appendix~\ref{appendix: proj_priors}.

\begin{table}
\centering
\begin{tabular}{cc|c|c}
\multirow{6}{*}{Non-linear} & Params & FM Standard ($\sigma$) & FM Analytic Marg ($\sigma$) \\
\hline
& $H_0$ &  67.67 (0.35) & 67.63 (0.34)\\[0.5ex]
& $\Omega_{\rm m}$ & 0.3139 (0.0023)& 0.3143 (${}^{+0.0026}_{-0.0023}$)\\[0.5ex]
&  $\log(10^{10} A_\mathrm{s})$& 2.998 (${}^{+0.017}_{-0.023}$)&3.001 (${}^{+0.017}_{-0.026}$)\\[0.5ex]
& $b\sigma_8$ & 1.642 (${}^{+0.019}_{-0.013}$)&1.644 (${}^{+0.022}_{-0.013}$)\\[0.5ex]
& $b_2$ & 0.8982 (${}^{+0.49}_{-0.32}$)&0.8705 (${}^{+0.51}_{-0.33}$)\\[0.5ex]
& $b_s$ & -0.7607 (${}^{+0.55}_{-0.87}$)&-0.8512 (${}^{+0.55}_{-0.95}$)\\[0.5ex]
\hline
\multirow{4}{*}{Linear} & $\alpha_0$ & 0.6987 (6.1)& 2.468\\
& $\alpha_2$ &-11.69 (5.7)& -13.08\\
& $SN_0$ & -890.3 (420)&-962.4\\
& $SN_2$ & -1.919e4 (4300)& -1.911e4
\end{tabular}
\caption{Comparison of Full-Modeling best-fit parameters with and without analytic marginalization. Uncertainties of the posterior distributions are given in parentheses for all sampled parameters.
}
\label{tab:am}
\end{table}

\section{Parameter projection effects and the role of priors}
\label{appendix: proj_priors}

\begin{figure}
\captionsetup[subfigure]{labelformat=empty}
\begin{subfigure}{.5\textwidth}
\centering
\includegraphics[height=8cm]{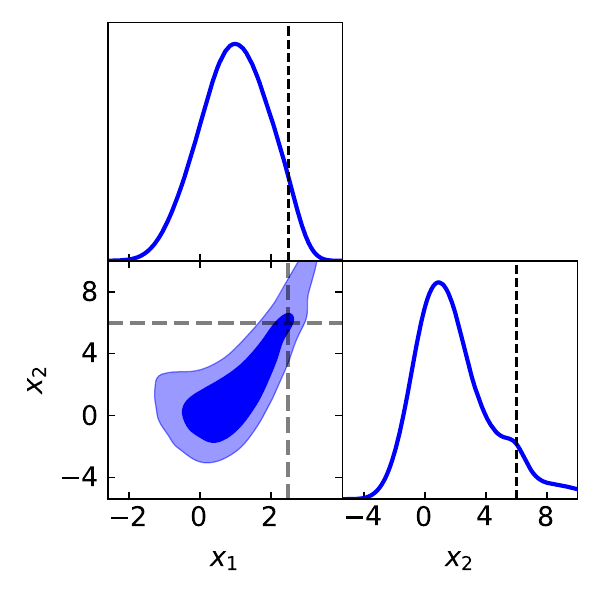}
\end{subfigure}%
\begin{subfigure}{.5\textwidth}
\centering
\includegraphics[height=8cm]{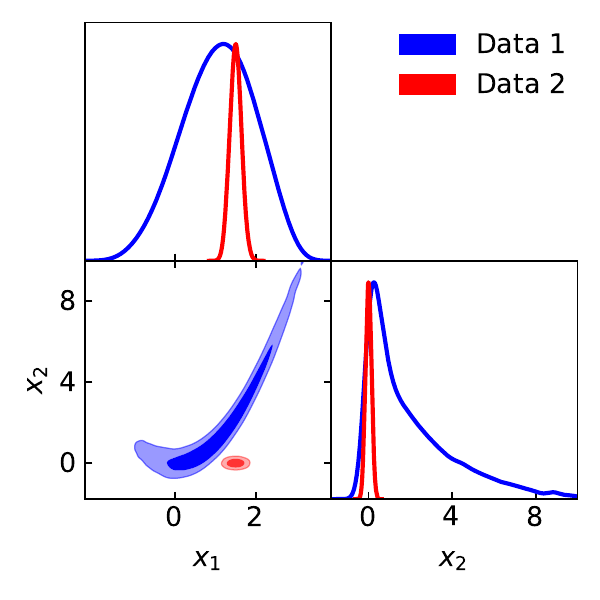}
\end{subfigure}%
    \caption{Toy model examples of information loss in projected posteriors. The left panel shows the posteriors from sampling from a likelihood distribution that is constructed out of the sum of a small Gaussian and a Rosenbrock function in 2D. The dashed lines label the maximum likelihood values of the two parameters $x_1$ and $x_2$.  The `truth', and likelihood maximum, appears to be in the tail of the 1D posteriors due to the large volume (area) at only slightly lower likelihood near $x_1\approx x_2\approx 0$. The right panel shows posteriors from two different ``data sets'' (different likelihood distributions). Data 1 is constructed from a Rosenbrock function and Data 2 is a Gaussian distribution. While in the full space (2D) it is clear the posteriors disagree, in projection (here 1D) the posteriors appear consistent.
    }
\label{fig: proj_ex}
\end{figure}

In this section we discuss the role of priors on the parameters of our model and the effect they can have on parameter projection effects -- defined here as shifts in the marginal posteriors away from the maximum likelihood regions due to a non-Gaussian posterior surface.  These effects frequently arise when there are several parameters in the model that are poorly constrained or partially degenerate.  If there are degeneracies between parameters in the model, regions of the parameter space far from the maximum likelihood point may have very little likelihood penalty compared to the best fit.  In spaces with large numbers of dimensions the ``parameter volume'' in such regions can be large, and integration over a subset of these parameters can shift the peaks or means of the marginal posterior distribution significantly away from the maximum likelihood values or the ``input cosmology'' in our tests.  In addition, when the data are not sufficiently powerful the constraints on the cosmological parameters can depend on the choice of priors and the parameterization.

It is notoriously difficult to visualize complex probability distributions in high-dimensional spaces, and unfortunately projections \emph{necessarily} remove information even if they are given from many viewpoints. For this reason marginal likelihoods can appear consistent (i.e.\ overlap in projection) when they are not and they can appear inconsistent when they are actually consistent.  Even linear changes of the projection axes can change the appearance of concordance. Such issues are by no means specific to our models: projection effects in high-dimensional parameter spaces have been encountered in many areas of cosmology and have been widely discussed in the literature (see e.g.\ refs.~\cite{Handley19b,Lemos21,GomezValent22,Hadzhiyska23,Sailer24} for recent discussions). 

In Fig.~\ref{fig: proj_ex} we show two toy model examples of projections, where the left plot is inspired by Fig.~1 of Ref.~\cite{GomezValent22} and the right plot is inspired by Fig.~1 of Ref.~\cite{Lemos21}. For the first example, we construct a fake likelihood distribution by adding a Rosenbrock function, $f(x_1,x_2) = (1.0-x_1)^2 + 0.5(x_2 - x_1^2)^2$, and a sharp 2D Gaussian centered at $(\bar{x}_1 = 2.5,\bar{x}_2 = 6)$ with a width of $\sigma=0.25$ along both parameter directions. The maximum of the total likelihood distribution is very close to the center of the Gaussian, and is labeled with grey dashed lines in the figure. However, the contribution of the Rosenbrock function peaks at $(1.0,1.0)$ but in a much more gradual way. The result is more likelihood ``volume'' for the MCMC to explore near $(1.0,1.0)$ than near the true maximum of the whole likelihood. As a result, the marginal posterior distributions for parameters $x_1$ and $x_2$ are significantly offset from the true best-fitting points.

The second cautionary example of projections is presented in the right panel of Fig.~\ref{fig: proj_ex} and shows posteriors from two ``data sets'', which we simulate by constructing two different fake likelihood distributions. For Data 1 we again use a Rosenbrock function, $f(x_1,x_2) = (1.0 - x_1)^2 + 10(x_2 - x_1^2)^2$ and for Data 2 we use a Gaussian with means $(\bar{x}_1 = 1.5,\bar{x}_2 = 0.0)$ and widths of 0.2. In this example we demonstrate how the constraints on $x_1$ and $x_2$ appear to agree for the two data sets when looking at the 1D posteriors, but in the 2D panel the two data sets are clearly in tension. This serves as a cautionary tale about interpreting constraints from a multi-dimensional posterior surface when looking at the projections onto lower dimensions. It is naturally difficult to visualize an N-dimensional volume, but looking only at 1D or 2D projections of the full distributions might lead one to misinterpret results.

Finally, as an honorable mention,  we refer readers to  Fig.~7 of Ref.~\cite{Handley19b} in which the authors show a toy model of posteriors from two different data sets with three sampled parameters, $x$, $y$, $z$. The posteriors for these three parameters are consistent between data sets. However, after performing a linear transformation to new coordinates, ($x + y -z$, $x + z -y$, $y + z -x$) one finds discrepant constraints on $x + y -z$. This shows that tensions can be hidden due to particular choices of parameterization, and that appropriate coordinate-independent metrics are necessary to measure the consistency between data sets or results.

\subsection{Projection effects for DESI}
\label{appendix: proj_desi}

\begin{figure}
\captionsetup[subfigure]{labelformat=empty}
\begin{subfigure}{.5\textwidth}
\centering
\includegraphics[height=8cm]{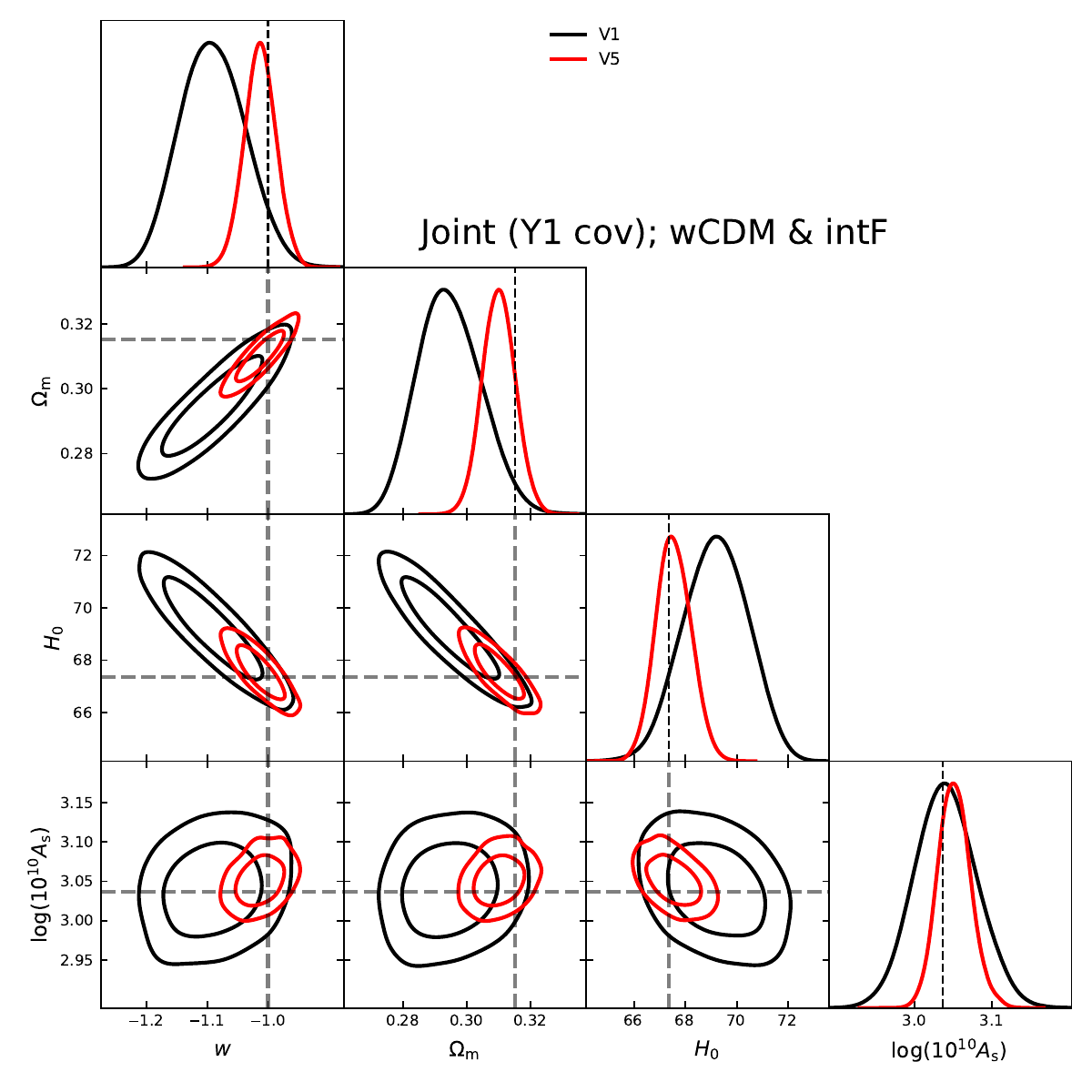}
\end{subfigure}%
\begin{subfigure}{.5\textwidth}
\centering
\includegraphics[height=8cm]{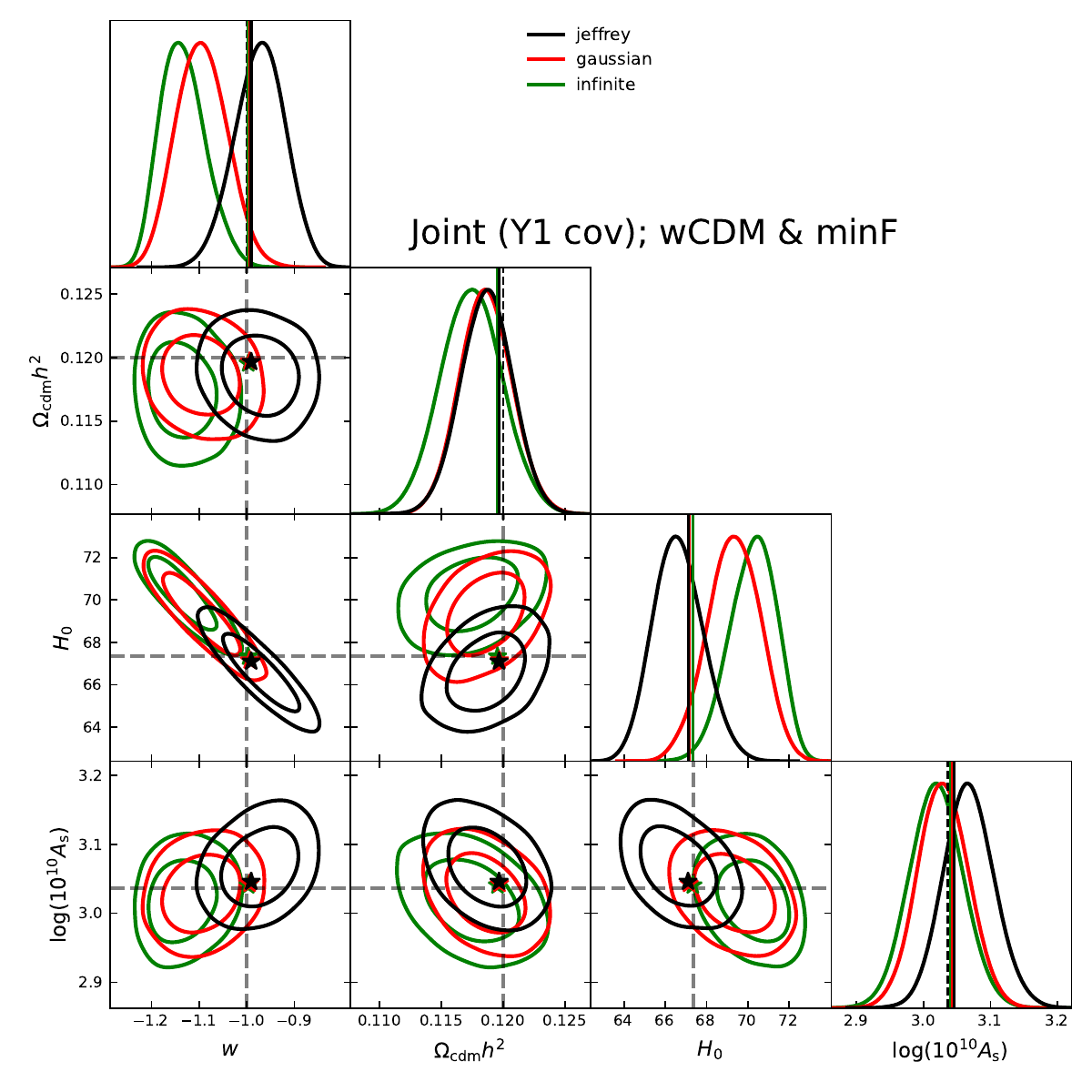}
\end{subfigure}%
    \caption{Full-Modeling $w$CDM fits to synthetic data created in each of the DESI Y1 redshift bins with corresponding analytic covariances. \texttt{Left:} Intermediate fits with the real Y1 volumes (black) compared to fits with the covariances rescaled by a factor of 1/5. \texttt{Right:} Minimal freedom fits with three different prior choices on the analytically marginalized counter and stochastic terms. The black contours correspond to a \textit{partial} Jeffrey's prior (only on linear parameters) discussed in the text, the red contours show the fit with our usual Gaussian priors described in Table~\ref{tab: param_priors}, and the green contours correspond to `infinite' priors. The stars in the 2D panels and solid vertical lines in the diagonal (1D posterior) panels denote the best-fit models obtained by running a minimizer starting at the MAP values of the chains.
    }
\label{fig: Y1_wCDM}
\end{figure}

To demonstrate the impact of projection effects in the specific case of DESI data with covariances similar to those expected from the first year we turn to synthetic data created with \velocileptors for each of the seven DESI Y1 redshift bins: BGS ($0.1<z<0.4$), LRG ($0.4<z<0.6$, $0.6<z<0.8$, $0.8<z<1.1$), ELG ($0.8<z<1.1$, $1.1<z<1.6$), and QSO ($0.8<z<2.1$).  Since the data we are fitting to have been generated from the model, with no noise added, the best-fit point occurs at ``truth'' and has $\chi^2=0$.  However $\chi^2$ may rise slowly along some directions which have significant volume, shifting the marginalized posteriors away from the best-fit point.  While the $\Lambda$CDM (with and without fixing $n_s$) and k$\Lambda$CDM  models do not exhibit significant projection effects, we do observe them for wCDM. We show the wCDM joint fits to the seven Y1 redshift bins in Fig.~\ref{fig: Y1_wCDM}.  Note that the marginal posteriors on several parameters (black lines in the left hand panels of Fig.~\ref{fig: Y1_wCDM}) peak way from the input model, even though the model is, by construction, a good fit to the (mock) data and the maximum likelihood point is (again by construction) at the true values of the parameters.  As the data become more constraining these projection effects are reduced -- shown as the red contours in the same figure where the errors have been scaled down by a factor of 5.  Note that some projection effects are still visible in the red contours.  The posterior for $\Omega_m$ is still offset by a non-trivial fraction of its ``new'' error bar, but the absolute value of the offset is reduced.  As we continue to reduce the error bars the contours shrink to eventually be $\delta$-functions at the true values.  It is also worth noting another feature of these projection effects.  They typically occur when there are many parameters, some of which are partially degenerate.  They also tend to lead to shifts that are $\mathcal{O}(1\,\sigma)$.  This is because the likelihood falls as $\exp[-\chi^2/2]$ moving away from the best-fit point, while the volume in parameter space grows as a power of the ``parameter distance''.  Eventually the Gaussian overcomes the impact of the volume. In the right panel of Fig.~\ref{fig: Y1_wCDM} we show wCDM constraints to the same synthetic data using three choices of priors on the linear parameters ($\alpha_0$,$\alpha_2$,SN$_0$,SN$_2$): infinite uniform, Gaussian, and the (partial) Jeffrey's prior. The stars and solid vertical lines denote the best-fit values obtained from running a minimizer, and demonstrate that the shift between marginal posteriors and maximum likelihood values are due to projection effects. We find that these projection effects are slightly reduced when switching from the flat to Gaussian prior, showing that the Gaussian priors on the linear parameters are not entirely uninformative. The projection effects are more significantly reduced when applying the Jeffrey's prior and we discuss the implications of using such a prior in the next section. 

\subsection{Jeffrey's prior and reparameterizations}

In addition to shifts in the posteriors such that they peak away from the `true' values, insufficiently constraining data in a high-dimensional parameter space can lead to increased sensitivity to priors and choice of parameterizations.  This is another manifestation of the likelihood not dominating the posterior and is a generic feature of inference in high dimensions.  If we had firm theoretical reasons to prefer one model parameterization over another this would not be a problem, but in practice there are several choices between which there is little theoretical preference.  We discuss some of these implications here -- first discussing the choice of parameters and then the Jeffrey's prior.

A natural\footnote{This is not the only choice.  One could imagine choosing e.g.\ log priors in the mass scale of the halos hosting the galaxies, or linear deviations from the peak-background split prediction (where the $b_{n>1}$ are non-linear functions of $b_1$), or many other choices.} set of parameters for the model would be the cosmological parameters (e.g.\ $\sigma_8$) and the bias parameters and counterterms ($b_i$ and $\alpha_i$).  However some of these are at least partially degenerate.  Lowering $A_s$ or $\sigma_8$ while raising $\alpha$ can leave $\alpha\, k^2P$ unchanged, and a similar upward adjustment of $b_i$ can reduce much of the impact from the other terms so that $\chi^2$ changes little.  Since, for linear priors on $b_i$ and $\alpha_i$, there is more ``volume'' at large values than small there is a natural tendency to shift the posterior to lower $\sigma_8$.  The quantities best-constrained from observation are the power spectrum multipoles, and in particular the monopole.  For this reason we use parameters that are closer to the data space, i.e.\ $b\sigma_8$ rather than $b$ (see Table \ref{tab: param_priors}).  While this is a natural choice, in terms of the $b_i$ it corresponds to a prior that rises with $\sigma_8$ \cite{Chen22_2}. For example, the Jacobian translating between $(b,\sigma_8)$ and $(b\sigma_8,\sigma_8)$ is simply $\sigma_8$.  Inference using the second set of parameters is thus equivalent to inference using the first, plus a prior $P(\sigma_8)\propto\sigma_8$.  When $\sigma_8$ is not well constrained by the data, this prior choice will shift the marginal posterior.  Similar comments hold for the other parameters of course.

\begin{figure}
    \centering
    \resizebox{\columnwidth}{!}{\includegraphics{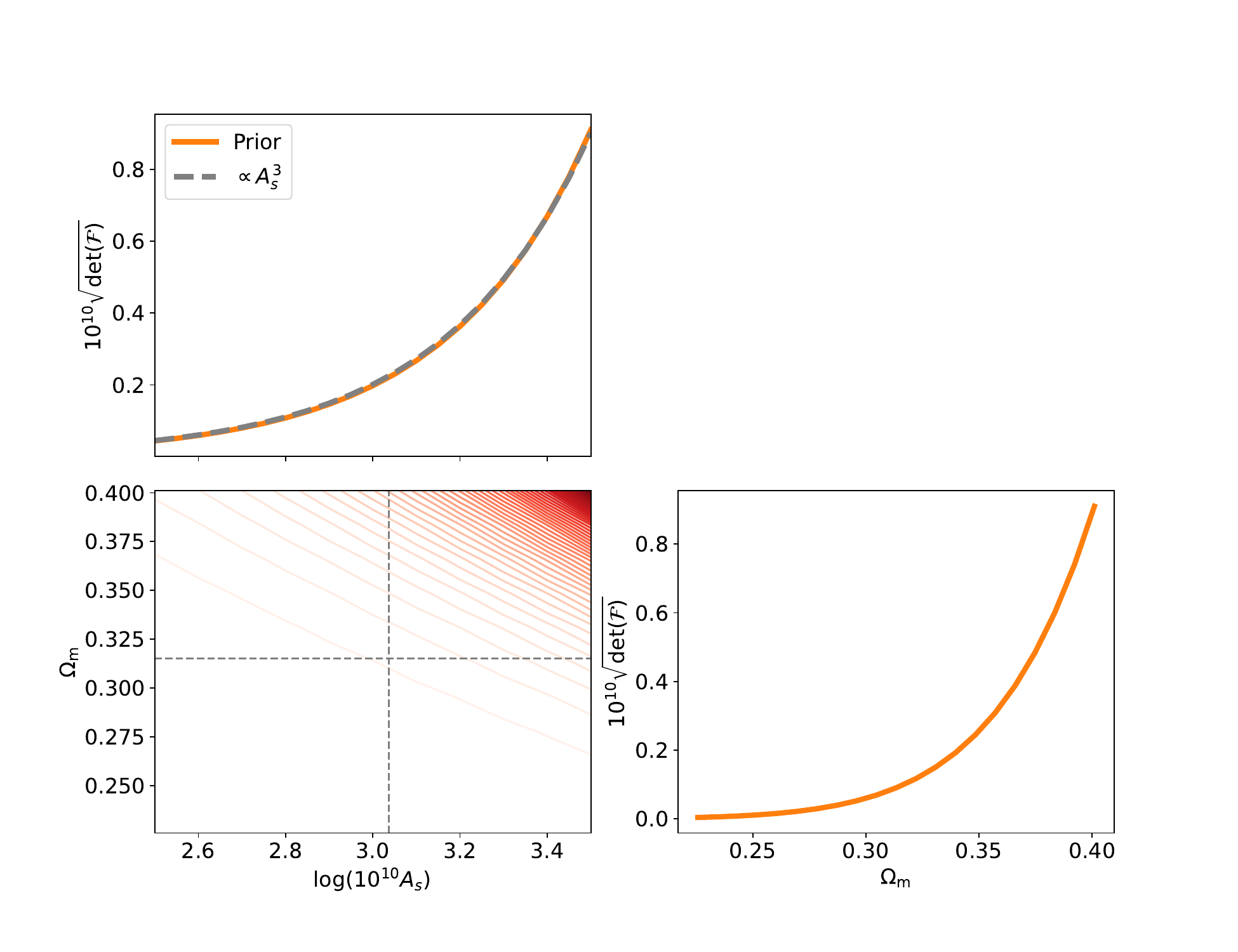}}
    \caption{A 2D slice through the \textit{partial} Jeffrey's prior for $\Lambda$CDM. We show the variation of the prior in the $\Omega_m$ and $A_s$ directions, with all other parameters fixed to their best fit.  The dashed grey lines in the lower left panel show the values of $\log(10^{10}A_s)$ and $\Omega_m$ along which $\det F$ was evaluated in each of the other panels. The dashed grey line in the upper left panel shows a power law $\propto A_s^3\propto\sigma_8^6$. Note the strong dependence of this prior on the cosmological parameters (see text).}
\label{fig:Jeffreys}
\end{figure}

A method that is sometimes used in the statistics literature to reduce the impact of parameter changes is to include a ``Jeffrey's prior''.  This corresponds to the square root of the determinant of the Fisher matrix, and has the same role as the familiar $\sqrt{-g}\,d^4x$ in General Relativity.  If implemented consistently, this removes the Jacobian from transformations of variables and so is sometimes termed\footnote{While common, this nomenclature is incorrect.  A much better term would be ``reparametersation invariant'' since in general -- and in our case -- the prior is ``informative'' from the point of view of inference.} ``uninformative''.  There are some concerns about taking this approach in our situation however\footnote{The Jeffrey's prior and problems with it are also discussed in ref.~\cite{Syversveen98}, including an example from ref.~\cite{BayesianTheory01}.}.  First, we do not believe that the physics indicates that e.g.\ $(\ln[1+b^{10}],\cosh\sigma_8)$ is as good a parameter set as $(b\sigma_8,\sigma_8)$ for example. Our parameters have at least some theoretical justification that we'd like to include as ``prior information'' in our model specification.  Secondly, as usually implemented, the Jeffrey's prior is a strong function of several key cosmological parameters.

To see this, let us consider the \textit{partial} Jeffrey's prior that is sometimes introduced.  This involves computing $\sqrt{\det F}$ for only those parameters that enter the model linearly (if all parameters enter linearly, then this is the ``full'' Jeffrey's prior, however in that limit the likelihood is Gaussian so the issue of projection effects does not arise).  The calculation in the previous appendix shows that introducing such a prior is equivalent to dropping the $\log ||L||$ term in Eq.~(\ref{eq: logP}) (see also ref.~\cite{Hadzhiyska23}), making this a very easy change to make.  That this prior is a strong function of the underlying cosmological parameters is most easily seen by again considering $\sigma_8$.  The Fisher matrix has the form
\begin{equation}
    F \sim \frac{\partial(\mathrm{theory})}{\partial(\mathrm{param})} C^{-1} \frac{\partial(\mathrm{theory})}{\partial(\mathrm{param})} 
    \sim (\mathrm{template})\, C^{-1}\, (\mathrm{template})
\end{equation}
where in the second step we have used the fact that for parameters entering linearly the derivative is just some linear-parameter-independent template -- e.g.\ for $\alpha k^2 P$ it would be $k^2P$.  In the case of our perturbative model, each of these `templates' is $P_{\rm lin}$ or some integral over one or more powers of $P_{\rm lin}$ and thus we expect the template to scale as a power of $A_s$ or $\sigma_8$.  The Fisher matrix is thus also a (high) power of $A_s$ or $\sigma_8$ and so including such a prior has the effect of shifting the marginal posterior to higher $\sigma_8$.

Fig.~\ref{fig:Jeffreys} shows a 2D slice through this (high-dimensional) prior to illustrate the previous points.  We have chosen to show the variation in the $\Omega_m$ and $A_s$ directions with all of the other parameters held fixed at their best-fit points.  The strong dependence on $A_s$ is clear ($\propto A_s^3\propto \sigma_8^6$), and has been described above. The $\Omega_m$ dependence can be understood similarly.  Raising $\Omega_m$, with all other parameters fixed, changes the shape of $P_{\rm lin}$ with more power on the quasi-linear scales of relevance to DESI (and less power at large scales).  The increase in the amplitude of $P_{\rm lin}$ increases $\det F$ in the same manner as for $A_s$ or $\sigma_8$.  The dependence on each of the other parameters can be similarly computed and understood, though they are not shown here for simplicity.  The introduction of such a prior is thus ``informative'' or ``strongly informative'' in the sense of introducing non-negligible shifts in the marginal posteriors given the size of the uncertainties. We note that in making Fig.~\ref{fig:Jeffreys} we used the more traditional form for the counterterms, e.g.\ $\alpha k^2 P_{\rm lin}$ instead of the parameterization of Eq.~\ref{eq: cterms}, since it is in that context that (partial) Jeffrey's priors have typically been discussed. For most of this paper we have chosen parameters scaling like $\alpha\sigma_8^2$, meaning that the ``template'' is closer to $k^2P_{\rm lin}/\sigma_8^2$ and is therefore largely independent of $\sigma_8$. Indeed, we find that in our preferred parameterization the (partial) Jeffrey's prior scales much more weakly with $\sigma_8$ than what is usually encountered. However, the strong dependence on $\Omega_{\rm m}$ and other cosmological parameters is unaffected by this particular reparameterization. 

There are two things to note about these examples.  First, in each case the shift in the marginal posterior was accomplished by the introduction of a what is effectively a prior, and not by any change in the model or the data.  It relies on the fact that the data are not sufficiently constraining such that such prior or parameterization choices are relevant.  Second, the two approaches change the prior through different parts of the theory.  In the first case we modified the biases while in the second we introduced a prior through the counterterms.  

Luckily the existing theoretical models are sufficiently accurate to model much more constraining data than DESI Y1 without the need to introduce additional free parameters (see the main body of the paper and refs.~\cite{Chen21,KP5s1-Maus}).  As the data become more constraining the impact of parameter choices and priors is expected to reduce, as shown earlier.  Combining the DESI data with other datasets that can break degeneracies is also expected to reduce the impact of these effects.  In this sense, the Y1 data may well be a ``worst case'' scenario.

\section{Connection to the halo model}
\label{appendix: halo}

It is sometimes helpful to establish the expected sizes of the terms in the theoretical model.  This can be done through arguments of self-consistency (see main text), and by comparing to other models.  In this appendix we compare the PT approach to a simplified, analytical halo model \cite{Seljak00,Peacock00} with the goal of understanding the expected size of the stochastic terms (see also the discussion in ref.~\cite{Chen_BOSSrecon2022}).  Since our goal is to gain insight, we shall deal with an analytically tractable version of the halo model in which galaxies reside in spherical, self-similar halos whose centers are distributed according to biased linear theory with scale-independent bias.  If $n(M)$ is the volume density of halos per unit mass, and each halo has a Fourier-space density profile $u(k,m,z)$, normalized to unity as $k\to 0$, then the power spectrum is (see e.g.\ ref.~\cite{Schaan21} for a recent, pedagogical discussion with references to the original literature)
\begin{equation}
    P_g(k,\mu,z) = P_g^{\rm 2-halo}(k,\mu,z) + P_g^{\rm 1-halo}(k,\mu,z) + P_g^{\rm shot}
    \quad .
\label{eqn:Phalomodel}
\end{equation}
If $N_\text{cen}(m)$ and $N_\text{sat}(m)$ denote the mean number of centrals and satellites in a halo of mass $m$ the mean number density of galaxies is simply $\bar{n}_g = \int dm\ n(m) \left[ N_\text{cen}(m) + N_\text{sat}(m) \right]$.  To compute the clustering we need to know the statistics of the galaxy occupation, and we shall follow standard practice in assuming the centrals are Bernoulli distributed while the satellites are Poisson distributed.

Under the above assumptions the 2-halo term in the power spectrum is given by:
\begin{equation}
    P_g^\text{2-halo} = \left( b_g + F \mu^2 \right)^2\;  P_\text{lin},
\label{eqn:2halo}
\end{equation}
where the bias is
\begin{equation}
  b_g(k, \mu, z) \equiv \frac{1}{\bar{n}_g} \int  dm\ n(m)\; b(m)  \; 
  \left[ N_\text{cen} + N_\text{sat} u(k,m,z) e^{-k^2 \mu^2\sigma_d^2(m) / 2} \right]
\end{equation}
and the effective growth rate of structure is
\begin{equation}
  F(k, \mu, z) \equiv f\; \int dm \; n(m)\;
  \left( \frac{m}{\bar{\rho}} \right) u(k,m)  e^{-k^2 \mu^2 \sigma_d^2(m) /2}
\end{equation}
which tends to $f$ as $k\to 0$.  In the above we have written the (linear) bias of a halo of mass $m$ as $b(m)$ and the mean matter density in the Universe as $\bar{\rho}$. We have also used the fact that in going into redshift-space, the density profile acquires a damping factor from the virial motions in halos: 
\begin{align}
    u(k,m,z)\rightarrow u(k,m,z)e^{-k^2\mu^2\sigma_d^2/2},
    \label{eqn:uk_fog}
\end{align} 
where $\sigma_d^2(m)$ is the velocity dispersion of such a halo in distance units. 
The 1-halo term has in its integrand the term $\langle N(N-1) \rangle$ which, when expanded is:
\begin{align}
    \langle N(N-1) \rangle &= \langle (N_\text{cen} + N_\text{sat})(N_\text{cen} + N_\text{sat} -1)\rangle \nonumber \\
    &= \langle N_\text{cen}^2 - N_\text{cen} + 2N_\text{cen}N_\text{sat} + N_\text{sat}(N_\text{sat}-1)\rangle \nonumber \\
    &= 2\langle N_\text{cen}N_\text{sat}\rangle + \langle N_\text{sat}(N_\text{sat}-1)\rangle \nonumber \\
    &= 2\langle N_\text{cen}\rangle \langle N_\text{sat}\rangle + \langle N_\text{sat} \rangle^2
\end{align}
where in going from the second to third line we used that $N_\text{cen}=0,1 \rightarrow \langle N_\text{cen}^2\rangle = \langle N_\text{cen} \rangle$. We obtain the last equality by assuming that the centrals and satellites are uncorrelated and that $N_\text{sat}$ follows a Poisson distribution, such that $\langle N_\text{sat}^2\rangle = \langle N_\text{sat}\rangle^2+\langle N_\text{sat}\rangle$. Using this, the 1-halo term becomes (dropping the $\langle\rangle$'s for simplicity):
\begin{equation}
  P_g^\text{1-halo} = \frac{1}{\bar{n}_g^2} \int dm \; n(m)\
  \left[ N_\text{sat}^2 \left| u(k,m,z) \right|^2 e^{-k^2 \mu^2\sigma_d^2(m)}
   + 2 N_\text{sat}N_\text{cen}u(k,m,z) e^{-k^2 \mu^2\sigma_d^2(m)/2} \right]   .
\end{equation}
Finally, the shot noise power spectrum is simply  $P^\text{shot}_g = \bar{n}_g^{-1}$ if we assume Poisson fluctuations for the galaxies and halos.

Our perturbative model should be able to describe any `complete' model of galaxy clustering, whether or not that model is correct in detail.  We can make the connection by considering the low-$k$ limit of the halo model.  To make our expressions slightly simpler we shall make an additional approximation that $u(k,m,z)\approx 1$ on the scales of interest, which corresponds to assuming that $k r_\text{vir}\ll 1$.  We shall further assume that $\sigma_d>r_\text{vir}$ so that the impact of virial velocities is more important than the fact that the satellites do not sit at the halo center.  Under these approximations, and for small $k$,
\begin{align}
    b_g(k,\mu,z) &\simeq  \frac{1}{\bar{n}_g} \int  dm\ n(m)\; b(m)
    \left[ N_\text{cen} + N_\text{sat} \left( 1 - \frac{1}{2}k^2 \mu^2\sigma_d^2(m) \right) \right] \\
    &= \frac{1}{\bar{n}_g} \int  dm\ n(m)\; b(m) N_\text{gal}  
    -\frac{1}{2}k^2\mu^2\ \frac{1}{\bar{n}_g} \int  dm\ n(m)\; b(m) N_\text{sat}  \sigma_d^2(m) \\
    &= b_{\rm eff}\left( 1 - \frac{1}{2}k^2\mu^2\sigma_{\rm 2,eff}^2 \right)
\end{align}
The $k^2\mu^2$ term above, combined with the $b$ or $f\mu^2$ term from the other power of $b_g$ in Eq.~(\ref{eqn:2halo}) contributes to the counterterms, $\alpha_i$.

Since the mass-integral in $F$ extends all the way to $m=0$, the $k^2\mu^2$ correction is smaller than for the bias and we shall neglect it, taking $F\to f$ henceforth.
The 1-halo term becomes
\begin{align}
    P_g^\text{1-halo}  &\simeq \frac{1}{\bar{n}_g^2} \int dm \; n(m)\
    N_{\rm sat} e^{-k^2\mu^2\sigma_d^2/2} \left[ N_{\rm cen}+ N_{\rm sat} e^{-k^2\mu^2\sigma_d^2/2} \right]  \\
    &\approx \frac{ f_{\rm sat} }{\bar{n}_g} \left( \cdots - \frac{1}{2}k^2\mu^2\sigma_{\rm 1,eff}^2 + \cdots \right)
\label{eqn:P1halo}
\end{align}
Thus we see that the halo model predicts that the stochastic terms are of order $\mathrm{SN}_0\sim\bar{n}_g^{-1}$ (from $P_g^{\rm shot}$ in Eq.~\ref{eqn:Phalomodel}) and $\mathrm{SN}_2\sim f_{\rm sat}\sigma_{1,{\rm eff}}^2/\bar{n}_g$ (from Eq.~\ref{eqn:P1halo}) as described in the main text.  Here $f_{\rm sat}$ is the satellite fraction such that $f_{\rm sat} \sigma^2_{1,\rm eff}$ is the mean velocity dispersion of halos weighted by $N_{\rm cen} N_{\rm sat}$, such that roughly speaking $\sigma^2_{1,\rm eff}$ is the mean velocity dispersion of the satellites in question. We often refer to $f_{\rm sat}^{1/2}\sigma_{1,{\rm eff}}$ as a ``characteristic halo velocity'' for simplicity.

The simple derivation above neglects several physical effects, including halo compensation and exclusion, correlations between the halo density and velocity profiles and between local environment and profile, correlations between mass bins in the halo shot noise, etc.  It is sufficient for order of magnitude estimates, since most of the neglected effects also have characteristic size set by the mean inter-galaxy separation or the virial or infall velocity of the halo but it should not be taken as a `complete' model of clustering.  As a single example of an effect missed by this simple treatment, let us further consider the effect of virial motions in Eq.~\ref{eqn:uk_fog}. Another way to account for the effect of FoG in the galaxy power spectrum is to introduce a random velocity field $\epsilon_i(\bs)$ to each galaxy, such that the observed position is $\bs + \hat{n} \cdot \epsilon_{\hat{n}}$. In this case the galaxy 2-point function with these additional velocities is \cite{Scoccimarro04,VlahWhite19}
\begin{align}
    P(k,\mu) &= \int d^3\bs\ e^{i\bk\cdot\bs} \Big\langle e^{i k\mu (\epsilon_{\hat{n}}(\bs) - \epsilon_{\hat{n}}(\bf{0})) } \big(1 + \delta_g(\bs) \big) \big(1 + \delta_g(\bf{0}) \big) \Big\rangle \nonumber \\
    &\approx \int d^3\bs\ e^{i\bk\cdot\bs} \Big\langle e^{i k\mu (\epsilon_{\hat{n}}(\bs) - \epsilon_{\hat{n}}(\bf{0})) } \Big\rangle \Big(1 + \xi_g(\bs) \Big) 
\end{align}
where in the second line we have made the (unphysical) assumption that the virial motions and galaxy densities are uncorrelated in order to isolate the pure effect of virial velocities usually called FoGs (in the literature models making this approximation are frequently referred to as ``dispersion'' or ``streaming'' models). The expectation value of the exponential can be expanded in powers of $k\mu$ as
\begin{equation}
    \ln\Big\langle e^{i k\mu (\epsilon_{\hat{n}}(\bs) - \epsilon_{\hat{n}}(\bf{0})) } \Big\rangle = 1 - k^2 \mu^2 \left[ \sigma_v^2 - \xi_\epsilon(\bs) \right] + \mathcal{O}(k^3\mu^3).
    \label{eqn:cumulant}
\end{equation}
where $\xi_\epsilon$ is the correlation function of the virial velocities projected along the line of sight. Since it describes virial motions, this correlation must fall rapidly to zero outside of the halo radius, $R_h$, and asymptote to the mean square velocity, $\sigma_v^2$, as $s \rightarrow 0$. Expanding this cumulant to first order we see that, in addition to the damping of the profile coming from $-k^2 \mu^2 \sigma^2_v$ in Eq.~\ref{eqn:cumulant} we also gain the contribution
\begin{equation}
    P(k,\mu) \supset k^2 \mu^2 \int d^3\bs\ e^{i\bk\cdot\bs}\ \xi_\epsilon(\bs) \big( 1 + \xi_g(\bs) \big) \approx k^2 \mu^2 (1 + \sigma^2_{g}) \int d^3\bs\ e^{i\bk\cdot\bs}\ \xi_{\epsilon}(\bs)
\end{equation}
where we have used that the linear galaxy density is smooth compared to the support of $\xi_\epsilon$ and $\sigma^2_g$ is its the mean on the halo scale. The integral in the final expression is simply the noise spectrum of the virial motions, which we expect to be positive and white on large ($> R_h$) scales and of order $\sim \sigma_v^2 R_h^3$. In order to differentiate between satellites and centrals we can simply set $\epsilon=0$ for central galaxies such that the cumulant in Equation~\ref{eqn:cumulant} is instead simply unity for the central-central correlation and $1 - \frac12 k^2\mu^2 \sigma^2_v$ for the central-satellite cross correlation. This gives the FoG prescription in the `analytic halo model', derived above, with the addition of a positive, scale-dependent noise along the line of sight.\footnote{We thank Misha Ivanov for pointing out that the sign of this effect in N-body simulations is often positive.}

We reiterate that our aim here was to motivate the scale of stochastic contributions and not to make claims about what numerical value (or even sign) they will take.  We see that the term discussed above, while missed by the halo model, did scale in the same manner as the included terms as we stated above. Other allowed parameter combinations, such as $R_h^4 \sigma_v$ for the stochastic piece, should be subdominant.

\section{Further tests}
\label{appendix: tests}
\subsection{Dependence on $\omega_{\rm b}$ prior}

\begin{figure}
\captionsetup[subfigure]{labelformat=empty}
\begin{subfigure}{.5\textwidth}
\centering
\includegraphics[height=8cm]{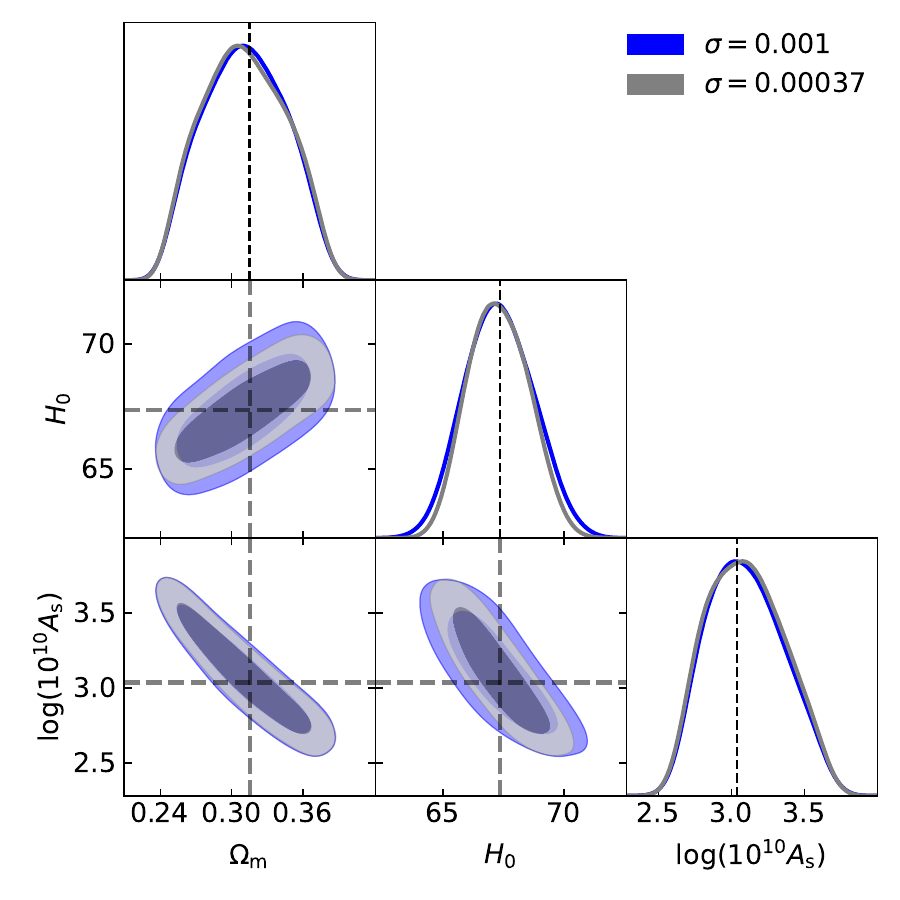}
\caption{Standard Template}
\end{subfigure}%
\begin{subfigure}{.5\textwidth}
\centering
\includegraphics[height=8cm]{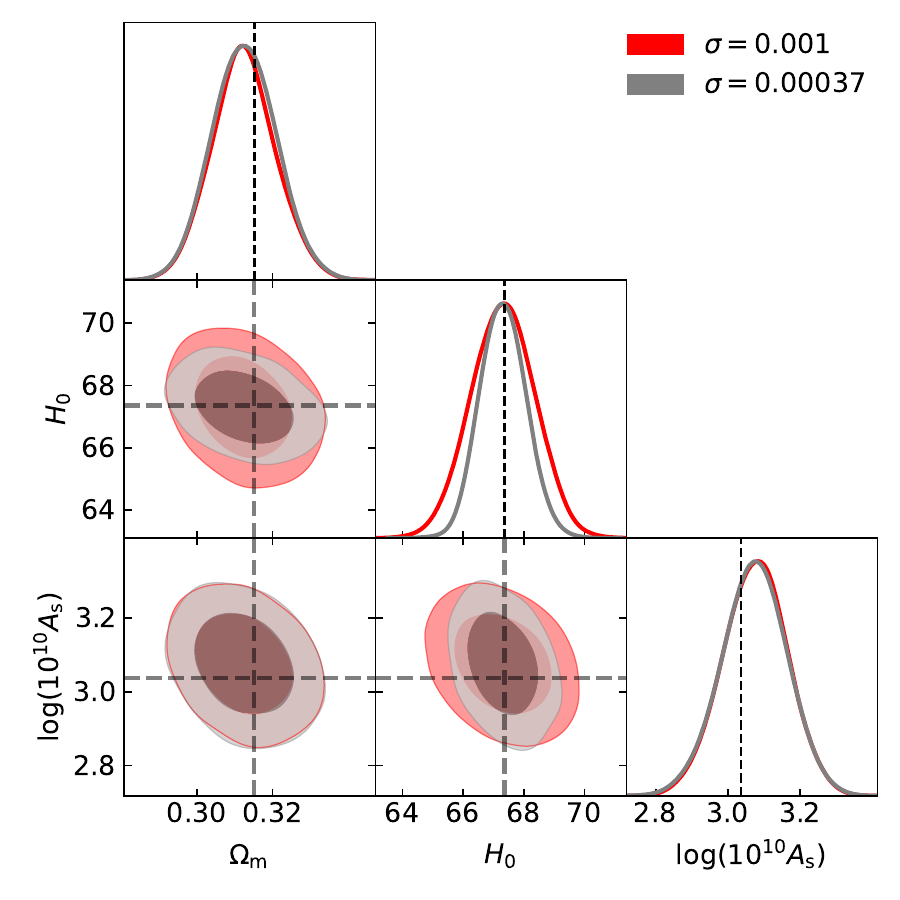}
\caption{ShapeFit}
\end{subfigure}%
\\
\begin{subfigure}{.5\textwidth}
\centering
\includegraphics[height=8cm]{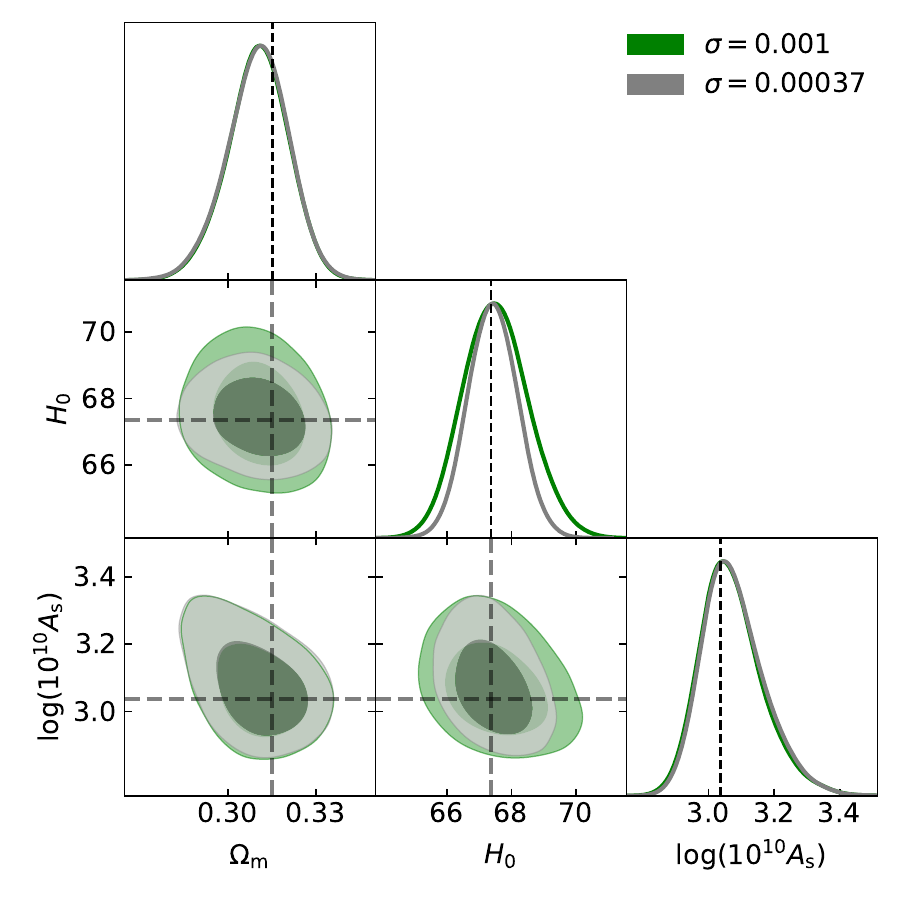}
\caption{Full-Modeling}
\end{subfigure}%
\begin{subfigure}{.5\textwidth}
\centering
\includegraphics[height=8cm]{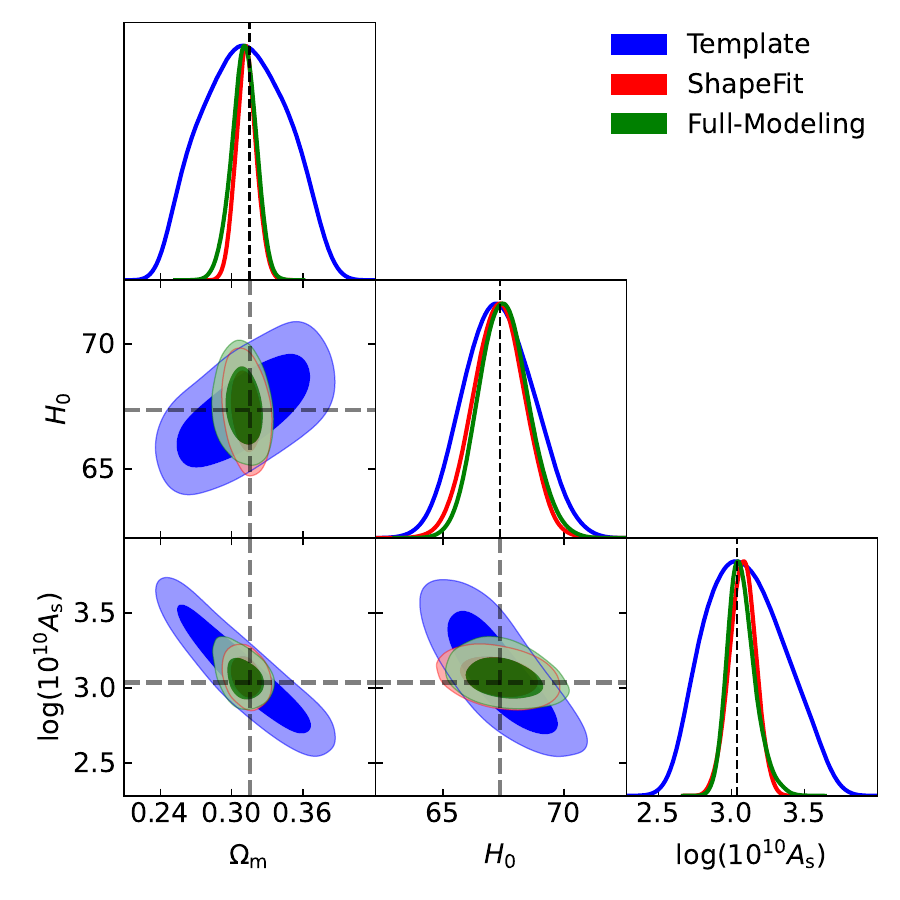}
\caption{All Methods}
\end{subfigure}%
\caption{Comparison of constraints when loosening the prior on $\omega_b$ from $\sigma= 0.00037$ to $\sigma=0.001$. In all cases we use the single box covariance. The bottom right plot shows a comparison of the three modeling methods while using a $\sigma=0.001$ prior on $\omega_b$
}
\label{fig: omb_prior}
\end{figure}

We next test the dependence of our constraints on the prior set on $\omega_{\rm b}$. The standard setting that we choose is a Gaussian prior centered on $\omega_{\rm b}^{\rm true} = 0.02237$ with a width of $\sigma = 0.00037$, which is based on the recentmost Big-Bang Nucleosynthesis (BBN) constraints on primordial deuterium abundance \cite{Cooke2018} which places stringent constraints on $\omega_{\rm b}$. We test the dependence on the prior by loosening it to $\sigma = 0.001$. The results are shown in Fig.~\ref{fig: omb_prior}. Within each individual method we show results for the covariance appropriate to the single-box volume. We find that for all three methods, $H_0$ becomes significantly less constrained. Meanwhile the $\Omega_{\rm m}$ constraints remain unchanged in all methods. 

In the Full-Modeling analysis, the measurement of $\Omega_{\rm m}$ is extracted from the shape of the power spectrum and scale of matter-radiation equality k$_{eq}$, and these depend on the full matter abundance rather than $\omega_b$ and $\omega_{cdm}$ separately. We thus do not see a degradation in the $\Omega_{\rm m}$ constraint when the prior on $\omega_b$ is relaxed. In the template and ShapeFit analyses $\Omega_{\rm m}$ is inferred from the compressed parameters, and because $f\simeq \Omega_{\rm m}^{0.55}$ we can extract a measurement of $\Omega_{\rm m}$ from the compressed amplitude parameter without any dependence on $\omega_b$ prior. In the ShapeFit case, additional constraining power on $\Omega_{\rm m}$ comes from the shape parameter $m$, but just like in the Full-Modeling case this power spectrum shape information translates to a measurement $\Omega_{\rm m}$ without any reliance on $\omega_b$ specifically. 

For the $H_0$ measurement we do observe a significant degradation in constraining power when the prior on $\omega_b$ is relaxed. In the template analysis, information about cosmological distances is extracted from the BAO feature and thus constrains $H(z)r_d$ and $D_A(z)/r_d$. Breaking the degeneracy between $H_0$ and $r_d$ requires a physical (dimensionful) length scale for the distance-redshift relation beyond just the angular size of the BAO feature \cite{Ivanov_2023}. This is accomplished with knowledge about $\omega_b$ (which determines $r_d$) from either BBN or CMB and then leads to a direct measurement of $H_0$. Therefore, relaxing the prior on $\omega_b$ worsens the constraint on $H_0$. The inclusion of the shape parameter $m$, while in general improving constraints when compared to the standard template, does not compensate for the changes in $\omega_b$ information and therefore ShapeFit also experiences worse $H_0$ constraint. The Full-Modeling method can in principle constrain $\omega_b$ (and by extension $r_d$) in the absence of an external prior because the amplitude of BAO wiggles depend on $\omega_b$ and $\omega_{\rm cdm}$ and can be modulated in Full-Modeling analyses, but this is still a much weaker constraint than what can be accomplished with a BBN prior \cite{Ivanov_2020}.

\subsection{Minimal and maximal freedom in the bias parameters}

\begin{figure}
\captionsetup[subfigure]{labelformat=empty}
\begin{subfigure}{0.5\textwidth}
\centering
\includegraphics[height=8cm]{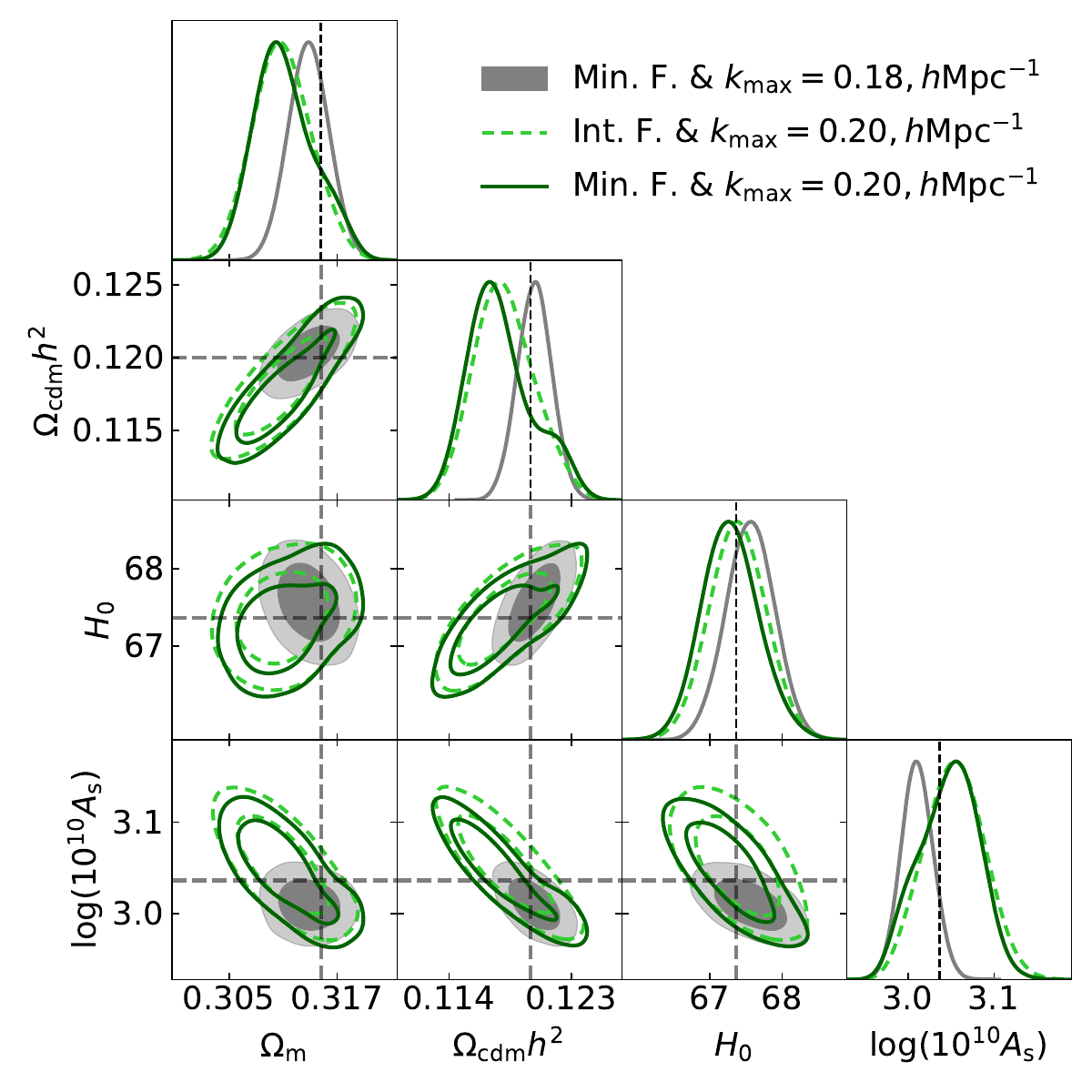}
\caption{Full Covariance (200 ($h^{-1}$Gpc)$^3$)}
\end{subfigure}%
\begin{subfigure}{0.5\textwidth}
\centering
\includegraphics[height=8cm]{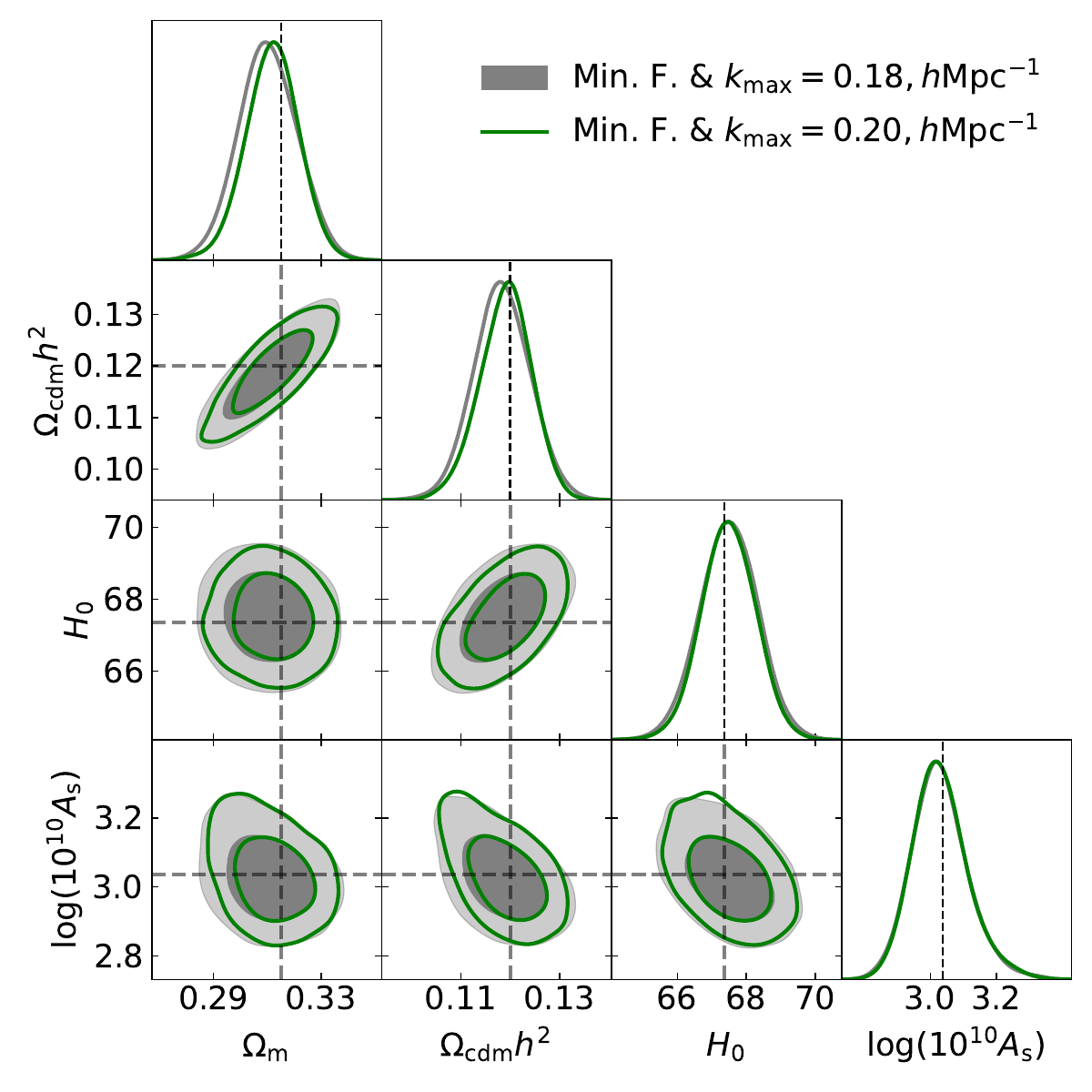}
\caption{Single box Covariance (8 ($h^{-1}$Gpc)$^3$)}
\end{subfigure}%
\caption{\textit{(left):} Full-Modeling constraints for the minimal freedom parametrization with $k_{\rm max} = 0.18 \, h$Mpc$^{-1}$ (grey), $k_{\rm max} = 0.20 \, h$Mpc$^{-1}$ (dark green), and intermediate freedom case with $k_{\rm max} = 0.20 \, h$Mpc$^{-1}$ (light green dashed). In the minimum freedom case there is a bimodal distribution (most pronounced in $\Omega_{cdm}h^2$) that appears when $k_{\rm max}$ is raised from 0.18 to 0.2 $h$Mpc$^{-1}$. The bi-modality disappears if the $b_s$ parameter is included, as in the intermediate freedom case.\textit{(right):} Full-Modeling constraints in the minimal freedom case with $k_{\rm max} = 0.18$ (grey) and $0.20 \, h$Mpc$^{-1}$ (green) using the single-box covariance.  }
\label{fig: minF_cov}
\end{figure}

\begin{figure}
\captionsetup[subfigure]{labelformat=empty}
\begin{subfigure}{0.5\textwidth}
\centering
\includegraphics[height=7.9cm]{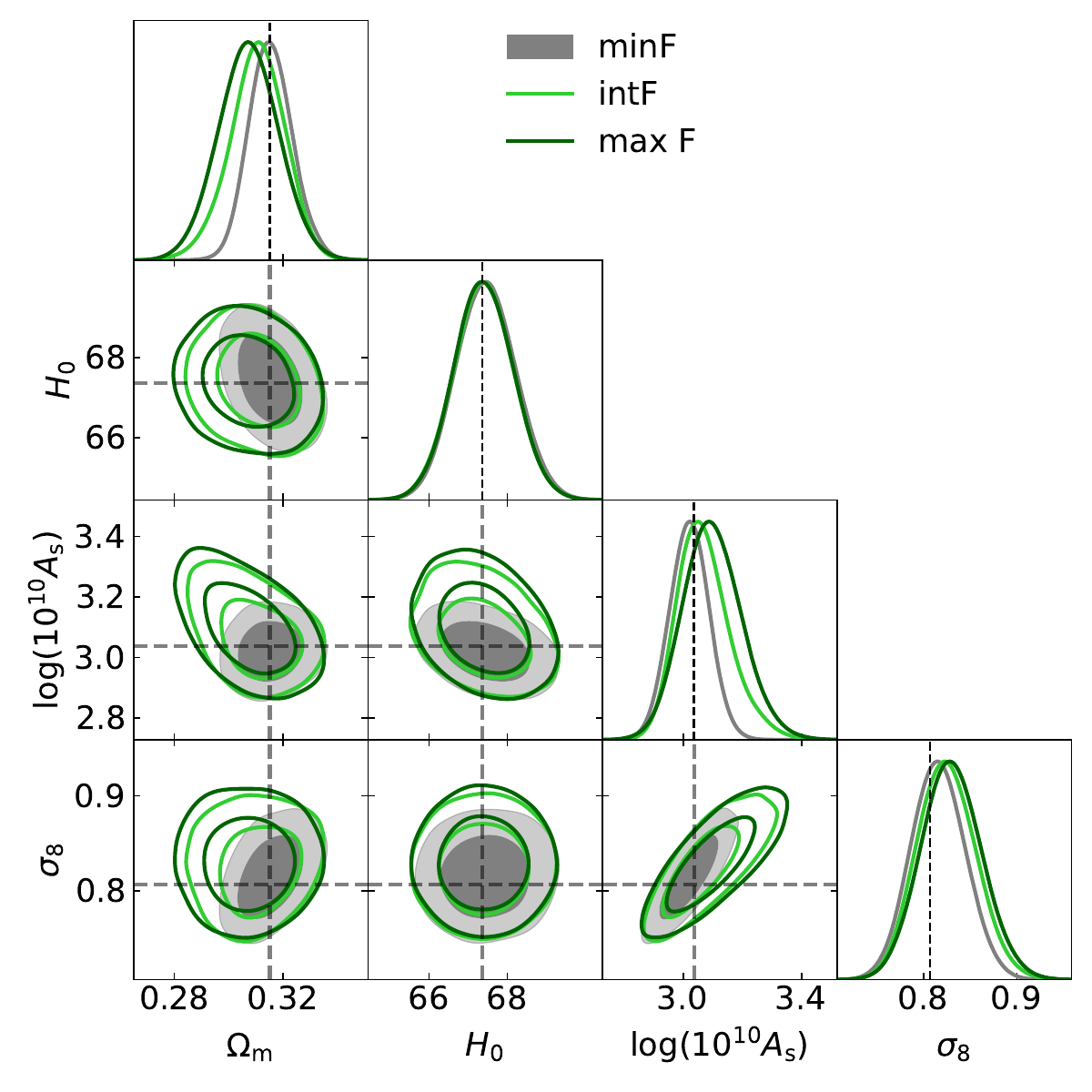}
\caption{Full-Modeling}
\end{subfigure}%
\begin{subfigure}{0.5\textwidth}
\centering
\includegraphics[height=7.7cm]{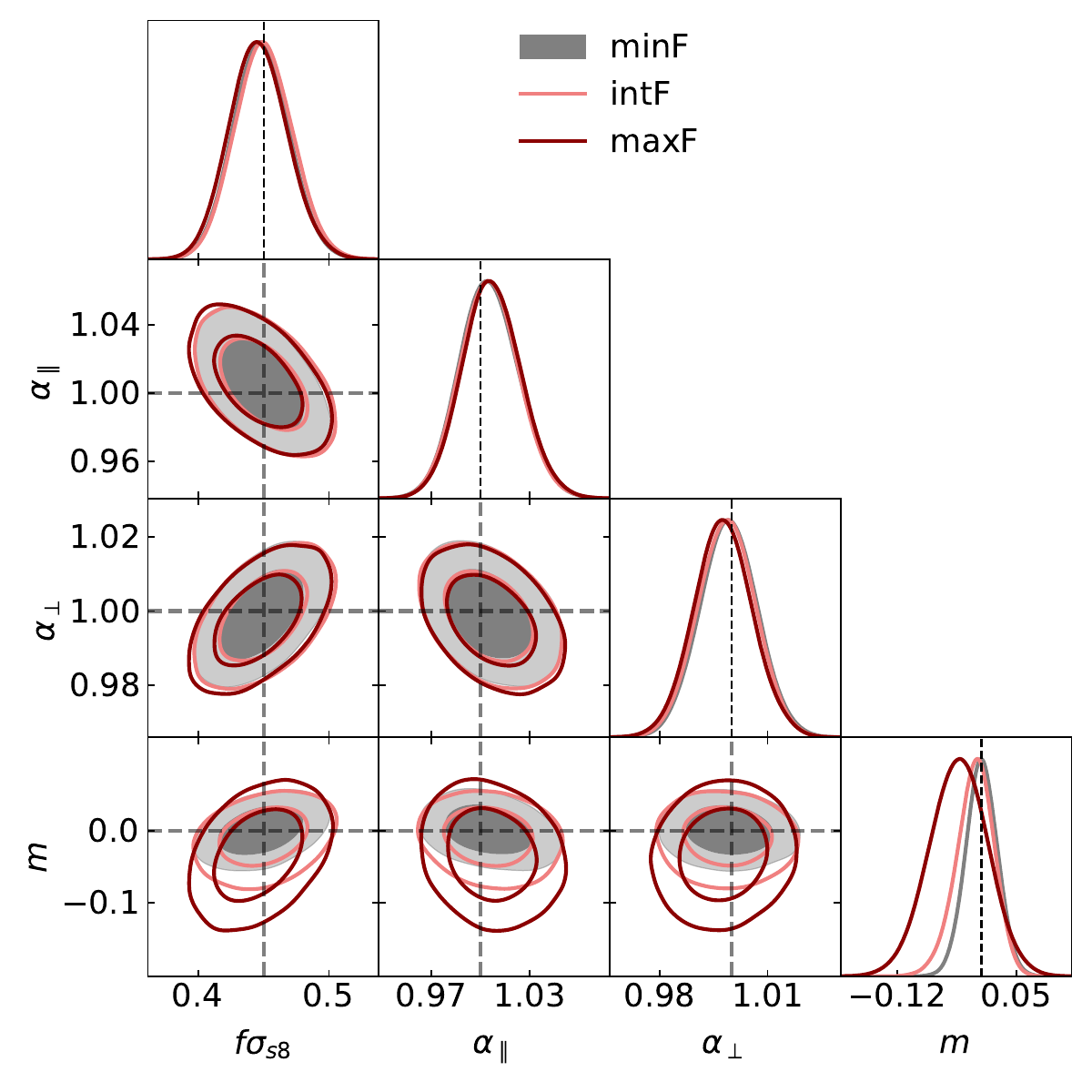}
\caption{ShapeFit}
\end{subfigure}%
\caption{Comparison of minimal ($b_s=b_3 = 0$), intermediate ($b_3 = 0$), and maximal ($b_s, b_3$ free) freedom parametrizations using the single box covariance and $k_{\rm max} = 0.20\, h$Mpc$^{-1}$. The Full-Modeling constraints are in the left figure and ShapeFit on the right.}
\label{fig: min_maxF}
\end{figure}

In this section we discuss three possible choices in freedom in the bias parameters. In total there are four bias parameters: $b_1$, $b_2$, $b_s$, and $b_3$. The first two parameters multiply the initial $\delta_0(q)$ and $\delta_0^2(q)$ overdensity fields in the bias expansion. The non-local tidal bias parameter, $b_s$ multiplies the initial shear field and, due to degeneracies between terms, the third order bias contributions are combined into a single operator with coefficient $b_3$. In the Lagrangian picture the bias contributions are evaluated at the initial positions $\bq$, whereas in the Eulerian framework the bias expansion is performed at observed coordinates $\bx$. This implies that the non-local bias terms in Eulerian PT are dependent on both the initial Lagrangian non-local contributions as well as gravitational evolution such that the Eulerian biases are affine transformations of the Lagrangian ones, with coefficients dependent on the definition of the bias operators in each space. Therefore, one commonly sees in the literature of Eulerian PT models (e.g.\ \cite{Beutler14,Briedan21}) a ``minimal'' and ``maximal'' freedom parametrization where the first assumes a local Lagrangian bias initially with no third-order contributions ($b_s^L=b_3^L=0$) and that tidal and 3rd order biases are induced entirely by gravitational nonlinearity \cite{Chan12}. In such a case, the tidal and third order Eulerian biases would coevolve with the linear bias terms, i.e.\ $b_{i}^E \propto b_1^L = b_1^E-1$. In the maximal freedom case, on the other hand, all bias parameters are allowed to vary independently. 

The two other Fourier space EFT models that will be used in the DESI collaboration, FOLPS$\nu$ and PyBird, are both based on the Eulerian frameworks and it has been shown that \velocileptors LPT and EPT agree closely with the other two models under a consistent choice of parametrization \cite{KP5s1-Maus}. For this reason we  are interested in comparing the three parameter choices within LPT. In the Lagrangian picture, it is not clear how well motivated the initially local bias assumption is, and for most of this paper we chose an intermediate option in which the tidal bias $b_s$ is allowed to vary along with $b_1$ and $b_2$, but the third order bias is kept fixed to zero, both because the cubic bias is expected to be small for intermediate mass halos and, more importantly, quite degenerate with the counterterms. We advise caution against restricting the parameter space further when fitting the high volume simulations with the 25 box covariance, as the tightness of the error bars can result in poor behavior of the model, which we demonstrate in the left panel of Fig.~\ref{fig: minF_cov}. While at $k_{\rm max}=0.18\, h$Mpc$^{-1}$ the constraints are fine, raising the scale cut to $k_{\rm max}=0.2\, h$Mpc$^{-1}$ results in a bimodal distribution appearing in the posteriors, most likely driven by some two-loop effects. However, including the $b_s$ parameter fixes the bimodal behavior and we instead recover more Gaussian posteriors. We also show that this problem is induced by the extremely tight covariance from the full 25- cubic box volume. In the right panel of Fig.~\ref{fig: minF_cov} we compare the Full-Modeling constraints between both $k_{\rm max}$ values with minimal freedom for the single box volume and find the two in agreement without any non-Gaussian behavior. 

Choosing the single-box covariance and a $k_{\rm max}=0.2\, h$Mpc$^{-1}$ we proceed with the comparison between the minimal, intermediate, and maximal freedom bias parametrizations. The results are shown in Fig.~\ref{fig: min_maxF} for the Full-Modelling and ShapeFit methods. We find that the parameters primarily controlling the shape of the linear power spectrum, i.e. $\Omega_{\rm m}$ in FM and $m$ in SF, are the most affected by the differences in parameterization. Meanwhile the amplitude $\sigma_8$ in FM is fairly resistant to these changes. We remind the reader that $\sigma_8$ is more directly constrained in LSS analyses than $\log(10^{10} A_\mathrm{s})$, suggesting it is a better way of quoting the normalization of the theory for these purposes. We find that fixing $b_3=0$ does not result in significant offsets away from the true cosmology, and mostly just tighten constraints. This is consistent with previous tests on the bias parametrization, and our standard choice of fixing $b_3$ in this paper mirrors that of previous analyses using \texttt{velocileptors}~\cite{Chen_BOSSrecon2022,Chen22_2}. We conclude this section by reiterating that despite the improvement in constraining power obtained in the minimal freedom case, fixing both $b_s$ and $b_3$ can lead to poor performance of the model in capturing the nonlinear effects that become increasingly important at very high simulation volumes, and it therefore is safer to use the intermediate freedom choice. In addition, depending on the method of galaxy sample-selection, larger values of $b_s$ than expected can occur  due to assembly bias (see e.g. Ref.~\cite{anzu21}). This further motivates keeping $b_s$ as a free parameter. While we have justification for the choice of fixing $b_3 = 0$, it is also a valid and more conservative option to allow $b_3$ to vary and we do not strongly discourage the maximal freedom choice in future analyses.

\subsection{Including hexadecapole}

\begin{figure}
\captionsetup[subfigure]{labelformat=empty}
\begin{subfigure}{.5\textwidth}
\centering
\includegraphics[height=8cm]{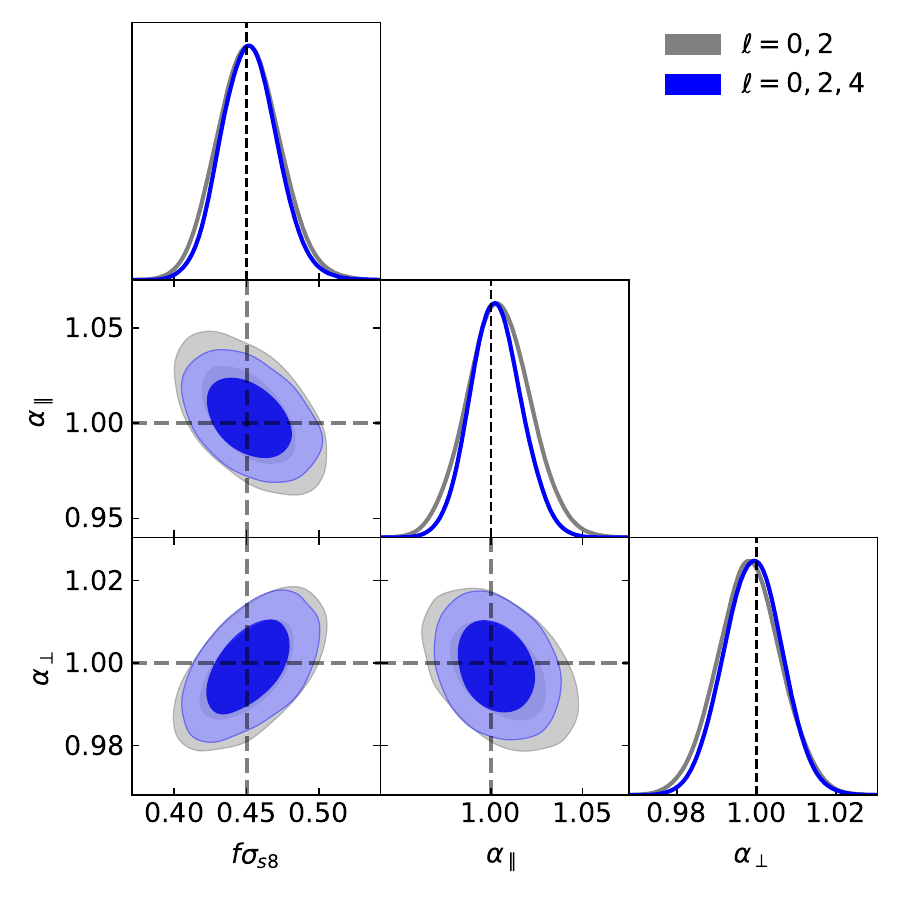}
\caption{Standard Template}
\end{subfigure}%
\\
\begin{subfigure}{.5\textwidth}
\centering
\includegraphics[height=8cm]{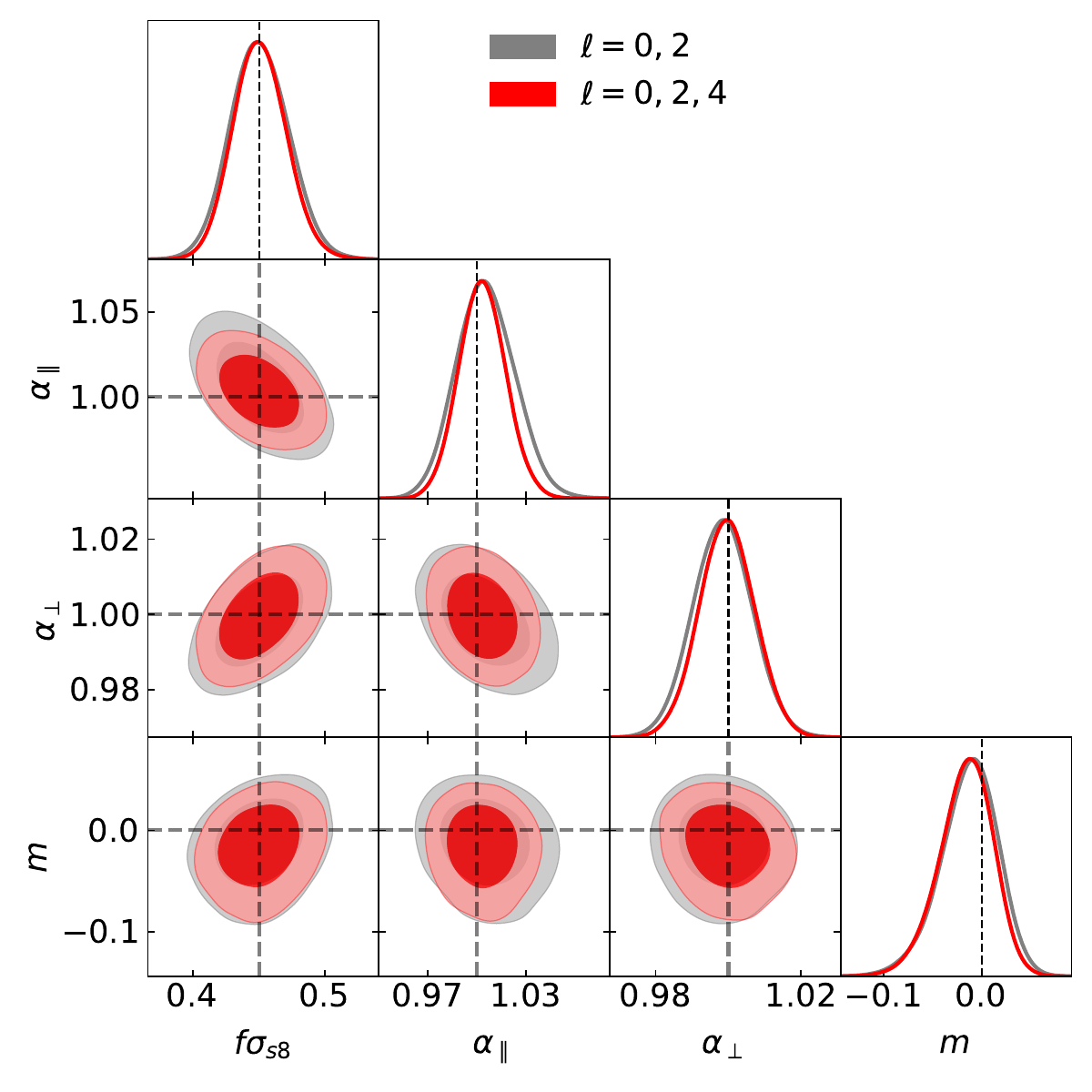}
\caption{ShapeFit}
\end{subfigure}%
\begin{subfigure}{.5\textwidth}
\centering
\includegraphics[height=8cm]{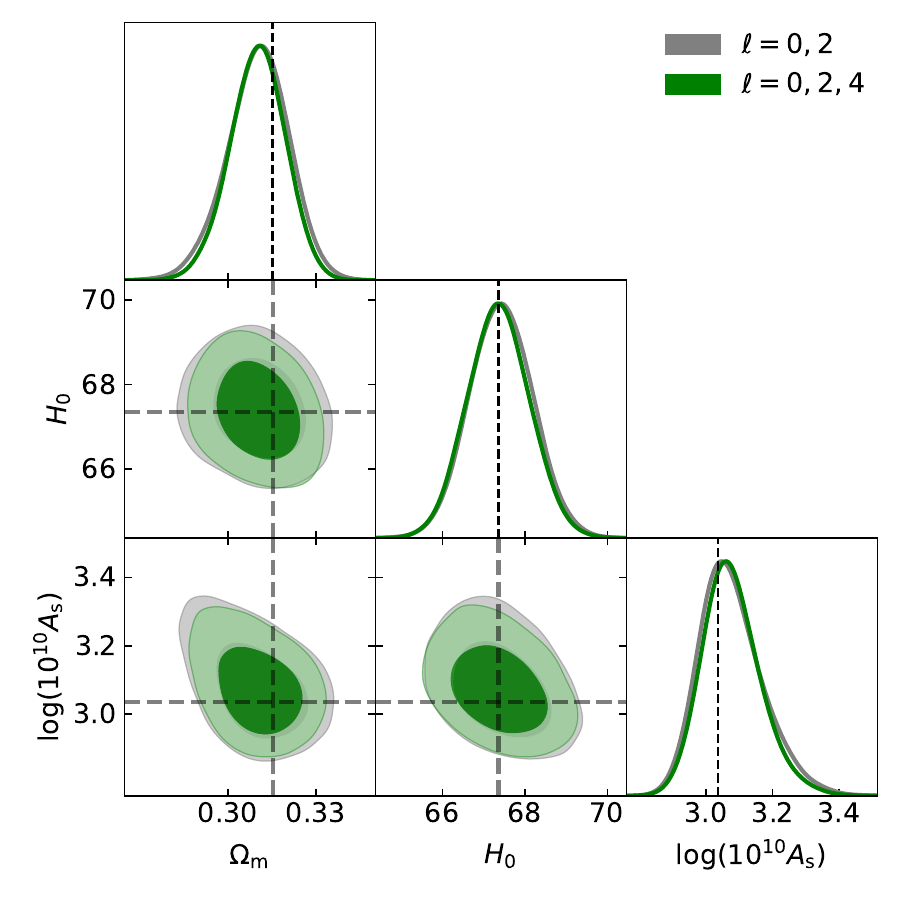}
\caption{Full-Modeling}
\end{subfigure}%
\caption{Comparison of constraints between $\ell=0,2$ and $\ell=0,2,4$. We present fits using the covariance for both the single-box volume ($1\cdot V$). 
}
\label{fig: hex_FM_SF}
\end{figure}

The 1-loop LPT model we use predicts the full angular dependence of the power spectrum $P(k,\mu)$ and therefore makes consistent predictions for the power spectrum hexadecapole and above in addition to the monopole and quadrupole. However, it should be noted that since the linear theory hexadecapole is substantially smaller than the monopole or quadrupole (there are no linear theory $\ell > 4$ multipoles) these higher multipoles will be more sensitive to nonlinear effects (e.g.\ Finger of God (FoG)), and thus the range of scales over which their 1-loop PT predictions is valid may be smaller. We present results of including the hexadecapole in Fig.~\ref{fig: hex_FM_SF} for the covariance of the single-box volume. We find a slight tightening of the constraints when including the hexadecapole.

\begin{figure}
\captionsetup[subfigure]{labelformat=empty}
\begin{subfigure}{.5\textwidth}
\centering
\includegraphics[height=8cm]{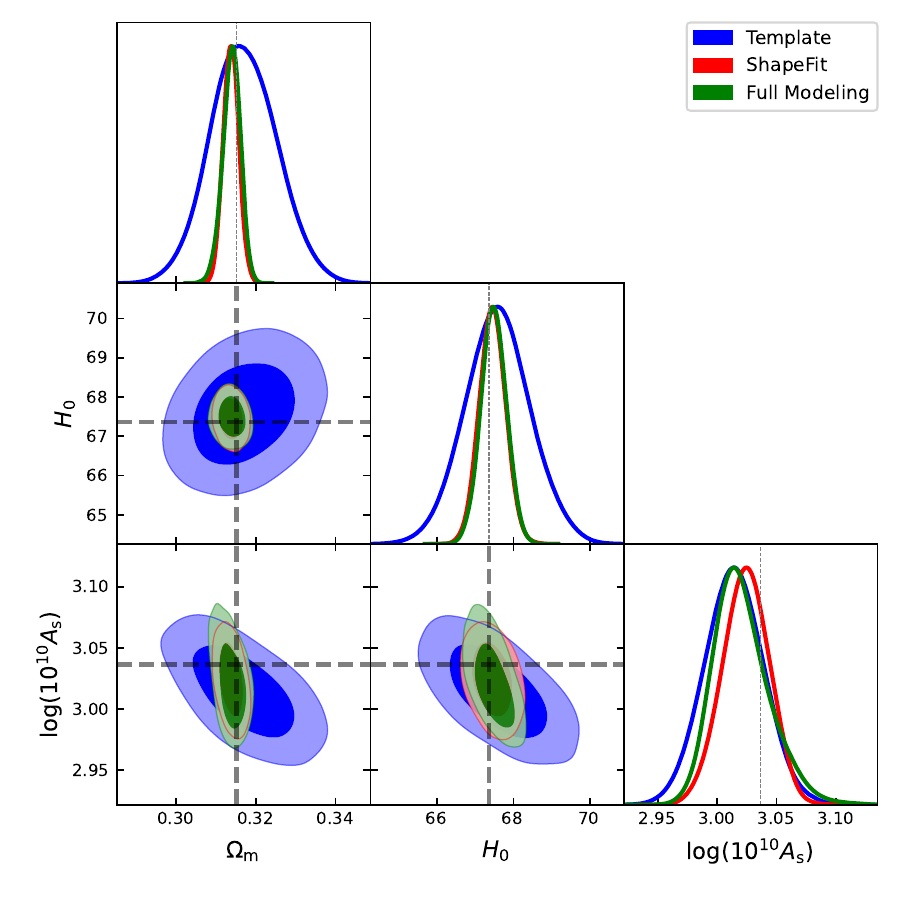}
\end{subfigure}%
\begin{subfigure}{.5\textwidth}
\centering
\includegraphics[height=8cm]{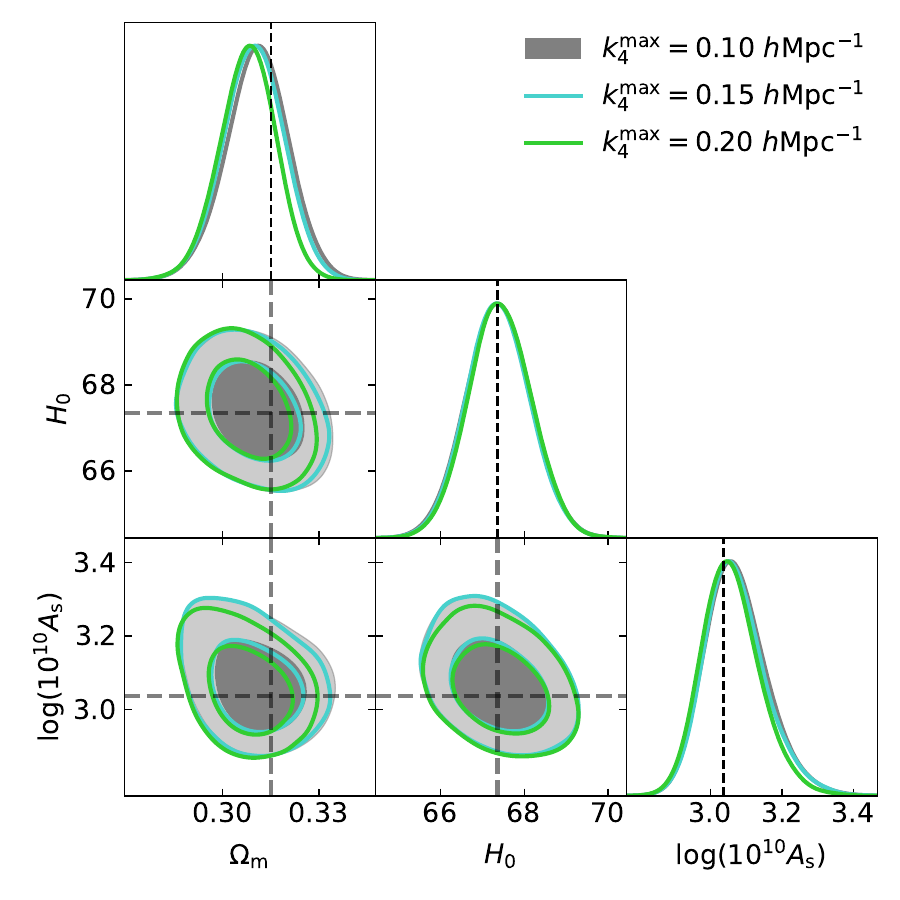}
\end{subfigure}%
\caption{\textit{(left):} Constraints on $\Lambda$CDM parameters for the three modeling methods with hexadecapole included. \textit{(right):} Comparison of constraints on $\Lambda$CDM parameters, varying k$_{\max}$ of the hexadecapole while keeping k$_{\max}$ of the monopole and quadrupole at 0.2 $h$Mpc$^{-1}$. 
}
\label{fig: hex}
\end{figure}

In Fig.~\ref{fig: hex} we show in the left panel the $\Lambda$CDM parameter constraints of all three methods when fitting $\ell=0,2,4$ instead of just $\ell=0,2$, using the covariance for the 25 box volume. As with the previous comparisons between methods, we find consistent constraints between ShapeFit and Full-Modeling and looser constraints for the standard template. We also test the dependence of the hexadecapole on it's $k$-range by lowering the upper bound from $k_{\rm max} = 0.2$ $h$Mpc$^{-1}$ down  to $0.15$ and $0.1$ $h$Mpc$^{-1}$, while keeping the range of scales of the monopole and quadrupole moments fixed at $0.2$ $h$Mpc$^{-1}$. While we see very little change in constraints in this case, other data sets may have significantly larger FoG effects (or observational systematics) that could affect the hexadecapole at $k \gtrsim 0.1 \ \ihmpc$. For this reason we still suggest using $k_{\rm max} = 0.1 \ \ihmpc$ for the hexadecapole and correspondingly widening the $\alpha_4$ prior to $\mathcal{N}[0,20]$ to maintain the 20$\%$ scaling at the new $k_{\rm max}$.

\section{Emulator error/performance}
\label{appendix: emulator}

\begin{figure}
\captionsetup[subfigure]{labelformat=empty}
\begin{subfigure}{.5\textwidth}
\centering
\includegraphics[height=8cm]{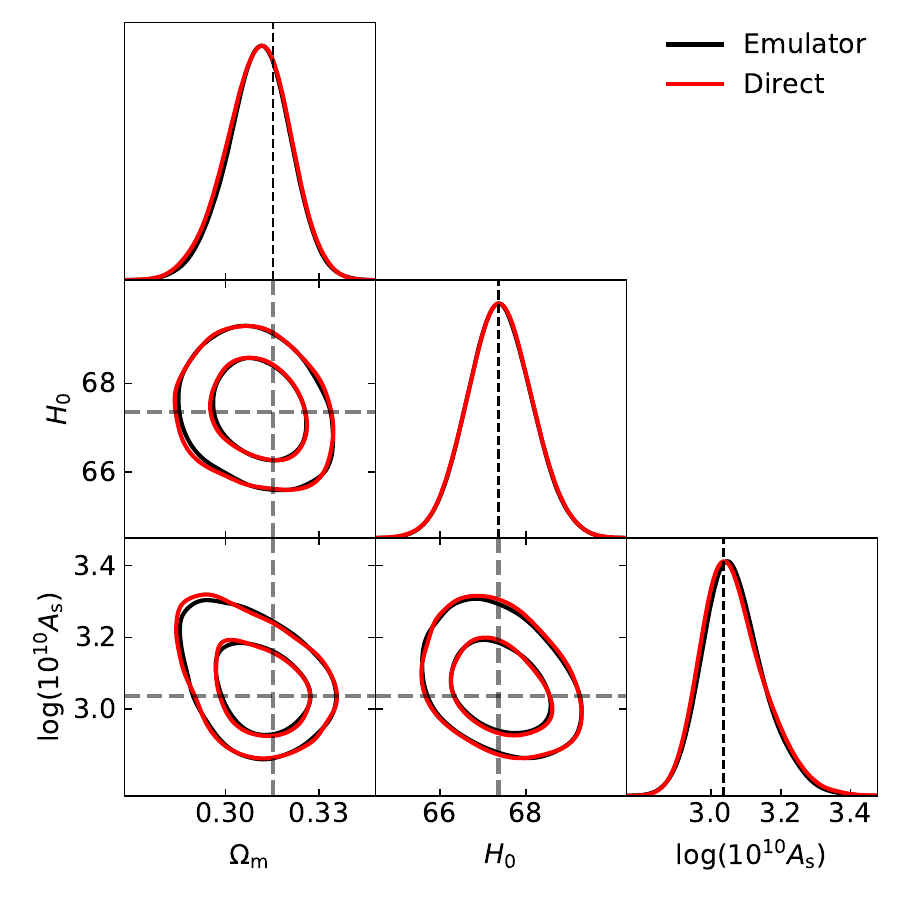}
\end{subfigure}%
\begin{subfigure}{.5\textwidth}
\centering
\includegraphics[height=8cm]{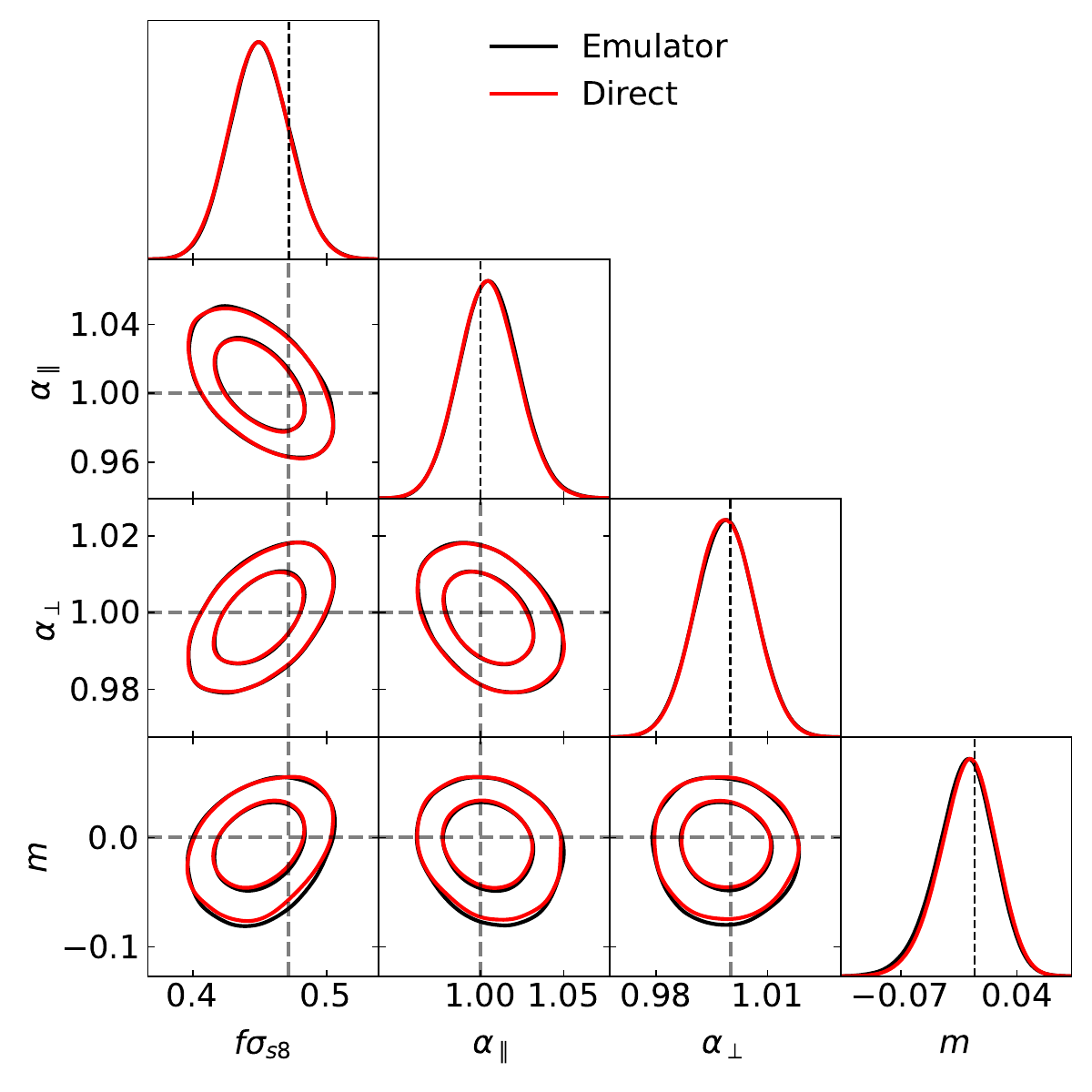}
\end{subfigure}%
\caption{Comparison of constraints from fitting LRG Cubic mock data using the Taylor series emulator vs. a direct fit in Full-Modeling \textit{(left)} and ShapeFit \textit{(right)}}
\label{fig: emulator}
\end{figure}

In order to speed up likelihood evaluations, we employ emulators that reproduce the theoretical power spectrum multipole predictions using a Taylor series centered on reasonably chosen values for the cosmological parameters, $\boldsymbol{\Omega}_0$, i.e. the Abacus fiducial values. The emulator is trained by evaluating the full \velocileptors\ prediction on a grid with $9$ points in each parameter direction, resulting in $9^N$ evaluations for $N$ cosmological parameters. For each training point, e.g. $\boldsymbol{\Omega}_n = (h, \omega_b,\omega_{cdm}, \log(10^{10}A_s))_n$ \velocileptors\ computes the power spectrum multipoles and separates the 19 terms within each multipole (i.e. the terms $P_{\ell,m}$ multiplied by $1,b_1, b_1^2, b_1b_2$, etc.) into a table. After the grid of $P_{\ell, m}(\boldsymbol{\Omega}_n,k)$ has been computed for every $n$'th set of cosmological parameters, we take numerical derivatives up to fourth order in each parameter using the finite differencing method\footnote{\texttt{findiff}; https://github.com/maroba/findiff \cite{findiff}}. These arrays of derivatives are then stored for later use. At each step of an MCMC, the emulated power spectrum multipole terms are produced for the proposal set of parameters $\boldsymbol{\Omega}_n$ by constructing the Taylor series: 
\begin{align}
    P_{\ell, m}^{\rm emu}(\boldsymbol{\Omega}_n,k) = P_{\ell,m}(\boldsymbol{\Omega}_0,k) + \sum_i^{N}\frac{\partial P_{\ell,m}}{\partial\boldsymbol{\Omega}_i}(\boldsymbol{\Omega}_{0,i} - \boldsymbol{\Omega}_{n,i}) + \nonumber \\
    + \frac{1}{2}\sum_{i,j}^{N}\frac{\partial^2 P_{\ell,m}}{\partial\boldsymbol{\Omega}_i\partial\boldsymbol{\Omega}_j}(\boldsymbol{\Omega}_{0,i} - \boldsymbol{\Omega}_{n,i})(\boldsymbol{\Omega}_{0,j} - \boldsymbol{\Omega}_{n,j}) + ...
\end{align}
where $\boldsymbol{\Omega}_0$ is the set of cosmological parameters that the Taylor series was centered around, $\boldsymbol{\Omega}_{0,i}$ is the $i$'th cosmological parameter in said vector, and $N$ is the number of parameters being varied in $\boldsymbol{\Omega}$. 
In order to demonstrate the accuracy of the emulator, we perform fits to the LRG cubic mocks both with the emulator and without. The results are shown in Fig.~\ref{fig: emulator} for ShapeFit and Full-Modeling. In both cases, the emulator reproduces the constraints of the direct computation exactly.

\section{Author Affiliations}
\label{sec:affiliations}

\begin{hangparas}{.5cm}{1}

$^{1}${Department of Physics, University of California, Berkeley, CA 94720, USA}

$^{2}${Lawrence Berkeley National Laboratory, 1 Cyclotron Road, Berkeley, CA 94720, USA}

$^{3}${Institute for Advanced Study, 1 Einstein Drive, Princeton, NJ 08540, USA}

$^{4}${}

$^{5}${Physics Dept., Boston University, 590 Commonwealth Avenue, Boston, MA 02215, USA}

$^{6}${Instituto Avanzado de Cosmolog\'{\i}a A.~C., San Marcos 11 - Atenas 202. Magdalena Contreras, 10720. Ciudad de M\'{e}xico, M\'{e}xico}

$^{7}${Instituto de Ciencias F\'{\i}sicas, Universidad Aut\'onoma de M\'exico, Cuernavaca, Morelos, 62210, (M\'exico)}

$^{8}${Institute for Astronomy, University of Edinburgh, Royal Observatory, Blackford Hill, Edinburgh EH9 3HJ, UK}

$^{9}${Department of Physics \& Astronomy, University College London, Gower Street, London, WC1E 6BT, UK}

$^{10}${Institute for Computational Cosmology, Department of Physics, Durham University, South Road, Durham DH1 3LE, UK}

$^{11}${Instituto de F\'{\i}sica, Universidad Nacional Aut\'{o}noma de M\'{e}xico,  Cd. de M\'{e}xico  C.P. 04510,  M\'{e}xico}

$^{12}${NSF NOIRLab, 950 N. Cherry Ave., Tucson, AZ 85719, USA}

$^{13}${University of California, Berkeley, 110 Sproul Hall \#5800 Berkeley, CA 94720, USA}

$^{14}${Institute of Cosmology and Gravitation, University of Portsmouth, Dennis Sciama Building, Portsmouth, PO1 3FX, UK}

$^{15}${Departamento de F\'isica, Universidad de los Andes, Cra. 1 No. 18A-10, Edificio Ip, CP 111711, Bogot\'a, Colombia}

$^{16}${Observatorio Astron\'omico, Universidad de los Andes, Cra. 1 No. 18A-10, Edificio H, CP 111711 Bogot\'a, Colombia}

$^{17}${Institut d'Estudis Espacials de Catalunya (IEEC), 08034 Barcelona, Spain}

$^{18}${Institute of Space Sciences, ICE-CSIC, Campus UAB, Carrer de Can Magrans s/n, 08913 Bellaterra, Barcelona, Spain}

$^{19}${Departament de F\'{\i}sica Qu\`{a}ntica i Astrof\'{\i}sica, Universitat de Barcelona, Mart\'{\i} i Franqu\`{e}s 1, E08028 Barcelona, Spain}

$^{20}${Institut de Ci\`encies del Cosmos (ICCUB), Universitat de Barcelona (UB), c. Mart\'i i Franqu\`es, 1, 08028 Barcelona, Spain.}

$^{21}${Department of Astrophysical Sciences, Princeton University, Princeton NJ 08544, USA}

$^{22}${Center for Cosmology and AstroParticle Physics, The Ohio State University, 191 West Woodruff Avenue, Columbus, OH 43210, USA}

$^{23}${Department of Physics, The Ohio State University, 191 West Woodruff Avenue, Columbus, OH 43210, USA}

$^{24}${The Ohio State University, Columbus, 43210 OH, USA}

$^{25}${School of Mathematics and Physics, University of Queensland, 4072, Australia}

$^{26}${Department of Physics, The University of Texas at Dallas, Richardson, TX 75080, USA}

$^{27}${Departament de F\'{i}sica, Serra H\'{u}nter, Universitat Aut\`{o}noma de Barcelona, 08193 Bellaterra (Barcelona), Spain}

$^{28}${Institut de F\'{i}sica d’Altes Energies (IFAE), The Barcelona Institute of Science and Technology, Campus UAB, 08193 Bellaterra Barcelona, Spain}

$^{29}${Instituci\'{o} Catalana de Recerca i Estudis Avan\c{c}ats, Passeig de Llu\'{\i}s Companys, 23, 08010 Barcelona, Spain}

$^{30}${Department of Physics and Astronomy, University of Sussex, Brighton BN1 9QH, U.K}

$^{31}${Department of Physics \& Astronomy, University  of Wyoming, 1000 E. University, Dept.~3905, Laramie, WY 82071, USA}

$^{32}${National Astronomical Observatories, Chinese Academy of Sciences, A20 Datun Rd., Chaoyang District, Beijing, 100012, P.R. China}

$^{33}${IRFU, CEA, Universit\'{e} Paris-Saclay, F-91191 Gif-sur-Yvette, France}

$^{34}${Department of Physics and Astronomy, University of Waterloo, 200 University Ave W, Waterloo, ON N2L 3G1, Canada}

$^{35}${Perimeter Institute for Theoretical Physics, 31 Caroline St. North, Waterloo, ON N2L 2Y5, Canada}

$^{36}${Waterloo Centre for Astrophysics, University of Waterloo, 200 University Ave W, Waterloo, ON N2L 3G1, Canada}

$^{37}${Space Sciences Laboratory, University of California, Berkeley, 7 Gauss Way, Berkeley, CA  94720, USA}

$^{38}${Department of Physics, Kansas State University, 116 Cardwell Hall, Manhattan, KS 66506, USA}

$^{39}${Ecole Polytechnique F\'{e}d\'{e}rale de Lausanne, CH-1015 Lausanne, Switzerland}

$^{40}${Department of Physics and Astronomy, Sejong University, Seoul, 143-747, Korea}

$^{41}${CIEMAT, Avenida Complutense 40, E-28040 Madrid, Spain}

$^{42}${Department of Physics, University of Michigan, Ann Arbor, MI 48109, USA}

$^{43}${University of Michigan, Ann Arbor, MI 48109, USA}

$^{44}${Department of Physics \& Astronomy, Ohio University, Athens, OH 45701, USA}

$^{45}${SLAC National Accelerator Laboratory, Menlo Park, CA 94305, USA}

$^{46}${Sorbonne Universit\'{e}, CNRS/IN2P3, Laboratoire de Physique Nucl\'{e}aire et de Hautes Energies (LPNHE), FR-75005 Paris, France}

\end{hangparas}

\bibliography{main}

\providecommand{\href}[2]{#2}\begingroup\raggedright\begin{thebibliography}{100}

\bibitem{Pee80}
P.J.E.~{Peebles}, \emph{{The large-scale structure of the universe}} (1980).

\bibitem{Peacock99}
J.A.~{Peacock}, \emph{{Cosmological Physics}} (Jan., 1999).

\bibitem{Dodelson03}
S.~{Dodelson}, \emph{{Modern Cosmology}} (2003).

\bibitem{Baumann22}
D.~Baumann, \emph{Cosmology}, Cambridge University Press (2022),
  \href{https://doi.org/10.1017/9781108937092}{10.1017/9781108937092}.

\bibitem{2dF2001}
M.~{Colless}, G.~{Dalton}, S.~{Maddox}, W.~{Sutherland}, P.~{Norberg},
  S.~{Cole} et~al., \emph{{The 2dF Galaxy Redshift Survey: spectra and
  redshifts}},
  \href{https://doi.org/10.1046/j.1365-8711.2001.04902.x}{\emph{\mnras}
  {\bfseries 328} (2001) 1039}
  [\href{https://arxiv.org/abs/astro-ph/0106498}{{\ttfamily
  astro-ph/0106498}}].

\bibitem{6dF2009}
D.H.~{Jones}, M.A.~{Read}, W.~{Saunders}, M.~{Colless}, T.~{Jarrett},
  Q.A.~{Parker} et~al., \emph{{The 6dF Galaxy Survey: final redshift release
  (DR3) and southern large-scale structures}},
  \href{https://doi.org/10.1111/j.1365-2966.2009.15338.x}{\emph{\mnras}
  {\bfseries 399} (2009) 683}
  [\href{https://arxiv.org/abs/0903.5451}{{\ttfamily 0903.5451}}].

\bibitem{GAMA}
S.P.~{Driver}, P.~{Norberg}, I.K.~{Baldry}, S.P.~{Bamford}, A.M.~{Hopkins},
  J.~{Liske} et~al., \emph{{GAMA: towards a physical understanding of galaxy
  formation}},
  \href{https://doi.org/10.1111/j.1468-4004.2009.50512.x}{\emph{Astronomy and
  Geophysics} {\bfseries 50} (2009) 5.12}
  [\href{https://arxiv.org/abs/0910.5123}{{\ttfamily 0910.5123}}].

\bibitem{Drinkwater10}
M.J.~{Drinkwater}, R.J.~{Jurek}, C.~{Blake}, D.~{Woods}, K.A.~{Pimbblet},
  K.~{Glazebrook} et~al., \emph{{The WiggleZ Dark Energy Survey: survey design
  and first data release}},
  \href{https://doi.org/10.1111/j.1365-2966.2009.15754.x}{\emph{\mnras}
  {\bfseries 401} (2010) 1429}
  [\href{https://arxiv.org/abs/0911.4246}{{\ttfamily 0911.4246}}].

\bibitem{SDSS2000}
D.G.~{York}, J.~{Adelman}, J.~{Anderson}, John~E., S.F.~{Anderson}, J.~{Annis},
  N.A.~{Bahcall} et~al., \emph{{The Sloan Digital Sky Survey: Technical
  Summary}}, \href{https://doi.org/10.1086/301513}{\emph{\aj} {\bfseries 120}
  (2000) 1579} [\href{https://arxiv.org/abs/astro-ph/0006396}{{\ttfamily
  astro-ph/0006396}}].

\bibitem{SDSSIII}
D.J.~{Eisenstein}, D.H.~{Weinberg}, E.~{Agol}, H.~{Aihara}, C.~{Allende
  Prieto}, S.F.~{Anderson} et~al., \emph{{SDSS-III: Massive Spectroscopic
  Surveys of the Distant Universe, the Milky Way, and Extra-Solar Planetary
  Systems}}, \href{https://doi.org/10.1088/0004-6256/142/3/72}{\emph{\aj}
  {\bfseries 142} (2011) 72} [\href{https://arxiv.org/abs/1101.1529}{{\ttfamily
  1101.1529}}].

\bibitem{BOSS_DR12}
S.~{Alam}, M.~{Ata}, S.~{Bailey}, F.~{Beutler}, D.~{Bizyaev}, J.A.~{Blazek}
  et~al., \emph{{The clustering of galaxies in the completed SDSS-III Baryon
  Oscillation Spectroscopic Survey: cosmological analysis of the DR12 galaxy
  sample}}, \href{https://doi.org/10.1093/mnras/stx721}{\emph{\mnras}
  {\bfseries 470} (2017) 2617}
  [\href{https://arxiv.org/abs/1607.03155}{{\ttfamily 1607.03155}}].

\bibitem{SDSSIII2015}
S.~{Alam}, F.D.~{Albareti}, C.~{Allende Prieto}, F.~{Anders}, S.F.~{Anderson},
  T.~{Anderton} et~al., \emph{{The Eleventh and Twelfth Data Releases of the
  Sloan Digital Sky Survey: Final Data from SDSS-III}},
  \href{https://doi.org/10.1088/0067-0049/219/1/12}{\emph{\apjs} {\bfseries
  219} (2015) 12} [\href{https://arxiv.org/abs/1501.00963}{{\ttfamily
  1501.00963}}].

\bibitem{eBOSS2020}
H.~{du Mas des Bourboux}, J.~{Rich}, A.~{Font-Ribera}, V.~{de Sainte Agathe},
  J.~{Farr}, T.~{Etourneau} et~al., \emph{{The Completed SDSS-IV Extended
  Baryon Oscillation Spectroscopic Survey: Baryon Acoustic Oscillations with
  Ly{\ensuremath{\alpha}} Forests}},
  \href{https://doi.org/10.3847/1538-4357/abb085}{\emph{\apj} {\bfseries 901}
  (2020) 153} [\href{https://arxiv.org/abs/2007.08995}{{\ttfamily
  2007.08995}}].

\bibitem{eBOSS2021}
A.~{Raichoor}, A.~{de Mattia}, A.J.~{Ross}, C.~{Zhao}, S.~{Alam}, S.~{Avila}
  et~al., \emph{{The completed SDSS-IV extended Baryon Oscillation
  Spectroscopic Survey: large-scale structure catalogues and measurement of the
  isotropic BAO between redshift 0.6 and 1.1 for the Emission Line Galaxy
  Sample}}, \href{https://doi.org/10.1093/mnras/staa3336}{\emph{\mnras}
  {\bfseries 500} (2021) 3254}
  [\href{https://arxiv.org/abs/2007.09007}{{\ttfamily 2007.09007}}].

\bibitem{SDSSIV2020}
B.W.~{Lyke}, A.N.~{Higley}, J.N.~{McLane}, D.P.~{Schurhammer}, A.D.~{Myers},
  A.J.~{Ross} et~al., \emph{{The Sloan Digital Sky Survey Quasar Catalog:
  Sixteenth Data Release}},
  \href{https://doi.org/10.3847/1538-4365/aba623}{\emph{\apjs} {\bfseries 250}
  (2020) 8} [\href{https://arxiv.org/abs/2007.09001}{{\ttfamily 2007.09001}}].

\bibitem{Euclid}
R.~{Laureijs}, J.~{Amiaux}, S.~{Arduini}, J..~{Augu{\`e}res}, J.~{Brinchmann},
  R.~{Cole} et~al., \emph{{Euclid Definition Study Report}}, {\emph{ArXiv
  e-prints} (2011) } [\href{https://arxiv.org/abs/1110.3193}{{\ttfamily
  1110.3193}}].

\bibitem{Amendola18}
L.~{Amendola}, S.~{Appleby}, A.~{Avgoustidis}, D.~{Bacon}, T.~{Baker},
  M.~{Baldi} et~al., \emph{{Cosmology and fundamental physics with the Euclid
  satellite}}, \href{https://doi.org/10.1007/s41114-017-0010-3}{\emph{Living
  Reviews in Relativity} {\bfseries 21} (2018) 2}
  [\href{https://arxiv.org/abs/1606.00180}{{\ttfamily 1606.00180}}].

\bibitem{DESI2016a.Science}
{DESI Collaboration}, A.~{Aghamousa}, J.~{Aguilar}, S.~{Ahlen}, S.~{Alam},
  L.E.~{Allen} et~al., \emph{{The DESI Experiment Part I: Science,Targeting,
  and Survey Design}}, {\emph{arXiv e-prints} (2016) arXiv:1611.00036}
  [\href{https://arxiv.org/abs/1611.00036}{{\ttfamily 1611.00036}}].

\bibitem{DESI2016_2}
{DESI Collaboration}, A.~{Aghamousa}, J.~{Aguilar}, S.~{Ahlen}, S.~{Alam},
  L.E.~{Allen} et~al., \emph{{The DESI Experiment Part II: Instrument Design}},
  \href{https://doi.org/10.48550/arXiv.1611.00037}{\emph{arXiv e-prints} (2016)
  arXiv:1611.00037} [\href{https://arxiv.org/abs/1611.00037}{{\ttfamily
  1611.00037}}].

\bibitem{DESI2022.KP1.Instr}
{DESI Collaboration}, B.~{Abareshi}, J.~{Aguilar}, S.~{Ahlen}, S.~{Alam},
  D.M.~{Alexander} et~al., \emph{{Overview of the Instrumentation for the Dark
  Energy Spectroscopic Instrument}},
  \href{https://doi.org/10.3847/1538-3881/ac882b}{\emph{\aj} {\bfseries 164}
  (2022) 207} [\href{https://arxiv.org/abs/2205.10939}{{\ttfamily
  2205.10939}}].

\bibitem{DESI2023}
{DESI Collaboration}, A.G.~{Adame}, J.~{Aguilar}, S.~{Ahlen}, S.~{Alam},
  G.~{Aldering} et~al., \emph{{Validation of the Scientific Program for the
  Dark Energy Spectroscopic Instrument}},
  \href{https://doi.org/10.48550/arXiv.2306.06307}{\emph{arXiv e-prints} (2023)
  arXiv:2306.06307} [\href{https://arxiv.org/abs/2306.06307}{{\ttfamily
  2306.06307}}].

\bibitem{DESI2023b.KP1.EDR}
{DESI Collaboration}, A.G.~{Adame}, J.~{Aguilar}, S.~{Ahlen}, S.~{Alam},
  G.~{Aldering} et~al., \emph{{The Early Data Release of the Dark Energy
  Spectroscopic Instrument}},
  \href{https://doi.org/10.48550/arXiv.2306.06308}{\emph{arXiv e-prints} (2023)
  arXiv:2306.06308} [\href{https://arxiv.org/abs/2306.06308}{{\ttfamily
  2306.06308}}].

\bibitem{DESI2024.I.DR1}
{DESI Collaboration}, \emph{{DESI 2024 I: Data Release 1 of the Dark Energy
  Spectroscopic Instrument}}, {\emph{in preparation} (2025) }.

\bibitem{DESI2024.II.KP3}
{DESI Collaboration}, \emph{{DESI 2024 II: Sample definitions, characteristics
  and two-point clustering statistics}}, {\emph{in preparation} (2024) }.

\bibitem{DESI2024.III.KP4}
{DESI Collaboration}, A.G.~Adame, J.~Aguilar, S.~Ahlen, S.~Alam, D.M.~Alexander
  et~al., \emph{{DESI 2024 III: Baryon Acoustic Oscillations from Galaxies and
  Quasars}}, \href{https://doi.org/10.48550/arXiv.2404.03000}{\emph{arXiv
  e-prints} (2024) arXiv:2404.03000}
  [\href{https://arxiv.org/abs/2404.03000}{{\ttfamily 2404.03000}}].

\bibitem{DESI2024.V.KP5}
{DESI Collaboration}, \emph{{DESI 2024 V: Analysis of the full shape of
  two-point clustering statistics from galaxies and quasars}}, {\emph{in
  preparation} (2024) }.

\bibitem{DESI2024.IV.KP6}
{DESI Collaboration}, A.G.~Adame, J.~Aguilar, S.~Ahlen, S.~Alam, D.M.~Alexander
  et~al., \emph{{DESI 2024 IV: Baryon Acoustic Oscillations from the Lyman
  Alpha Forest}}, \href{https://doi.org/10.48550/arXiv.2404.03001}{\emph{arXiv
  e-prints} (2024) arXiv:2404.03001}
  [\href{https://arxiv.org/abs/2404.03001}{{\ttfamily 2404.03001}}].

\bibitem{DESI2024.VI.KP7A}
{DESI Collaboration}, A.G.~Adame, J.~Aguilar, S.~Ahlen, S.~Alam, D.M.~Alexander
  et~al., \emph{{DESI 2024 VI: Cosmological Constraints from the Measurements
  of Baryon Acoustic Oscillations}},
  \href{https://doi.org/10.48550/arXiv.2404.03002}{\emph{arXiv e-prints} (2024)
  arXiv:2404.03002} [\href{https://arxiv.org/abs/2404.03002}{{\ttfamily
  2404.03002}}].

\bibitem{DESI2024.VII.KP7B}
{DESI Collaboration}, \emph{{DESI 2024 VII: Cosmological constraints from
  full-shape analyses of the two-point clustering statistics measurements}},
  {\emph{in preparation} (2024) }.

\bibitem{DESI2024.VIII.KP7C}
{DESI Collaboration}, \emph{{DESI 2024 VIII: Constraints on Primordial
  Non-Gaussianities}}, {\emph{in preparation} (2024) }.

\bibitem{Des18}
V.~{Desjacques}, D.~{Jeong} and F.~{Schmidt}, \emph{{Large-scale galaxy bias}},
  \href{https://doi.org/10.1016/j.physrep.2017.12.002}{\emph{\physrep}
  {\bfseries 733} (2018) 1} [\href{https://arxiv.org/abs/1611.09787}{{\ttfamily
  1611.09787}}].

\bibitem{Kaiser87}
N.~{Kaiser}, \emph{{Clustering in real space and in redshift space}},
  \href{https://doi.org/10.1093/mnras/227.1.1}{\emph{\mnras} {\bfseries 227}
  (1987) 1}.

\bibitem{Hamilton92}
A.J.S.~{Hamilton}, \emph{{Measuring Omega and the real correlation function
  from the redshift correlation function}},
  \href{https://doi.org/10.1086/186264}{\emph{\apjl} {\bfseries 385} (1992)
  L5}.

\bibitem{CHS12}
J.J.M.~{Carrasco}, M.P.~{Hertzberg} and L.~{Senatore}, \emph{{The effective
  field theory of cosmological large scale structures}},
  \href{https://doi.org/10.1007/JHEP09(2012)082}{\emph{Journal of High Energy
  Physics} {\bfseries 9} (2012) 82}
  [\href{https://arxiv.org/abs/1206.2926}{{\ttfamily 1206.2926}}].

\bibitem{PorSenZal14}
R.A.~{Porto}, L.~{Senatore} and M.~{Zaldarriaga}, \emph{{The Lagrangian-space
  Effective Field Theory of large scale structures}},
  \href{https://doi.org/10.1088/1475-7516/2014/05/022}{\emph{\jcap} {\bfseries
  5} (2014) 022} [\href{https://arxiv.org/abs/1311.2168}{{\ttfamily
  1311.2168}}].

\bibitem{VlaWhiAvi15}
Z.~{Vlah}, M.~{White} and A.~{Aviles}, \emph{{A Lagrangian effective field
  theory}}, \href{https://doi.org/10.1088/1475-7516/2015/09/014}{\emph{\jcap}
  {\bfseries 9} (2015) 014} [\href{https://arxiv.org/abs/1506.05264}{{\ttfamily
  1506.05264}}].

\bibitem{Chen20}
S.-F.~{Chen}, Z.~{Vlah} and M.~{White}, \emph{{Consistent modeling of velocity
  statistics and redshift-space distortions in one-loop perturbation theory}},
  \href{https://doi.org/10.1088/1475-7516/2020/07/062}{\emph{\jcap} {\bfseries
  2020} (2020) 062} [\href{https://arxiv.org/abs/2005.00523}{{\ttfamily
  2005.00523}}].

\bibitem{Chen21}
S.-F.~{Chen}, Z.~{Vlah}, E.~{Castorina} and M.~{White}, \emph{{Redshift-space
  distortions in Lagrangian perturbation theory}},
  \href{https://doi.org/10.1088/1475-7516/2021/03/100}{\emph{\jcap} {\bfseries
  2021} (2021) 100} [\href{https://arxiv.org/abs/2012.04636}{{\ttfamily
  2012.04636}}].

\bibitem{Pybird21}
G.~{D'Amico}, L.~{Senatore} and P.~{Zhang}, \emph{{Limits on wCDM from the
  EFTofLSS with the PyBird code}},
  \href{https://doi.org/10.1088/1475-7516/2021/01/006}{\emph{\jcap} {\bfseries
  2021} (2021) 006} [\href{https://arxiv.org/abs/2003.07956}{{\ttfamily
  2003.07956}}].

\bibitem{dAmico20}
G.~{d'Amico}, J.~{Gleyzes}, N.~{Kokron}, K.~{Markovic}, L.~{Senatore},
  P.~{Zhang} et~al., \emph{{The cosmological analysis of the SDSS/BOSS data
  from the Effective Field Theory of Large-Scale Structure}},
  \href{https://doi.org/10.1088/1475-7516/2020/05/005}{\emph{\jcap} {\bfseries
  2020} (2020) 005} [\href{https://arxiv.org/abs/1909.05271}{{\ttfamily
  1909.05271}}].

\bibitem{Colas20}
T.~{Colas}, G.~{d'Amico}, L.~{Senatore}, P.~{Zhang} and F.~{Beutler},
  \emph{{Efficient cosmological analysis of the SDSS/BOSS data from the
  Effective Field Theory of Large-Scale Structure}},
  \href{https://doi.org/10.1088/1475-7516/2020/06/001}{\emph{\jcap} {\bfseries
  2020} (2020) 001} [\href{https://arxiv.org/abs/1909.07951}{{\ttfamily
  1909.07951}}].

\bibitem{Noriega22}
H.E.~Noriega, A.~Aviles, S.~Fromenteau and M.~Vargas-Maga\~na, \emph{{Fast
  computation of non-linear power spectrum in cosmologies with massive
  neutrinos}},  \href{https://arxiv.org/abs/2208.02791}{{\ttfamily
  2208.02791}}.

\bibitem{Ramirez23}
S.~{Ramirez}, M.~{Icaza-Lizaola}, S.~{Fromenteau}, M.~{Vargas-Maga{\~n}a} and
  A.~{Aviles}, \emph{{Full Shape Cosmology Analysis from BOSS in configuration
  space using Neural Network Acceleration}},
  \href{https://doi.org/10.48550/arXiv.2310.17834}{\emph{arXiv e-prints} (2023)
  arXiv:2310.17834} [\href{https://arxiv.org/abs/2310.17834}{{\ttfamily
  2310.17834}}].

\bibitem{KP5s3-Noriega}
H.E.~Noriega, A.~Aviles, H.~Gil-Marín, S.~Ramirez-Solano, S.~Fromenteau,
  M.~Vargas-Magaña et~al., \emph{Comparing compressed and full-modeling
  analyses with folps: Implications for desi 2024 and beyond},
  \href{https://doi.org/10.48550/arXiv.2404.07269}{\emph{arXiv e-prints} (2024)
  arXiv:2404.07269} [\href{https://arxiv.org/abs/2404.07269}{{\ttfamily
  2404.07269}}].

\bibitem{KP5s4-Lai}
Y.~Lai, C.~Howlett, M.~Maus, H.~Gil-Marín, H.E.~Noriega, S.~Ramírez-Solano
  et~al., \emph{{A comparison between Shapefit compression and Full-Modelling
  method with PyBird for DESI 2024 and beyond}},
  \href{https://doi.org/10.48550/arXiv.2404.07283}{\emph{arXiv e-prints} (2024)
  arXiv:2404.07283} [\href{https://arxiv.org/abs/2404.07283}{{\ttfamily
  2404.07283}}].

\bibitem{KP5s5-Ramirez}
S.~Ramirez-Solano, M.~Icaza-Lizaola, H.E.~Noriega, M.~Vargas-Magaña,
  S.~Fromenteau, A.~Aviles et~al., \emph{Full modeling and parameter
  compression methods in configuration space for desi 2024 and beyond},
  \href{https://doi.org/10.48550/arXiv.2404.07268}{\emph{arXiv e-prints} (2024)
  arXiv:2404.07268} [\href{https://arxiv.org/abs/2404.07268}{{\ttfamily
  2404.07268}}].

\bibitem{KP5s1-Maus}
M.~Maus, Y.~Lai, H.E.~Noriega, S.~Ramirez-Solano, A.~Aviles, S.~Chen et~al.,
  \emph{A comparison of effective field theory models of redshift space galaxy
  power spectra for desi 2024 and future surveys},
  \href{https://doi.org/10.48550/arXiv.2404.07272}{\emph{arXiv e-prints} (2024)
  arXiv:2404.07272} [\href{https://arxiv.org/abs/2404.07272}{{\ttfamily
  2404.07272}}].

\bibitem{Chen22}
S.-F.~{Chen}, Z.~{Vlah} and M.~{White}, \emph{{A new analysis of galaxy 2-point
  functions in the BOSS survey, including full-shape information and
  post-reconstruction BAO}},
  \href{https://doi.org/10.1088/1475-7516/2022/02/008}{\emph{\jcap} {\bfseries
  2022} (2022) 008} [\href{https://arxiv.org/abs/2110.05530}{{\ttfamily
  2110.05530}}].

\bibitem{ChenVlahWhite19}
S.-F.~{Chen}, Z.~{Vlah} and M.~{White}, \emph{{The reconstructed power spectrum
  in the Zeldovich approximation}},
  \href{https://doi.org/10.1088/1475-7516/2019/09/017}{\emph{\jcap} {\bfseries
  2019} (2019) 017} [\href{https://arxiv.org/abs/1907.00043}{{\ttfamily
  1907.00043}}].

\bibitem{Maksimova21}
N.A.~Maksimova, L.H.~Garrison, D.J.~Eisenstein, B.~Hadzhiyska, S.~Bose and
  T.P.~Satterthwaite, \emph{{AbacusSummit: a massive set of high-accuracy,
  high-resolution N-body simulations}},
  \href{https://doi.org/10.1093/mnras/stab2484}{\emph{Monthly Notices of the
  Royal Astronomical Society} {\bfseries 508} (2021) 4017}
  [\href{https://arxiv.org/abs/https://academic.oup.com/mnras/article-pdf/508/3/4017/40811763/stab2484.pdf}{{\ttfamily
  https://academic.oup.com/mnras/article-pdf/508/3/4017/40811763/stab2484.pdf}}].

\bibitem{Briedan21}
S.~{Brieden}, H.~{Gil-Mar{\'\i}n} and L.~{Verde}, \emph{{ShapeFit: extracting
  the power spectrum shape information in galaxy surveys beyond BAO and RSD}},
  \href{https://doi.org/10.1088/1475-7516/2021/12/054}{\emph{\jcap} {\bfseries
  2021} (2021) 054} [\href{https://arxiv.org/abs/2106.07641}{{\ttfamily
  2106.07641}}].

\bibitem{Maus23}
M.~{Maus}, S.-F.~{Chen} and M.~{White}, \emph{{A comparison of template vs.
  direct model fitting for redshift-space distortions in BOSS}},
  \href{https://doi.org/10.1088/1475-7516/2023/06/005}{\emph{\jcap} {\bfseries
  2023} (2023) 005} [\href{https://arxiv.org/abs/2302.07430}{{\ttfamily
  2302.07430}}].

\bibitem{Garrison21}
L.H.~Garrison, D.J.~Eisenstein, D.~Ferrer, N.A.~Maksimova and P.A.~Pinto,
  \emph{{The abacus cosmological N-body code}},
  \href{https://doi.org/10.1093/mnras/stab2482}{\emph{Monthly Notices of the
  Royal Astronomical Society} {\bfseries 508} (2021) 575}
  [\href{https://arxiv.org/abs/https://academic.oup.com/mnras/article-pdf/508/1/575/40458823/stab2482.pdf}{{\ttfamily
  https://academic.oup.com/mnras/article-pdf/508/1/575/40458823/stab2482.pdf}}].

\bibitem{Chuang2015}
C.-H.~{Chuang}, F.-S.~{Kitaura}, F.~{Prada}, C.~{Zhao} and G.~{Yepes},
  \emph{{EZmocks: extending the Zel'dovich approximation to generate mock
  galaxy catalogues with accurate clustering statistics}},
  \href{https://doi.org/10.1093/mnras/stu2301}{\emph{\mnras} {\bfseries 446}
  (2015) 2621}.

\bibitem{Grove22}
C.~{Grove}, C.-H.~{Chuang}, N.C.~{Devi}, L.~{Garrison}, B.~{L'Huillier},
  Y.~{Feng} et~al., \emph{{The DESI N-body simulation project - I. Testing the
  robustness of simulations for the DESI dark time survey}},
  \href{https://doi.org/10.1093/mnras/stac1947}{\emph{\mnras} {\bfseries 515}
  (2022) 1854} [\href{https://arxiv.org/abs/2112.09138}{{\ttfamily
  2112.09138}}].

\bibitem{Angulo22}
R.E.~{Angulo} and O.~{Hahn}, \emph{{Large-scale dark matter simulations}},
  \href{https://doi.org/10.1007/s41115-021-00013-z}{\emph{Living Reviews in
  Computational Astrophysics} {\bfseries 8} (2022) 1}
  [\href{https://arxiv.org/abs/2112.05165}{{\ttfamily 2112.05165}}].

\bibitem{Hartlap07}
J.~{Hartlap}, P.~{Simon} and P.~{Schneider}, \emph{{Why your model parameter
  confidences might be too optimistic. Unbiased estimation of the inverse
  covariance matrix}},
  \href{https://doi.org/10.1051/0004-6361:20066170}{\emph{\aap} {\bfseries 464}
  (2007) 399} [\href{https://arxiv.org/abs/astro-ph/0608064}{{\ttfamily
  astro-ph/0608064}}].

\bibitem{Abidi18}
M.M.~{Abidi} and T.~{Baldauf}, \emph{{Cubic halo bias in Eulerian and
  Lagrangian space}},
  \href{https://doi.org/10.1088/1475-7516/2018/07/029}{\emph{\jcap} {\bfseries
  2018} (2018) 029} [\href{https://arxiv.org/abs/1802.07622}{{\ttfamily
  1802.07622}}].

\bibitem{CLPT}
J.~{Carlson}, B.~{Reid} and M.~{White}, \emph{{Convolution Lagrangian
  perturbation theory for biased tracers}},
  \href{https://doi.org/10.1093/mnras/sts457}{\emph{\mnras} {\bfseries 429}
  (2013) 1674} [\href{https://arxiv.org/abs/1209.0780}{{\ttfamily 1209.0780}}].

\bibitem{SenZal15}
L.~{Senatore} and M.~{Zaldarriaga}, \emph{{The IR-resummed Effective Field
  Theory of Large Scale Structures}},
  \href{https://doi.org/10.1088/1475-7516/2015/02/013}{\emph{\jcap} {\bfseries
  2} (2015) 13} [\href{https://arxiv.org/abs/1404.5954}{{\ttfamily
  1404.5954}}].

\bibitem{Blas16}
D.~{Blas}, M.~{Garny}, M.M.~{Ivanov} and S.~{Sibiryakov}, \emph{{Time-sliced
  perturbation theory II: baryon acoustic oscillations and infrared
  resummation}},
  \href{https://doi.org/10.1088/1475-7516/2016/07/028}{\emph{\jcap} {\bfseries
  2016} (2016) 028} [\href{https://arxiv.org/abs/1605.02149}{{\ttfamily
  1605.02149}}].

\bibitem{Vlah16}
Z.~{Vlah}, U.~{Seljak}, M.~{Yat Chu} and Y.~{Feng}, \emph{{Perturbation theory,
  effective field theory, and oscillations in the power spectrum}},
  \href{https://doi.org/10.1088/1475-7516/2016/03/057}{\emph{Journal of
  Cosmology and Astro-Particle Physics} {\bfseries 2016} (2016) 057}
  [\href{https://arxiv.org/abs/1509.02120}{{\ttfamily 1509.02120}}].

\bibitem{VlaCasWhi16}
Z.~{Vlah}, E.~{Castorina} and M.~{White}, \emph{{The Gaussian streaming model
  and convolution Lagrangian effective field theory}},
  \href{https://doi.org/10.1088/1475-7516/2016/12/007}{\emph{\jcap} {\bfseries
  12} (2016) 007} [\href{https://arxiv.org/abs/1609.02908}{{\ttfamily
  1609.02908}}].

\bibitem{Schmittfull2021}
M.~{Schmittfull}, M.~{Simonovi{\'c}}, M.M.~{Ivanov}, O.H.E.~{Philcox} and
  M.~{Zaldarriaga}, \emph{{Modeling galaxies in redshift space at the field
  level}}, \href{https://doi.org/10.1088/1475-7516/2021/05/059}{\emph{\jcap}
  {\bfseries 2021} (2021) 059}
  [\href{https://arxiv.org/abs/2012.03334}{{\ttfamily 2012.03334}}].

\bibitem{Alcock79}
C.~{Alcock} and B.~{Paczynski}, \emph{{An evolution free test for non-zero
  cosmological constant}}, \href{https://doi.org/10.1038/281358a0}{\emph{\nat}
  {\bfseries 281} (1979) 358}.

\bibitem{Chen_BOSSrecon2022}
S.-F.~{Chen}, Z.~{Vlah} and M.~{White}, \emph{{A new analysis of galaxy 2-point
  functions in the BOSS survey, including full-shape information and
  post-reconstruction BAO}},
  \href{https://doi.org/10.1088/1475-7516/2022/02/008}{\emph{\jcap} {\bfseries
  2022} (2022) 008} [\href{https://arxiv.org/abs/2110.05530}{{\ttfamily
  2110.05530}}].

\bibitem{KP5s8-Findlay}
{N.~Findlay, R.~Gsponer, F.~Rodr{\'\i}guez-Mart{\'\i}nez et al.},
  \emph{{Fiducial cosmology impact for DESI 2024 full shape analysis}},
  {\emph{in preparation} (2024) }.

\bibitem{Cooke2018}
R.J.~{Cooke}, M.~{Pettini} and C.C.~{Steidel}, \emph{{One Percent Determination
  of the Primordial Deuterium Abundance}},
  \href{https://doi.org/10.3847/1538-4357/aaab53}{\emph{\apj} {\bfseries 855}
  (2018) 102} [\href{https://arxiv.org/abs/1710.11129}{{\ttfamily
  1710.11129}}].

\bibitem{Brieden23}
S.~{Brieden}, H.~{Gil-Mar{\'\i}n} and L.~{Verde}, \emph{{A tale of two (or
  more) h's}},
  \href{https://doi.org/10.1088/1475-7516/2023/04/023}{\emph{\jcap} {\bfseries
  2023} (2023) 023} [\href{https://arxiv.org/abs/2212.04522}{{\ttfamily
  2212.04522}}].

\bibitem{ESSS07}
D.J.~{Eisenstein}, H.-J.~{Seo}, E.~{Sirko} and D.N.~{Spergel}, \emph{{Improving
  Cosmological Distance Measurements by Reconstruction of the Baryon Acoustic
  Peak}}, \href{https://doi.org/10.1086/518712}{\emph{\apj} {\bfseries 664}
  (2007) 675} [\href{https://arxiv.org/abs/astro-ph/0604362}{{\ttfamily
  astro-ph/0604362}}].

\bibitem{Noh_recon2009}
Y.~{Noh}, M.~{White} and N.~{Padmanabhan}, \emph{{Reconstructing baryon
  oscillations}}, \href{https://doi.org/10.1103/PhysRevD.80.123501}{\emph{\prd}
  {\bfseries 80} (2009) 123501}
  [\href{https://arxiv.org/abs/0909.1802}{{\ttfamily 0909.1802}}].

\bibitem{Padmanabhan_recon2009}
N.~{Padmanabhan}, M.~{White} and J.D.~{Cohn}, \emph{{Reconstructing baryon
  oscillations: A Lagrangian theory perspective}},
  \href{https://doi.org/10.1103/PhysRevD.79.063523}{\emph{\prd} {\bfseries 79}
  (2009) 063523} [\href{https://arxiv.org/abs/0812.2905}{{\ttfamily
  0812.2905}}].

\bibitem{White_recon2015}
M.~{White}, \emph{{Reconstruction within the Zeldovich approximation}},
  \href{https://doi.org/10.1093/mnras/stv842}{\emph{\mnras} {\bfseries 450}
  (2015) 3822} [\href{https://arxiv.org/abs/1504.03677}{{\ttfamily
  1504.03677}}].

\bibitem{Chen_recon2019}
S.-F.~{Chen}, Z.~{Vlah} and M.~{White}, \emph{{The reconstructed power spectrum
  in the Zeldovich approximation}},
  \href{https://doi.org/10.1088/1475-7516/2019/09/017}{\emph{\jcap} {\bfseries
  2019} (2019) 017} [\href{https://arxiv.org/abs/1907.00043}{{\ttfamily
  1907.00043}}].

\bibitem{KP4s2-Chen}
S.-F.~{Chen}, C.~{Howlett}, M.~{White}, P.~{McDonald}, A.J.~{Ross}, H.-J.~{Seo}
  et~al., \emph{{Baryon Acoustic Oscillation Theory and Modelling Systematics
  for the DESI 2024 results}},
  \href{https://doi.org/10.48550/arXiv.2402.14070}{\emph{arXiv e-prints} (2024)
  arXiv:2402.14070} [\href{https://arxiv.org/abs/2402.14070}{{\ttfamily
  2402.14070}}].

\bibitem{Wallisch2018}
B.~{Wallisch}, \emph{{Cosmological probes of light relics}}, Ph.D. thesis,
  University of Cambridge, UK, Jan., 2018.

\bibitem{Sugiyama24}
N.~{Sugiyama}, \emph{{Developing a Theoretical Model for the Resummation of
  Infrared Effects in the Post-Reconstruction Power Spectrum
  (youtu.be/u1-xx3\_4xCg)}},
  \href{https://doi.org/10.48550/arXiv.2402.06142}{\emph{arXiv e-prints} (2024)
  arXiv:2402.06142} [\href{https://arxiv.org/abs/2402.06142}{{\ttfamily
  2402.06142}}].

\bibitem{Hockney88}
R.W.~{Hockney} and J.W.~{Eastwood}, \emph{{Computer simulation using
  particles}} (1988).

\bibitem{Jeong10}
D.~{Jeong}, \emph{{Cosmology with high (z>1) redshift galaxy surveys}}, Ph.D.
  thesis, University of Texas, Austin, Aug., 2010.

\bibitem{Planck18-I}
{Planck Collaboration}, N.~{Aghanim}, Y.~{Akrami}, F.~{Arroja}, M.~{Ashdown},
  J.~{Aumont} et~al., \emph{{Planck 2018 results. I. Overview and the
  cosmological legacy of Planck}},
  \href{https://doi.org/10.1051/0004-6361/201833880}{\emph{\aap} {\bfseries
  641} (2020) A1} [\href{https://arxiv.org/abs/1807.06205}{{\ttfamily
  1807.06205}}].

\bibitem{Aghanim2020}
{Planck Collaboration}, N.~{Aghanim}, Y.~{Akrami}, M.~{Ashdown}, J.~{Aumont},
  C.~{Baccigalupi} et~al., \emph{{Planck 2018 results. VI. Cosmological
  parameters}}, \href{https://doi.org/10.1051/0004-6361/201833910}{\emph{\aap}
  {\bfseries 641} (2020) A6}
  [\href{https://arxiv.org/abs/1807.06209}{{\ttfamily 1807.06209}}].

\bibitem{McDRoy09}
P.~{McDonald} and A.~{Roy}, \emph{{Clustering of dark matter tracers:
  generalizing bias for the coming era of precision LSS}},
  \href{https://doi.org/10.1088/1475-7516/2009/08/020}{\emph{\jcap} {\bfseries
  8} (2009) 020} [\href{https://arxiv.org/abs/0902.0991}{{\ttfamily
  0902.0991}}].

\bibitem{ChenVlahWhite20}
S.-F.~{Chen}, Z.~{Vlah} and M.~{White}, \emph{{Consistent modeling of velocity
  statistics and redshift-space distortions in one-loop perturbation theory}},
  \href{https://doi.org/10.1088/1475-7516/2020/07/062}{\emph{\jcap} {\bfseries
  2020} (2020) 062} [\href{https://arxiv.org/abs/2005.00523}{{\ttfamily
  2005.00523}}].

\bibitem{Bridle02}
S.L.~{Bridle}, R.~{Crittenden}, A.~{Melchiorri}, M.P.~{Hobson}, R.~{Kneissl}
  and A.N.~{Lasenby}, \emph{{Analytic marginalization over CMB calibration and
  beam uncertainty}},
  \href{https://doi.org/10.1046/j.1365-8711.2002.05709.x}{\emph{\mnras}
  {\bfseries 335} (2002) 1193}
  [\href{https://arxiv.org/abs/astro-ph/0112114}{{\ttfamily
  astro-ph/0112114}}].

\bibitem{Taylor10}
A.N.~{Taylor} and T.D.~{Kitching}, \emph{{Analytic methods for cosmological
  likelihoods}},
  \href{https://doi.org/10.1111/j.1365-2966.2010.17201.x}{\emph{\mnras}
  {\bfseries 408} (2010) 865}
  [\href{https://arxiv.org/abs/1003.1136}{{\ttfamily 1003.1136}}].

\bibitem{Handley19b}
W.~{Handley} and P.~{Lemos}, \emph{{Quantifying tensions in cosmological
  parameters: Interpreting the DES evidence ratio}},
  \href{https://doi.org/10.1103/PhysRevD.100.043504}{\emph{\prd} {\bfseries
  100} (2019) 043504} [\href{https://arxiv.org/abs/1902.04029}{{\ttfamily
  1902.04029}}].

\bibitem{Lemos21}
P.~{Lemos}, M.~{Raveri}, A.~{Campos}, Y.~{Park}, C.~{Chang}, N.~{Weaverdyck}
  et~al., \emph{{Assessing tension metrics with dark energy survey and Planck
  data}}, \href{https://doi.org/10.1093/mnras/stab1670}{\emph{\mnras}
  {\bfseries 505} (2021) 6179}
  [\href{https://arxiv.org/abs/2012.09554}{{\ttfamily 2012.09554}}].

\bibitem{GomezValent22}
A.~{G{\'o}mez-Valent}, \emph{{Fast test to assess the impact of marginalization
  in Monte Carlo analyses and its application to cosmology}},
  \href{https://doi.org/10.1103/PhysRevD.106.063506}{\emph{\prd} {\bfseries
  106} (2022) 063506} [\href{https://arxiv.org/abs/2203.16285}{{\ttfamily
  2203.16285}}].

\bibitem{Hadzhiyska23}
B.~{Hadzhiyska}, K.~{Wolz}, S.~{Azzoni}, D.~{Alonso},
  C.~{Garc{\'\i}a-Garc{\'\i}a}, J.~{Ruiz-Zapatero} et~al., \emph{{Cosmology
  with 6 parameters in the Stage-IV era: efficient marginalisation over
  nuisance parameters}},
  \href{https://doi.org/10.21105/astro.2301.11895}{\emph{The Open Journal of
  Astrophysics} {\bfseries 6} (2023) 23}
  [\href{https://arxiv.org/abs/2301.11895}{{\ttfamily 2301.11895}}].

\bibitem{Sailer24}
N.~{Sailer}, ``Cosmological constraints from the cross-correlation of desi
  luminous red galaxies with cmb lensing from planck pr4 and act dr6.'' 2024.

\bibitem{Chen22_2}
S.-F.~{Chen}, M.~{White}, J.~{DeRose} and N.~{Kokron}, \emph{{Cosmological
  analysis of three-dimensional BOSS galaxy clustering and Planck CMB lensing
  cross correlations via Lagrangian perturbation theory}},
  \href{https://doi.org/10.1088/1475-7516/2022/07/041}{\emph{\jcap} {\bfseries
  2022} (2022) 041} [\href{https://arxiv.org/abs/2204.10392}{{\ttfamily
  2204.10392}}].

\bibitem{Syversveen98}
A.~Syversveen, \emph{Noninformative bayesian priors. interpretation and
  problems with construction and applications.}, .

\bibitem{BayesianTheory01}
J.M.~Bernardo and A.F.M.~Smith, \emph{Bayesian theory},
  \href{https://doi.org/10.1088/0957-0233/12/2/702}{\emph{Measurement Science
  and Technology} {\bfseries 12} (2001) 221}.

\bibitem{Seljak00}
U.~{Seljak}, \emph{{Analytic model for galaxy and dark matter clustering}},
  \href{https://doi.org/10.1046/j.1365-8711.2000.03715.x}{\emph{\mnras}
  {\bfseries 318} (2000) 203}
  [\href{https://arxiv.org/abs/astro-ph/0001493}{{\ttfamily
  astro-ph/0001493}}].

\bibitem{Peacock00}
J.A.~{Peacock} and R.E.~{Smith}, \emph{{Halo occupation numbers and galaxy
  bias}}, \href{https://doi.org/10.1046/j.1365-8711.2000.03779.x}{\emph{\mnras}
  {\bfseries 318} (2000) 1144}
  [\href{https://arxiv.org/abs/astro-ph/0005010}{{\ttfamily
  astro-ph/0005010}}].

\bibitem{Schaan21}
E.~{Schaan} and M.~{White}, \emph{{Multi-tracer intensity mapping:
  cross-correlations, line noise \& decorrelation}},
  \href{https://doi.org/10.1088/1475-7516/2021/05/068}{\emph{\jcap} {\bfseries
  2021} (2021) 068} [\href{https://arxiv.org/abs/2103.01964}{{\ttfamily
  2103.01964}}].

\bibitem{Scoccimarro04}
R.~{Scoccimarro}, \emph{{Redshift-space distortions, pairwise velocities, and
  nonlinearities}},
  \href{https://doi.org/10.1103/PhysRevD.70.083007}{\emph{\prd} {\bfseries 70}
  (2004) 083007} [\href{https://arxiv.org/abs/astro-ph/0407214}{{\ttfamily
  astro-ph/0407214}}].

\bibitem{VlahWhite19}
Z.~{Vlah} and M.~{White}, \emph{{Exploring redshift-space distortions in
  large-scale structure}},
  \href{https://doi.org/10.1088/1475-7516/2019/03/007}{\emph{\jcap} {\bfseries
  2019} (2019) 007} [\href{https://arxiv.org/abs/1812.02775}{{\ttfamily
  1812.02775}}].

\bibitem{Ivanov_2023}
M.M.~{Ivanov} and O.H.E.~{Philcox}, \emph{{Measuring $H_0$ with Spectroscopic
  Surveys}}, \href{https://doi.org/10.48550/arXiv.2305.07977}{\emph{arXiv
  e-prints} (2023) arXiv:2305.07977}
  [\href{https://arxiv.org/abs/2305.07977}{{\ttfamily 2305.07977}}].

\bibitem{Ivanov_2020}
M.M.~{Ivanov}, M.~{Simonovi{\'c}} and M.~{Zaldarriaga}, \emph{{Cosmological
  parameters from the BOSS galaxy power spectrum}},
  \href{https://doi.org/10.1088/1475-7516/2020/05/042}{\emph{\jcap} {\bfseries
  2020} (2020) 042} [\href{https://arxiv.org/abs/1909.05277}{{\ttfamily
  1909.05277}}].

\bibitem{Beutler14}
F.~{Beutler}, S.~{Saito}, H.-J.~{Seo}, J.~{Brinkmann}, K.S.~{Dawson},
  D.J.~{Eisenstein} et~al., \emph{{The clustering of galaxies in the SDSS-III
  Baryon Oscillation Spectroscopic Survey: testing gravity with redshift space
  distortions using the power spectrum multipoles}},
  \href{https://doi.org/10.1093/mnras/stu1051}{\emph{\mnras} {\bfseries 443}
  (2014) 1065} [\href{https://arxiv.org/abs/1312.4611}{{\ttfamily 1312.4611}}].

\bibitem{Chan12}
K.C.~{Chan}, R.~{Scoccimarro} and R.K.~{Sheth}, \emph{{Gravity and large-scale
  nonlocal bias}},
  \href{https://doi.org/10.1103/PhysRevD.85.083509}{\emph{\prd} {\bfseries 85}
  (2012) 083509} [\href{https://arxiv.org/abs/1201.3614}{{\ttfamily
  1201.3614}}].

\bibitem{anzu21}
N.~Kokron, J.~DeRose, S.-F.~Chen, M.~White and R.H.~Wechsler, \emph{The
  cosmology dependence of galaxy clustering and lensing from a hybrid
  n-body-perturbation theory model},
  \href{https://doi.org/10.1093/mnras/stab1358}{\emph{\mnras} {\bfseries 505}
  (2021) 1422}.

\bibitem{findiff}
M.~Baer, \emph{{findiff} software package},  2018.

\end{thebibliography}\endgroup
\bibliographystyle{jhep}
\end{document}